\documentclass[a4paper,10pt]{article}
\usepackage{graphicx}
\usepackage{amsmath}
\usepackage{citesort}
\usepackage[sort&compress,numbers]{natbib}
\usepackage{amsfonts}
\usepackage{amssymb}
\usepackage{amsmath}
\usepackage{latexsym}
\usepackage{color}
\usepackage{ntheorem}
\usepackage{braket}
\usepackage{url}

\newcommand{\coneg}{\check g}
\newcommand{\coneGamma}{\check \Gamma}
\newcommand{\conenabla}{\check \nabla}
\newcommand{\coneR}{\check R}

\newcommand{\mnote}[1]
{\protect{\stepcounter{mnotecount}}$^{\mbox{\footnotesize
$
\bullet$\themnotecount}}$ \marginpar{
\raggedright\tiny\em
$\!\!\!\!\!\!\,\bullet$\themnotecount: #1} }

\newcommand{\gamman}{h^{(n)}}

\newcommand{\ourdoteq}{\sim}

\newcommand{\Oinfty}{O}

\newcounter{mnotecount}[section]

\renewcommand{\themnotecount}{\thesection.\arabic{mnotecount}}

\newtheorem{theorem}{\sc  Theorem\rm}[section]

\newtheorem{definition}[theorem]{\sc  Definition\rm}

\newtheorem{lemma}[theorem]{\sc Lemma\rm}

\newtheorem{Proposition}[theorem]{\sc Proposition\rm}
\newtheorem{remark}[theorem]{\sc Remark\rm}
\newtheorem{Remark}[theorem]{\sc Remark\rm}

\newcommand{\ol}[1]{\overline{#1}{}}

\newcommand{\jlcax}[1]{}
\newcommand{\eean}{\nonumber\end{eqnarray}}

\newcommand{\kk}[1]{}

\newcommand{\mcH}{{\mycal H}}

\newcommand{\beq}{\begin{equation}}

\newcommand{\FS}       
                  {F}

\newcommand{\HS} 
       {H_{\mbox{\scriptsize volume}}}

{\ptc{this should be removed in the oberwolfach version}}%

\newcommand{\eeal}[1]{\label{#1}\end{eqnarray}}
\newcommand{\bed}{\begin{deqarr}}
\newcommand{\eed}{\end{deqarr}}
\newcommand{\bedl}[1]{\begin{deqarr}\label{#1}}
\newcommand{\eedl}[2]{\arrlabel{#1}\label{#2}\end{deqarr}}

\newcommand{\mcN}{{\mycal N}}

\newcommand{\bel}[1]{\begin{equation}\label{#1}}
\newcommand{\bea}{\begin{eqnarray}}
\newcommand{\bean}{\begin{eqnarray}\nonumber}
\newcommand{\beal}[1]{\begin{eqnarray}\label{#1}}
\newcommand{\eea}{\end{eqnarray}}

\newcommand{\be}{\begin{equation}}
\newcommand{\eeq}{\end{equation}}
\newcommand{\ee}{\end{equation}}
\newcommand{\beqa}{\begin{eqnarray}}
\newcommand{\eeqa}{\end{eqnarray}}
\newcommand{\beqan}{\begin{eqnarray*}}
\newcommand{\eeqan}{\end{eqnarray*}}
\newcommand{\ba}{\begin{array}}
\newcommand{\ea}{\end{array}}

\newcommand{\mcM}{{\mycal M}}

\newcommand{\scri}{{\mycal I}}%
\newcommand{\scrip}{\scri^{+}}%
\newcommand{\Scri}{\scri}

\newcommand{\warn}[1]
{\protect{\stepcounter{mnotecount}}$^{\mbox{\footnotesize
$
\bullet$\themnotecount}}$ \marginpar{
\raggedright\tiny\em
$\!\!\!\!\!\!\,\bullet$\themnotecount: {\bf Warning:} #1} }

\newcommand{\R}{\mathbb R}

\newcommand{\N}{\mathbb N}

\newcommand{\eq}[1]{(\ref{#1})}

\newcommand{\ptc}[1]{\mnote{{\bf ptc:}#1}}

\newcommand{\beqar}{\begin{deqarr}}
\newcommand{\eeqar}{\end{deqarr}}

\newcommand{\beaa}{\begin{eqnarray*}}
\newcommand{\eeaa}{\end{eqnarray*}}

\DeclareFontFamily{OT1}{rsfs}{}
\DeclareFontShape{OT1}{rsfs}{m}{n}{ <-7> rsfs5 <7-10> rsfs7 <10-> rsfs10}{}
\DeclareMathAlphabet{\mycal}{OT1}{rsfs}{m}{n}

{\catcode `\@=11 \global\let\AddToReset=\@addtoreset}
\AddToReset{equation}{section}

{\catcode `\@=11 \global\let\AddToReset=\@addtoreset}
\AddToReset{figure}{section}

{\catcode `\@=11 \global\let\AddToReset=\@addtoreset}
\AddToReset{table}{section}

\begin{document}

\title{Characteristic initial data and smoothness of Scri.\\ II. Asymptotic expansions and construction of conformally smooth data sets%
\thanks{Preprint UWThPh-2014-9.  
}}
\author{
Tim-Torben Paetz{}\thanks{Email  Tim-Torben.Paetz@univie.ac.at}   
 \vspace{0.5em}\\  \textit{Gravitational Physics, University of Vienna}  \\ \textit{Boltzmanngasse 5, 1090 Vienna, Austria }}
\maketitle

\vspace{-0.2em}

\begin{abstract}
We derive necessary-and-sufficient conditions on characteristic
 initial data for Friedrich's conformal field equations
in $3+1$ dimensions to have no logarithmic terms in an asymptotic expansion at  null infinity.
\end{abstract}

\noindent
\hspace{2.1em} PACS: 02.30.Hq, 02.30.Mv, 04.20.Ex, 04.20.Ha

\tableofcontents

\section{Introduction}

In this work we continue the work initiated in \cite{ChPaetz2} to analyze the occurrence of logarithmic terms in the asymptotic expansion
of the metric tensor and some other fields at null infinity.
In part I of this work, where we also described the setting,  it has been shown that the harmonic coordinate condition is not compatible with a smooth asymptotic structure at the conformal boundary at infinity, but has to be replaced by a wave-map gauge condition with non-vanishing gauge-source functions.

However, it is expected \cite{bondi,TorrenceCouch} (compare also \cite{ChoquetBruhat73})
that  even for smooth initial data the asymptotic expansion of the space-time metric at null infinity will generically be polyhomogeneous and involve logarithmic terms which do not have their origin in an inconvenient choice of coordinates.
One main object of this note, treated in  Section~\ref{solutions_constraints},  is to study  thoroughly  the asymptotic behavior of solutions of the  Einstein's vacuum constraint equations and analyze
under which conditions a smooth conformal completion of the \emph{restriction
of the space-time metric to the characteristic initial surface} across null infinity is possible.
As announced in \cite[Theorem~5.1]{ChPaetz2} we intend to provide necessary-and-sufficient conditions on the initial data and the gauge source functions
which permit such extensions.
In doing so it will become manifest that many, though not all, of the logarithmic terms which arise at infinity
are gauge artifacts.
The remaining non-gauge logarithmic terms can be eliminated by imposing  restrictions on the
asymptotic behavior of the initial data, captured by what we call  \emph{no-logs-condition}.

In Section~\ref{sec_sufficient},
we will show that solutions of the characteristic vacuum constraint equations satisfying the no-logs-condition lead to
smooth initial data for Friedrich's conformal field equations.
The data will be computed in a new
gauge scheme developed in Section~\ref{sec_mgauge}
and  will provide the basis  to solve the evolution problem and construct
 space-times with a ``piece of smooth~$\scri^+$''.

In Section~\ref{sec_preliminaries} we give a summary of \cite{ChPaetz2} where we briefly describe the framework and recall the most important definitions and results of part I.
Finally, in Appendix~\ref{asymptotic_solutions} our proceeding in Section~\ref{solutions_constraints}-\ref{sec_sufficient}
to solve the constraint equations in terms of polyhomogeneous expansions will be rigorously justified, while in Appendix~\ref{meaning_kappa}
we  compare the peculiarities of  different gauge schemes.

\section{Preliminaries}
\label{sec_preliminaries}

We use all the notation, terminology and conventions introduced in part I  \cite{ChPaetz2}.
For the convenience of the reader, though,  let us briefly recall the most essential ingredients and definitions of our framework.

\subsection{Notation}
\label{sec_notation}

Consider a smooth function
\begin{equation*}
 f : (0,\infty)\times S^{2} \longrightarrow \mathbb{R} \;, \quad (r,x^A) \longmapsto f(r,x^A)
 \;.
\end{equation*}
If this function permits an asymptotic expansion  as a power series in $r$,
we denote by $f_n$, or $(f)_n$, the coefficient of $r^{-n}$ in the corresponding expansion.
Be aware that sometimes a lower index might denote both the component of a vector, and the $n$-th order term in an expansion of the corresponding object.
If both indices need to appear simultaneously we use brackets and place the index corresponding to the $n$-th order expansion term outside the brackets.
We write
 \[   f(r,x^A) \ourdoteq
   \sum_{k=-N}^{\infty} f_k(x^A) r^{-k} \]
if the right-hand side  
 is the polyhomogeneous expansion at $x=0$ of the function $x\mapsto r^{-N} f(r, \cdot)|_{r=1/x}$.
 Moreover, we write $f=\mathcal{O}(r^N)$ (or $f=\mathcal{O}(x^{-N})$, $x\equiv1/r$), $N\in\mathbb{Z}$,   if
the function $x\mapsto r^{-N} f(r, \cdot)|_{r=1/x}$ is smooth at $x=0$.

\subsection{Setting}

We consider a $3+1$-dimensional $C^{\infty}$-manifold $\mcM $.
For definiteness we take~as~initial surface either a globally smooth light-cone $C_O\subset\mcM$ or two null hypersurfaces $\mcN_1,\,\mcN_2\subset\mcM$ intersecting transversally along a smooth submanifold $S\cong S^2$.
Suppose that the closure (in the completed
space-time) $\ol \mcN$ of $\mcN\in\{C_O\setminus\{O\},\mcN_1\}$ meets $\scrip$ transversally   in a smooth  spherical cross-section.

We introduce  
 \emph{adapted null coordinates} $(u\equiv x^0, r\equiv x^1,x^A)$ on $\mcN$
(i.e.\ $\mcN=\{u=0\}$, where $r$ parameterizes the null rays generating $\mcN$, and $(x^A)$ are local coordinates on $\Sigma_r\equiv \{r=\mathrm{const},u=0\}\cong S^2$).
Then the trace $\ol g$ of the metric on $\mcN$ becomes
\begin{equation}
 \overline g = \overline g_{00}\mathrm{d}u^2 + 2\nu_0\mathrm{d}u\mathrm{d}r + 2\nu_A\mathrm{d}u\mathrm{d}x^A + \coneg
 \;,
 \label{null2}
\end{equation}
where
\begin{equation}
 \coneg = \coneg_{AB}\mathrm{d}x^A\mathrm{d}x^B := \overline g_{AB}\mathrm{d}x^A\mathrm{d}x^B
\end{equation}
is a degenerate quadratic form induced by $g$ on $\mcN$ which induces on each slice $\Sigma_r$ an $r$-dependent Riemannian metric  $\coneg_{\Sigma_r}$ (coinciding with $\coneg (r,\cdot)$ in the coordinates above).
While the components $\overline g_{00}$, $\nu_0$ and $\nu_A$ depend upon the choice of coordinates off $\mcN$,
the quadratic form $\coneg$ is intrinsically defined on $\mcN$.

In fact, we will be interested merely in the asymptotic behavior of the restriction of the space-time metric to $\mcN$.
A regular light-cone or two transversally intersecting null hypersurfaces with appropriately specified initial data, though,  guarantee
that the  vacuum constraint equations have unique solutions.

Throughout this work
we use an overline  to denote a space-time object restricted to~$\mcN$.
The symbol ``$\,\check\enspace\,$" will be used to denote objects associated with
the Riemannian metric $\coneg_{\Sigma_r}$.

\begin{definition}[cf.\ \cite{ChPaetz2}]
\label{definition_smooth}
\rm{
We say that a smooth metric tensor $\overline g_{\mu\nu}$ defined on a null hypersurface $\mcN$
given in adapted null coordinates  has a
\textit{smooth conformal completion at infinity}
if
the unphysical metric tensor $\overline{\tilde g}_{\mu\nu}$
obtained via the coordinate transformation $r\mapsto 1/r=: x$ and the conformal rescaling $\overline g\mapsto  \overline{\tilde g} \equiv x^2 \overline g$
is, as a Lorentzian metric, smoothly extendable at $\{x=0\}$.

The components of a smooth tensor field on  $\mcN$  will be said to be \textit{smooth at infinity} whenever they admit a smooth expansion in the conformally rescaled space-time at $\{x=0\}$ and expressed in the $(x,x^A)$-coordinates.
}
\end{definition}

As remarked in \cite{ChPaetz2},  Definition~\ref{definition_smooth} concerns only fields on $\mcN$ and is not tied to the existence of an associated space-time.
Moreover, it  concerns  both conditions on the metric and on the coordinate system.

The Einstein equations split into a set of evolution equations  and a set of constraint equations which need to be satisfied on the initial surface.
In the characteristic case the data for the evolution equations are provided by the trace of the metric on the initial surface.
Due to the constraints not all of its components can be prescribed freely.
There are various ways of choosing free  data \cite{ChPaetz,ChPaetz2}.
Here we focus on Rendall's scheme \cite{Rendall}, where the free data are formed by the conformal class $[\gamma]$  of the
tensor $\coneg_{\Sigma_r}$ (and the function $\kappa$ introduced below).%
\footnote{
In the case of two transversally intersecting null hypersurfaces these data need to be supplemented by corresponding data on $\mcN_2$ and
certain data on the intersection manifold $S$.
}
By choosing a representative they can be viewed as a one-parameter family, parameterized by $r$, of Riemannian metrics  $\gamma(r,\cdot)$ on $S^{2}$.
A major advantage of  this scheme in particular in view of Section~\ref{sec_mgauge}
 is that it permits a separation of physical and gauge degrees of freedom.
Some comments on how things change for other  approaches to prescribe characteristic initial data are given in \cite{ChPaetz2}, cf.\ Remark~\ref{sec_other_settings}.

\emph{Einstein's vacuum constraint equations in a $\hat g$-generalized wave-map gauge}~are obtained
from Einstein's vacuum  equations assuming that the \emph{wave-gauge vector}
\begin{equation}
 H^{\lambda} :=\Gamma^{\lambda}-\hat \Gamma^{\lambda} - W^{\lambda}=0
\;, \quad
\Gamma^{\lambda}:= g^{\alpha\beta}\Gamma^{\lambda}_{\alpha\beta}\;, \quad
\hat \Gamma^{\lambda}:= g^{\alpha\beta}\hat \Gamma^{\lambda}_{\alpha\beta}
\;.
\end{equation}
vanishes.
We use the hat-symbol ``$\,\hat\enspace\,$" to indicate objects associated with some \textit{target metric $\hat g$},
which we assume for convenience to be of the  form
%
\begin{eqnarray}
& \ol{\hat g}_{1i}=0\;, \quad \ol{\hat g}_{AB} = r^2s_{AB} + \mathcal{O}(1)  \;, \quad \hat\nu_0 = 1 + \mathcal{O}(r^{-3})\;,\quad \hat\nu_A = \mathcal{O}(r^{-2})\;, &
\nonumber
\\
&  \ol{\hat g}_{00} = -1 + \mathcal{O}(r^{-2})\;, \quad \ol{\partial_0 \hat g_{1i}} = \mathcal{O}(r^{-3})\;, \quad \ol{\partial_0 \hat g_{AB}} = \mathcal{O}(r^{-1})
&
 \label{Minktarget_asympt}
\end{eqnarray}
on $\mcN$, where $s=s_{AB}\mathrm{d}x^A\mathrm{d}x^B$ is the  unit round metric on the sphere $S^{2}$.
By $W^{\lambda}=W^{\lambda}(x^{\mu},g_{\mu\nu})$  we denote  the components of a vector field, the \emph{gauge source functions}, which can be
arbitrarily prescribed.
They reflect the freedom to choose coordinates off the initial surface, and thus
allow us to analyze smoothness of the metric tensor at infinity in arbitrary coordinates.

For given initial data $\gamma=\gamma_{AB}\mathrm{d}x^A\mathrm{d}x^B$ the wave-map gauge constraints form a hierarchical system of ODEs along the null generators of the cone (cf.\ \cite{CCM2}):
%
\begin{eqnarray}
 \partial^2_{rr}\varphi -\kappa\partial_r\varphi + \frac{1}{2}|\sigma|^2\varphi &=&0
 \;,
 \label{constraint_phi}
\\
( \partial_r + \frac{1}{2}\tau + \kappa)\nu^0 + \frac{1}{2}(\overline W{}^0+  \ol {\hat\Gamma}^0)  &=& 0
 \;,
 \label{constraint_nu0}
\\
  (\partial_r + \tau)\xi_A  - 2\conenabla_B \sigma_A^{\phantom{A}B} +  \partial_A\tau + 2\partial_A \kappa
&=&0
 \;,
 \label{eqn_nuA_general}
\\
  2\nu^0(\partial_r\nu_A - 2\nu_B\chi_A{}^B) - \nu_A(\overline W{}^0+ \ol {\hat\Gamma}^0) - \overline g_{AB}( \overline W{}^B
+ \ol {\hat\Gamma}^B)
&&
 \nonumber
\\
 + \gamma_{AB} \gamma^{CD} \coneGamma{}^B_{CD} +  \xi_A  &=& 0
  \;,
 \label{eqn_xiA}
\\
 (\partial_r + \tau + \kappa)\zeta +  \coneR - \frac{1}{2}\xi_A\xi^A +\conenabla_A\xi^A &=&0
 \;,
 \label{zeta_constraint}
\\
  (2\partial_r + \tau + 2\kappa)\overline g^{rr} + 2\overline W{}^r + 2 \ol {\hat\Gamma}^r - \zeta &=& 0
 \;,
 \label{dfn_zeta}
\end{eqnarray}
where $\tau$ and $\sigma_A{}^B$ denote the expansion and the shear of $\mcN$, respectively,
\begin{eqnarray}
 \kappa \,=\, \ol\Gamma^r_{rr}\;,
\quad
\xi_A \,=\, -2\ol\Gamma^r_{rA}\;,
\quad \zeta \,=\, 2\ol g^{AB}\ol\Gamma^r_{AB} + \tau \ol g^{rr}
\:.
\end{eqnarray}
Apart from $W^{\lambda}$ the function $\kappa$ turns out to be another gauge degree of freedom, reflecting
the freedom to choose $r$, which can be prescribed
conveniently.

Integrating these equations successively one determines all the components of $\overline g$ (note that $\nu_0=(\nu^0)^{-1}$ and $\ol g_{00}  =\ol g^{AB}\nu_A\nu_B - (\nu_0)^2\ol  g^{rr}$).
The relevant boundary conditions follow either from regularity conditions at the vertex \cite{CCM2} when $\mcN$ is a cone, or from the
remaining data specified on $\mcN_2$ and $S$ (cf.\ e.g.\ \cite{ChPaetz, Rendall}) in the case of two characteristic surfaces intersecting transversally.
%
%

Our ultimate goal is to find necessary-and-sufficient conditions on the initial data given on $\mcN$, such that the resulting space-time has a smooth conformal completion at infinity \emph{\`a la Penrose}.
This requires the following 
ingredients:
\begin{enumerate}
\item To exclude the appearance of conjugate points or coordinate singularities on $\mcN$,
the constraint equations need to admit a non-degenerated global solution $\ol g$ on  $\mcN$. This will be the case if and only if  the functions $\varphi$ and $\nu^0$ are of constant sign, 
\begin{equation}
 \varphi\ne 0\;, \quad \nu^0\ne 0 \quad \text{on $\mcN$}
\label{in_condition1}
\;.
\end{equation}
\item The metric $\ol g$ needs to be smoothly extendable as a Lorentzian metric, which means that the functions $\varphi_{-1}$ and $(\nu^0)_0$
need to have a sign 
%
\begin{equation}
 \varphi_{-1}\ne 0\;, \quad (\nu^0)_0\ne 0 \quad \text{on $S^2$}
\label{in_condition2}
\;.
\end{equation}
This assumption excludes conjugate points at the intersection of $\mcN$ with~$\scrip$.
\item The components of $\ol g$ need to be smooth  at $\scri^+$. For this one has to make sure that their asymptotic expansions
contain no logarithmic terms and have the correct order in terms of powers of $r$.
\item All  the fields which appear in Friedrich's conformal field equations (which provide an evolution system which, in contrast to Einstein's field equations, is
well-behaved at $\scrip$) need to be  smooth at $\scri^+$.
\item Finally, an appropriate well-posedness result for the conformal field equations is needed.
\end{enumerate}
Point 1 and 2 have been addressed in \cite{ChPaetz2}, cf.\ Proposition~\ref{P6XII13.1} below.
Point 3 will be the subject of Section~\ref{solutions_constraints}, while point 4 will be investigated in Section~\ref{sec_sufficient}. Point 5 will be addressed elsewhere.

From now on we shall consider exclusively conformal data $\gamma_{AB}(r,\cdot)dx^A dx^B$ and gauge  functions $\kappa$ and $\ol W^{\lambda}$ for which \eq{in_condition1} and \eq{in_condition2} hold.
Let us summarize some of the results established in \cite{ChPaetz2} (adapted to the smooth setting on which we focus here)
which
provide sufficient conditions
such that \eq{in_condition1} and \eq{in_condition2} hold  in the case where $\mcN$ represents a regular light-cone $C_O$:%
\footnote{
In the conventions of \cite{ChPaetz2} the functions $\varphi$, $\nu^0$, $\varphi_{-1}$ and $(\nu^0)_0$ need to be positive.
}
\begin{Proposition}
 \label{P6XII13.1}
\begin{enumerate}
 \item Solutions of the Raychaudhuri equation \eq{constraint_phi} with prescribed $\kappa=\mathcal{O}(r^{-3})$ and $\sigma_A{}^B=\mathcal{O}(r^{-2})$
 lead to a globally positive $\varphi$ on $C_O\setminus\{O\}$ with $\varphi_{-1}>0$ on $S^2$ if
\begin{eqnarray*}
 \int_0^{\infty}\Big( \int_0^re^{H(\hat r)}\mathrm{d}\hat r \Big) e^{-H(r)}   |\sigma|^2(r) \,\mathrm{d} r < 2 \;, \enspace \text{where} \enspace
H(r,x^A) := \int_0 ^r  {\kappa(\tilde r,x^A)}   d\tilde r
\;.
\end{eqnarray*}
 \item Assuming a Minkowski target and
\begin{equation*}
\ol W^{0}=\mathcal{O}(r^{-1}) \enspace \text{with} \enspace \ol W{}^0< r\varphi^{-2}\sqrt{\frac{\det\gamma}{\det s} }\gamma^{AB} s_{AB}
\enspace \text{and} \enspace  (\ol W{}^0)_1< 2(\varphi_{-1})^{-2}
\;,
\end{equation*}
 any positive solution $\varphi$  of \eq{constraint_phi}  leads to a globally defined positive  function $\nu_0$ on $C_O$ with $0<(\nu_0)_0<\infty$.
\end{enumerate}
\end{Proposition}

\subsection{A priori restrictions}
\label{sec_a_priori}

Before  we analyze thoroughly the asymptotic behavior of
the vacuum Einstein constraint equations
and derive necessary-and-sufficient conditions concerning smoothness of the solutions at infinity
it is convenient to have some a priori knowledge regarding the lowest admissible orders
 of the gauge functions,
and to exclude the appearance of log terms in the expansion of ``auxiliary'' fields such as $\xi_A=\ol \Gamma^r_{rA}$.
In \cite{ChPaetz2} we have shown that the following equations need to be necessarily fulfilled in some adapted null coordinate system
to end up with a trace of a metric on $\mcN$ which admits a smooth conformal completion and infinity, and connection coefficients
which are smooth  at $\scri^+$:
%
%
\begin{equation}
 \kappa= \mathcal{O}(r^{-3})\;, \quad \ol W^0 = \mathcal{O}(r^{-1})\;,\quad \xi_A = \mathcal{O}(r^{-1})\;, \quad \ol W^A =  \mathcal{O}(r^{-1})
\;.
\end{equation}
Moreover,  we may assume the initial data to be of the asymptotic form
\begin{equation}
 \gamma_{AB} \ourdoteq  r^2 \Big( s_{AB}+ \sum_{n=1}^{\infty} \gamman _{AB} r^{-n}\Big)
 \;,
 \label{initial_data}
\end{equation}
for some smooth tensor fields $ \gamman _{AB}$ on $S^2$.
If  $\gamma$ is not  of the form \eq{initial_data}, it can either be brought to  it  via a conformal rescaling and an suitable choice of $r$,~or it leads to a metric $\ol g$ which does not have a smooth conformal completion at~$\scrip$.

At this stage we do not know whether a space-time which admits a conformal completion at infinity is compatible
with polyhomogeneous rather than smooth expansions of the functions $\zeta$ and $\ol W{}^r$.
It will turn out that this is not the case.
However, we note that it follows  from the constraint equations and the above a priori restrictions that
\begin{equation}
\mbox{if $\zeta=\mathcal{O}(r^{-1})$ then $\ol W{}^r = \mathcal O(r)$.}
\label{a_priori_zeta}
\end{equation}

\section{Asymptotic solutions of Einstein's characteristic vacuum constraint equations}
\label{solutions_constraints}


It is useful to  introduce some notation first:
 Let $w_{AB}$ be a rank-2-tensor on the initial surface $\mcN$. We denote by $\breve w_{AB}$, or $(w_{AB})\breve{}$, its trace-free part
w.r.t.\ the metric $\ol g_{AB}\mathrm{d}x^A\mathrm{d}x^B$.
Consider now the expansion coefficients $(w_{AB})_n$ at infinity, which are tensors on $S^2$. We denote by $(\breve w_{AB})_n$, or $(w_{AB})\breve{}_n$,  the trace-free part
w.r.t.\ the unit round metric $s=s_{AB}\mathrm{d}x^A\mathrm{d}x^B$.
Finally, we set
\begin{equation*}
|w|^2:=\ol g^{AC}\ol g^{BD}w_{AB} w_{CD}\;, \quad \text{and} \quad
|w_n|^2:= s^{AC}s^{BD} (w_{AB})_n (w_{CD})_n\;.
\end{equation*}

Let us make the convention
to raise indices of the expansion coefficients $h^{(n)}_{AB}$  with the  standard metric. Moreover, we set
\begin{eqnarray*}
 \gamman  &:=&
s^{AB}\gamman _{AB}
 \;.
\end{eqnarray*}
A ring $\mathring{}$ on a covariant derivative operator or a connection coefficient indicates  that the corresponding object is associated  to
$s=s_{AB}\mathrm{d}x^A\mathrm{d}x^B$.

\subsection{Asymptotic solutions of the constraint equations}

The object of this section is twofold:
First of all we will show that the Einstein wave-map gauge constraints  \eq{constraint_phi}-\eq{dfn_zeta} can be solved asymptotically in terms of polyhomogeneous expansions at infinity of the  solution.
This is done by rewriting the equations in a form to which Appendix~\ref{asymptotic_solutions}
applies.
The second aim is to make some general considerations concerning the appearance of logarithmic terms in the asymptotic solutions of  \eq{constraint_phi}-\eq{dfn_zeta} for initial data
of the form (\ref{initial_data}). We want to extract necessary-and-sufficient conditions
leading to the trace $\overline g$  of a metric on $\mcN$ which admits a smooth conformal
completion at infinity in the sense of Definition~\ref{definition_smooth}.

Our starting point are initial data $\gamma$ with  an asymptotic behavior of the form \eq{initial_data} and gauge functions
\begin{equation}
\kappa=\mathcal{O}(r^{-3})\;, \quad  \ol W{}^{0}=\mathcal{O}(r^{-1})\;, \quad  \ol W{}^{A}=\mathcal{O}(r^{-1})\;, \quad \ol W{}^{r}=\mathcal{O}(r)
\;,
\label{asymp_gauge_source_functions}
\end{equation}
for which  $\varphi$, $\nu^0$, $\varphi_{-1}$ and $(\nu^0)_0$ have a sign on $\mcN$ and $S^2$,
respectively.
We~further require that the asymptotic expansion of $\xi_A$ contains no logarithmic terms, i.e.\ $\xi_A=\mathcal{O}(r^{-1})$.
A violation of one of these assumptions would not be compatible with a space-time which admits a smooth conformal completion at infinity
as follows from the a priori restriction and the following fact:
The considerations below reveal that \eq{initial_data}, $\kappa=\mathcal{O}(r^{-3})$ and $\xi_A=\mathcal{O}(r^{-1})$ \emph{imply}
 $\zeta=\mathcal{O}(r^{-1})$, and that \eq{a_priori_zeta} applies.
So we are imposing no restrictions when assuming $\ol W{}^{r}=\mathcal{O}(r)$.

Consider the shear tensor,
\begin{equation}
 \sigma_A{}^B\,=\,  \frac{1}{2}\overline g^{BC}(\partial_r\overline g_{AC})\breve{}
 \,=\, \frac{1}{2}\gamma^{BC}(\partial_r\gamma_{AC})\breve{}
 \;,
 \label{formula_sigma}
\end{equation}
 whose asymptotic expansion we express in terms  of the expansion coefficients of the initial data $\gamma$
\begin{eqnarray}
 \sigma_A{}^B &=&- \frac{1}{2}\breve h_A^{(1)B}  r^{-2}
 +\Big(\frac{1}{2}h^{(1)} \breve h_A^{(1)B} - \breve h_A^{(2)B}  \Big) r^{-3}
  + \mathcal{O}(r^{-4})
 \;,
 \label{expansion_sigma_A^B}
\\
 |\sigma|^2 
&=&  \frac{1}{4}|\breve h^{(1)}|^2  r^{-4}
+\Big( \breve h_A^{(1)B}  \breve h_B^{(2)A}  - \frac{1}{2}h^{(1)}|\breve h^{(1)}|^2  \Big) r^{-5}
  + \mathcal{O}(r^{-6})
\;.
 \label{expansion_sigma}
\end{eqnarray}

A global solution, and thereby also the value of the ``asymptotic integration functions'', to each of the constraint ODEs is determined by regularity conditions at the
vertex $O$ of a light-cone, or by the data on the intersection manifold $S$ for two intersecting characteristic surfaces.
However,  the integration functions, which
depend on the initial data $\gamma$, the gauge functions  and the  boundary conditions at the vertex or the intersection manifold, appear difficult to control.

\subsubsection{Expansion of $\varphi$}
\label{sec_expan_conf}

We start with the constraint equation (\ref{constraint_phi}) for the function $\varphi$,
\begin{equation}
 \partial_{rr}^2\varphi - \kappa\partial_r\varphi + \frac{1}{2}|\sigma|^2\varphi = 0
 \;.
 \label{eqn_varphi}
\end{equation}
In order to enable an easier comparison to Appendix~\ref{asymptotic_solutions} we make the transformation $r\mapsto 1/r\equiv x$, with $\varphi$, $\kappa$ and $|\sigma|^2$ treated as scalars,
and rewrite the ODE as a first-order system.
The equation then reads with $\varphi^{(1)}:=\varphi$ and $\varphi^{(2)}:=x\partial_x\varphi$,
\begin{eqnarray*}
 \left[  x\partial_x+ \begin{pmatrix} 0 & -1 \\ \frac{|\sigma|^2}{2x} & 1+\frac{\kappa}{x} \end{pmatrix}\right] \begin{pmatrix}\varphi^{(1)} \\  \varphi^{(2)} \end{pmatrix} =0
 \;,
\end{eqnarray*}
or, when the leading order term is diagonalized, 
\begin{eqnarray*}
 \Bigg[ x\partial_x + \begin{pmatrix} 1 & 0 \\
 0 & 0 \end{pmatrix}
+ \underbrace{\begin{pmatrix} \frac{\kappa}{x}-\frac{|\sigma|^2}{2x} & -\frac{|\sigma|^2}{\sqrt{2}x} \\
 \frac{|\sigma|^2}{2^{3/2}x}-\frac{\kappa}{\sqrt{2}x} & \frac{|\sigma|^2}{2x} \end{pmatrix}}_{=\mathcal{O}(x)}\Bigg] \begin{pmatrix}\tilde\varphi^{(1)} \\  \tilde\varphi^{(2)} \end{pmatrix}
=0
 \;,
\end{eqnarray*}
with
\begin{eqnarray*}
 \begin{pmatrix} \tilde\varphi^{(1)} \\  \tilde\varphi^{(2)} \end{pmatrix} :=  \begin{pmatrix} 0 & -\sqrt{2} \\ 1 & 1 \end{pmatrix} \begin{pmatrix}\varphi^{(1)}\\  \varphi^{(2)} \end{pmatrix}
 \;.
\end{eqnarray*}
The results  in Appendix~\ref{asymptotic_solutions} (we have, in the notation used there, $\lambda_1=-1$, $\lambda_2=0$, and, since there is no source, $\hat\ell=-1$)
shows that this ODE can be solved asymptotically via a polyhomogeneous expansion
with $\tilde{\varphi}^{(n)} =O (x^{-1})$.
It also reveals that the coefficients $(\tilde\varphi^{(n)})_{\lambda_n}$, $n=1,2$, i.e. $(\tilde\varphi^{(1)})_{-1}$ and $(\tilde\varphi^{(2)})_{0}$,
can be regarded as integrations functions, and that
logarithmic terms do not appear
if and only if
(cf. condition \eq{no_log_terms_diagsystem})
\begin{eqnarray}
 \Big[ \Big(\frac{|\sigma|^2}{2^{3/2}x}-\frac{\kappa}{\sqrt{2}x}\Big) \tilde\varphi^{(1)} - \frac{|\sigma|^2}{2x} \tilde\varphi^{(2)}\Big]_0 = 0
 \;.
\label{cond_kappa_phi}
\end{eqnarray}
Recall that $\cdot\,_n$ denotes the $r^{-n}$-term ($x^n$-term) in the asymptotic expansion of the
corresponding field.
Using $\kappa=\mathcal{O}(r^{-3})$ and $|\sigma|^2=\mathcal{O}(r^{-4})$ we observe that \eq{cond_kappa_phi} holds automatically.
%
%
In particular we  have $\varphi=\mathcal{O}(r)$. 
Furthermore, since
\begin{eqnarray*}
 (\tilde\varphi^{(1)})_{-1} &=& -\sqrt{2} (\varphi^{(2)})_{-1} = \sqrt{2}\varphi_{-1}
 \;,
\\
 (\tilde\varphi^{(2)})_{0} &=& (\varphi^{(1)})_0 + (\varphi^{(2)})_0 = \varphi_0
 \;,
\end{eqnarray*}
the coefficients $\varphi_{-1}$ and $\varphi_0$ can be identified as  the  integration functions. As indicated above, though being left undetermined by the equation itself, they have global character.


In the following we shall set for convenience
\begin{equation}
 \sigma_n:=(|\sigma|^2)_n
\;.
\end{equation}
%
Inserting the expansion $\varphi\ourdoteq  \sum_{n=\hat \ell}^{\infty}\varphi_nr^{-n}$ into (\ref{eqn_varphi}) and equating coefficients gives the expansion coefficients $\varphi_n$ by a hierarchy of equations,
\begin{eqnarray}
 \varphi_1 &=& \Big(\frac{1}{2}\kappa_3 - \frac{1}{4}\sigma_4\Big) \varphi_{-1}
 \;,
\\
 \varphi_2 &=& \Big(\frac{1}{6}\kappa_4 - \frac{1}{12}\sigma_5\Big)\varphi_{-1} - \frac{1}{12}\sigma_4\varphi_0
 \;,
 \label{expansion_coeff_phi}
\end{eqnarray}
while
\begin{eqnarray}
 \tau \,\equiv\, 2\partial_r\log\varphi &=& 2r^{-1} -2\varphi_0 (\varphi_{-1})^{-1} r^{-2}
 \nonumber
\\
 &&+ [2(\varphi_0)^2(\varphi_{-1})^{-2} +\sigma_4 - 2\kappa_3 ]r^{-3} + \mathcal{O}(r^{-4})
 \;.
 \label{expansion_tau}
\end{eqnarray}

Consider  the conformal factor relating the $r$-dependent Riemannian metric $\check g_{\Sigma_r}$ and $\gamma$, $\overline g_{AB}\mathrm{d}x^A\mathrm{d}x^B= \Omega^2 \gamma_{AB}\mathrm{d}x^A\mathrm{d}x^B$.
We find
%
\begin{eqnarray}
 \det\gamma 
 &=& \det s\Big[ r^4 + h^{(1)}r^3 + \Big(h^{(2)} + \frac{1}{2}(h^{(1)})^2 - \frac{1}{2}|h^{(1)}|^2\Big)r^2 \Big] +\mathcal{O}(r)
\;,
\\
 \Omega &\equiv& \varphi \Big( \frac{\det s}{\det\gamma}\Big)^{1/4} = \varphi_{-1} -\frac{1}{4} \varphi_{-1}\Big(2\tau_2 + h^{(1)}\Big)r^{-1}
\\
 && + \frac{1}{4} \varphi_{-1}\Big[ 2 \kappa_3 + \frac{1}{4}(h^{(1)})^2 + \frac{1}{4}|h^{(1)}|^2 -  h^{(2)}+ \frac{1}{2}\tau_2 h^{(1)}  \Big] r^{-2}
 + \mathcal{O}(r^{-3})
 \;.
\nonumber
\end{eqnarray}
%
%
We conclude that
\begin{eqnarray}
 \overline g_{AB} &=& (\varphi_{-1})^2\Big[ s_{AB}r^2 + (\breve h^{(1)}_{AB} -\tau _2s_{AB}  )r
\nonumber
\\
 && \hspace{-3em}+ \breve h^{(2)}_{AB} -(\tau_2+   \frac{1}{2}h^{(1)}) \breve h^{(1)}_{AB}  + [\frac{1}{4}(\tau_2)^2 + \kappa_3 +  \frac{1}{2}\sigma_4] s_{AB}\Big] + \mathcal{O}(r^{-1})
 \;.
\label{expansion_gAB}
\end{eqnarray}
%
To sum it up, \eq{initial_data} and $\kappa=\mathcal{O}(r^{-3})$ imply that no logarithmic terms appear in the conformal factor relating $\gamma_{AB}$ and $\overline g_{AB}$,
the latter one thus being smoothly extendable at $\scri^+$ as a Riemannian metric on $S^2$ whenever a global solution of the Raychaudhuri equation
exists on $\mcN$ with $\varphi_{-1}\ne 0$.

\subsubsection{Expansion of $\nu^0$}

We consider the constraint equation (\ref{constraint_nu0}) which determines $\nu^0$,
\begin{eqnarray}
 \partial_r\nu^0 + \nu^0(\frac{1}{2}\tau + \kappa ) + \frac{1}{2}(\overline W{}^0+  \ol {\hat\Gamma}^0)  &=& 0
 \;,
 \label{eqn_hatphi}
\end{eqnarray}
where $\overline W{}^0=\mathcal{O}(r^{-1})$  and, using \eq{Minktarget_asympt},
\begin{eqnarray}
\hspace{-1em}
 \ol {\hat\Gamma}{}^0&=& 2\nu^0(\hat\nu^0\partial_r\hat\nu_0-\hat\kappa) - \hat\nu^0 \Omega^{-2} \gamma^{AB} \hat\chi_{AB}
\nonumber
\\
&=&  \mathcal{O}(r^{-3})\nu^0 -2(\varphi_{-1})^{-2}r^{-1} -2(\varphi_{-1})^{-2}   \tau_2 r^{-2}  +  \mathcal{O}(r^{-3})
 \;.
\end{eqnarray}
Again, we express the ODE by  $x\equiv 1/r$. Its asymptotic form reads
%
%
\begin{eqnarray}
&& \hspace{-4em} [x\partial_x - 1 - \frac{\tau_2}{2}x +\mathcal{O}(x^{2}) ]\nu^0
\nonumber
\\
&=& \underbrace{\frac{1}{2} (\ol W{}^0)_1  - (\varphi_{-1})^{-2} }_{=: -\Phi^{-2}}
+ [\frac{1}{2}(\ol W{}^0)_2
- (\varphi_{-1})^{-2}   \tau_2 ]x + \mathcal{O}(x^2)
 \;.
\end{eqnarray}
We want to make sure that the  asymptotic solution of $\nu^0$ can be written as power series.
In the notation of Appendix~\ref{asymptotic_solutions} we have $ \lambda =1$ and $  \hat\ell=0$,
%
and condition (\ref{cond_no_logs}) which characterizes the absence of logarithmic terms reads
\begin{eqnarray}
 (\overline W{}^0)_2
&=& \Big[\frac{1}{2} (\overline W{}^0)_1 + (\varphi_{-1})^{-2}\Big]\tau_2
 \;.
 \label{0gaugecond}
 \label{cond_Y0}
\end{eqnarray}
Assuming \eq{cond_Y0} we
insert the expansion $\nu^0\sim \sum_{n=\hat\ell}^{\infty} (\nu^0)_n r^{-n}$  into (\ref{eqn_hatphi})
to obtain the expansion coefficients  $(\nu^0)_n$ in terms of $\kappa$, $\overline W{}^0$, $\varphi$ and $\gamma$,
%
\begin{eqnarray}
 \nu^0  &=&\Phi^{-2} - D^{(\nu_0)}\Phi^{-4}r^{-1} + \mathcal{O}(r^{-2})
 \;,
 \label{expansion_nu0}
\end{eqnarray}
where $D^{(\nu_0)}$ denotes the  globally defined integration function.

As mentioned above, for $\ol g_{\mu\nu}$ to have a smooth conformal completion at infinity
the function $(\nu^0)_0$ needs to be of constant sign
which is the case if and only if the gauge source function
is chosen so that (compare Proposition~\ref{P6XII13.1})
\begin{equation}
 (\overline W{}^0)_1 \ne 2(\varphi_{-1})^{-2} \enspace \forall\,x^A
 \;.
 \label{nonvanishing_nu0}
\end{equation}
For the inverse of $\nu^0$ we then find
\begin{eqnarray}
\nu_0 &=& \Phi^{2}  +    D^{(\nu_0)} r^{-1} + \mathcal{O}(r^{-2})
\;.
\end{eqnarray}
Note that in the special case where $(\overline W{}^0)_1 =0$ we have $\Phi=\varphi_{-1}$ and the positivity
  of $(\nu^0)_0$ follows from that of $\varphi_{-1}$.

Since $(\overline W{}^0)_1$ can be prescribed arbitrarily and the value of $\varphi_{-1}$ does not depend on that choice, 
 \eq{nonvanishing_nu0} is not  a geometric restriction.
Similarly,  \eq{cond_Y0} can  be fulfilled by an appropriate choice of  $(\overline W{}^0)_2$.
The gauge freedom associated with the choice of $\overline W{}^0$ can  be used to control the  behavior of $\nu^0$
and to get rid of the log terms in its asymptotic expansion (only the two leading order terms in the expansion of $\ol W{}^{0}$ are affected, compare the discussion in Section~\ref{sec_discussion}).

\subsubsection{Expansion of $\xi_A$}
\label{sec_exp_xiA}

The connection coefficients $\xi_A = -2\ol \Gamma^r_{rA}$ are determined by (\ref{eqn_nuA_general}),
\begin{eqnarray}
 (\partial_r + \tau)\xi_A - 2 \conenabla_B \sigma_A^{\phantom{A}B} +\partial_A\tau +2 \partial_A \kappa &=& 0
\;.
 \label{system_xi}
\end{eqnarray}
To compute the covariant derivative of $\sigma_A{}^{B}$ associated to  $\overline g_{AB}\mathrm{d}x^A\mathrm{d}x^B$,
we first determine the asymptotic form of the Christoffel symbols,
\begin{equation*}
  \coneGamma^C_{AB}  \,= \,\frac{1}{2}\gamma^{CD}(2\partial_{(A} \gamma_{B)D}  - \partial_D \gamma_{AB})
 + 2\delta_{(B}{}^C\partial_{A)}\log\Omega
- \gamma^{CD}\gamma_{AB}\partial_D \log\Omega
\,=\, \mathcal{O}(1)
\;,
\end{equation*}
 with
\begin{eqnarray}
 (\coneGamma^C_{AB})_0
 &= &\mathring\Gamma ^C_{AB} + 2\delta_{(B}{}^C\mathring\nabla_{A)}\log\varphi_{-1}  - s_{AB}\mathring\nabla^C \log\varphi_{-1}
 \;,
\label{component_christoffel}
\\
 (\coneGamma^C_{AB})_1
 &= & \mathring\nabla_{(A} \breve h_{B)}^{(1)C}  - \frac{1}{2} \mathring\nabla^C\breve h^{(1)}_{AB}
  - \delta_{(A}{}^C\mathring\nabla_{B)}\tau_2  + \frac{1}{2} s_{AB}\mathring\nabla^C\tau_2
\nonumber
\\
 && -( s^{CD}h^{(1)}_{AB} - h^{(1)CD}s_{AB})\mathring\nabla_D\log\varphi_{-1}
\label{component_christoffel2}
\;.
\end{eqnarray}
Invoking \eq{expansion_sigma_A^B} that yields
\begin{eqnarray}
 \conenabla_B\sigma_A{}^B &= &
\Xi_A^{(2)}r^{-2} + \Xi_A^{(3)}r^{-3} + \mathcal{O}(r^{-4})
 \;,
\end{eqnarray}
%
where 
\begin{eqnarray}
\Xi_A^{(2)}=-\frac{1}{2}\mathring\nabla_B\breve h^{(1)B}_A - \breve h_A^{(1)B}\mathring\nabla_B\log\varphi_{-1}
=\mathring\nabla_B(\sigma_A{}^B)_2 + 2(\sigma_A{}^B)_2\mathring\nabla_B\log\varphi_{-1}
\;,
\label{expansion_div_sigma2}
\\
\Xi_A^{(3)}
  = \mathring\nabla_B(\sigma_A{}^B)_3 + 2(\sigma_A{}^B)_3\mathring\nabla_B \log\varphi_{-1}
-(\sigma_A{}^B)_2\mathring\nabla_B\tau_2
+ \frac{1}{2}\mathring \nabla_A\sigma_4
\label{expansion_div_sigma3}
 \;. \hspace{4em}
\end{eqnarray}

Substituting now the coefficients by their asymptotic expansions
we observe that \eq{system_xi} has the asymptotic structure,
%
\begin{eqnarray*}
&& (\partial_r + 2r^{-1} + \tau_2 r^{-2} + \mathcal{O}(r^{-3}))\xi_A
\\
 && \qquad  = ( 2\Xi^{(2)}_A -\partial_A\tau_2 )r^{-2} + [2\Xi^{(3)}_A -\partial_A(\tau_3+2\kappa_3 )]r^{-3} +  \mathcal{O}(r^{-4})
 \;.
\end{eqnarray*}
Nicely enough,  the ODEs for $\xi_A$, $A=2,3$, are decoupled.
For comparison with the formulae in Appendix~\ref{asymptotic_solutions} we rewrite them in terms of $x\equiv1/r$,
\begin{eqnarray*}
 && (x\partial_x - 2 -\tau_2 x + \mathcal{O}(x^2))\xi_A
\\
 && \qquad= (\partial_A\tau_2-2\Xi^{(2)}_A)x   -  [ 2\Xi^{(3)}_A -\partial_A(\tau_3+2\kappa_3  ) ]x^2+  \mathcal{O}(x^3)
 \;.
\end{eqnarray*}
Appendix~\ref{asymptotic_solutions} tells us (with $\lambda=2$ and $\hat\ell=1$) that there are no logarithmic terms in the expansion of $\xi_A$ if and only if (\ref{cond_no_logs}) holds,
\begin{equation}
 \tau_2 (2\Xi^{(2)}_A - \partial_A\tau_2) = 2\Xi^{(3)}_A -\mathring\nabla_A\left(\tau_3+2\kappa_3  \right)
 \;.
 \label{particular_case}
\end{equation}
The asymptotic expansion  (\ref{expansion_tau}) of $\tau$ implies
\begin{equation}
 \tau_3 = \frac{1}{2}(\tau_2)^2 + \sigma_4 - 2\kappa_3
 \;,
\end{equation}
such that (\ref{particular_case}) can be written as
\begin{equation}
 2\tau_2  \Xi^{(2)}_A = 2\Xi^{(3)}_A  - \mathring\nabla_A \sigma_4
 \;.
 \label{log-condition}
\end{equation}
Note that $\kappa_3$, on which we have not imposed conditions yet, drops out, so there is no
 gauge freedom left which could be appropriately adjusted to fulfill this equation.
The impact of \eq{log-condition} will be analyzed  in Section~\ref{sec_no_log},
where it becomes manifest that it \emph{does} impose geometric restrictions on the initial data.
We refer to \eq{log-condition} as \textit{no-logs-condition}.

Whenever the no-logs-condition holds, which we assume henceforth, the covector field  $\xi_A$ can be expanded as a power series,
\begin{equation}
 \xi_A = (2\Xi^{(2)}_A-\mathring\nabla_A\tau_2)r^{-1} + C^{(\xi_B)}_Ar^{-2} + \mathcal{O}(r^{-3})
 \;.
 \label{solution_xiA}
\end{equation}
The coefficients $C^{(\xi_B)}_A=C^{(\xi_B)}_A(x^C)$, $A=2,3$, represent  globally defined  integration functions.

\subsubsection{Expansion of $\nu_A$}

We analyze  (\ref{eqn_xiA}) to compute the asymptotic behavior of $\nu_A$,
\begin{equation}
 \big[ 2\partial_r  - \nu_0(\overline W{}^0 + \ol{\hat\Gamma}^0) \big]\nu_A - 4\nu_B\chi_A{}^{B}
+ \nu_0[\xi_A   -  \overline g_{AB}(\overline W{}^B+ \ol {\hat\Gamma}^B) + \gamma_{AB}\gamma^{CD}\coneGamma^B_{CD}  ] = 0
 \;.
\label{eqn_xiA2}
\end{equation}
Here
\begin{eqnarray*}
\nu_0\ol g_{AB}\ol {\hat\Gamma}^B &=& 2\ol g_{AB}\ol{\hat \Gamma}^B_{01} - 2\ol g_{AB}\nu^C \hat \chi_C{}^B + \nu_0\gamma_{AB}\gamma^{CD}(\hat{\tilde\Gamma}^B_{CD} -\ol{\hat g}^{1B} \hat \chi_{CD})
\\
&=&  - \hat\tau\nu_A + v_A{}^B\nu_B
+ \nu_0\gamma_{AB}\gamma^{CD}\mathring\Gamma^B_{CD} + \mathcal{O}(r^{-2})
\;,
\end{eqnarray*}
with $v_C{}^B= \mathcal{O}(r^{-2} ) \in \mathrm{Mat}(2,2)$,
as follows from \eq{Minktarget_asympt}.
Recall that  $\overline W{}^0=\mathcal{O}(r^{-1})$ and $\overline W{}^A=\mathcal{O}(r^{-1})$, or, by  \eq{expansion_gAB}, $\overline W_A := \overline g_{AB}\overline W{}^B = \mathcal{O}(r)$.
We  determine the asymptotic expansions  of the coefficients involved,
\begin{eqnarray}
  \nu_B\chi_A^{\phantom{A}B} &\equiv& 
 \frac{1}{2} \nu_B\overline{g}^{BC}\partial_r \overline{g}_{AC}
 \,=\, \nu_A\partial_r\log\Omega  + \frac{1}{2}\nu_B\gamma^{BC}\partial_r\gamma_{AC}
 \nonumber
\\
 &=&
 \nu_Ar^{-1} + \frac{1}{2}( \tau_2\nu_A -  \breve h^{(1)B}_{A}\nu_B ) r^{-2}  +  \mathcal{O}(r^{-3})
 \;.
 \label{nu_chi}
\end{eqnarray}
For the source term involving  Christoffel symbols of the metrics
$s_{AB}$ and $\overline g_{AB}=\Omega^2\gamma_{AB}$ we obtain,
\begin{eqnarray}
\gamma_{AB} \gamma^{CD} (\mathring\Gamma^B_{CD} - \coneGamma^B_{CD})
 &=&-\mathring\nabla_B\breve h_A^{(1)B} r^{-1} + \mathcal{O}(r^{-2})
 \;.
 \label{dfn_cA}
\end{eqnarray}
Combining this with what we found for $\nu_0$, $\ol{\hat\Gamma}^0$ and $\xi_A$ (assuming that the no-logs-condition holds) the ODE for $\nu_A$ takes the asymptotic form
\begin{eqnarray*}
\partial_r \nu_A + w_A{}^B\nu_B
 = \frac{\Phi^{2}}{2} (\overline W_A)_{-1}r + \frac{1}{2}\Big[\Phi^{2}(\overline W_A)_{0}  +D^{(\nu_0)}(\overline W_A)_{-1}  \Big]
+ \frac{1}{2}\Big[(\nu_0)_2(\overline W_A)_{-1}
\\
+D^{(\nu_0)}(\overline W_A)_0 + \Phi^{2}\big( \mathring\nabla_A \tau_2
 + 2\breve h^{(1)B}_A\mathring\nabla_B\log\varphi_{-1}
 + (\overline W_A)_1  \big) \Big] r^{-1}
+ \mathcal{O}(r^{-2})
 \;,
\end{eqnarray*}
where
\begin{eqnarray*}
w_A{}^B&=& \Big[ \frac{\hat \tau}{2}- \frac{1}{2}\nu_0(\ol W{}^0 + \ol{\hat\Gamma}^0)\Big] \delta_A{}^B -2 \chi_A{}^B
\\
&=&\Big[ [\Phi^{-2}D^{(\nu_0)}  -\frac{1}{2}\tau_2]\delta_A{}^B -2(\sigma_A{}^B)_2\Big]r^{-2} + \mathcal{O}(r^{-3}) \in \mathrm{Mat}(2,2) \;.
\end{eqnarray*}
Note that this first-order system is \textit{not} decoupled. 
In terms of $x\equiv1/r$ the equation becomes (set $\tilde w_A{}^B(x,x^A) =-x^{-1} w_A{}^B (x^{-1},x^A) = \mathcal{O}(x) $)
\begin{eqnarray*}
  x\partial_x \nu_A + \tilde w_A{}^B\nu_B  &=& -\frac{1}{2} \Phi^{2}(\overline W_A)_{-1}x^{-2} - \frac{1}{2}\Big[\Phi^{2}(\overline W_A)_{0}   +D^{(\nu_0)}(\overline W_A)_{-1}  \Big]x^{-1}
\\
&&\hspace{-10em}- \frac{1}{2}\Big[(\nu_0)_2(\overline W_A)_{-1}  +D^{(\nu_0)}(\overline W_A)_0 + \Phi^{2}\big( \mathring\nabla_A \tau_2
+ 2\breve h^{(1)B}_A\mathring\nabla_B\log\varphi_{-1}   + (\overline W_A)_1  \big) \Big] + \mathcal{O}(x)
 \;.
\end{eqnarray*}
Again, we consult Appendix~\ref{asymptotic_solutions};  inspection of (\ref{no_log_terms_diagsystem}) tells us (with $\lambda_1=\lambda_2=0$, $\hat \ell=-2$) that
no logarithmic terms appear 
whenever the source satisfies,
%
\begin{eqnarray}
 (\overline W_A)_1 &=& 4(\sigma_A{}^{B})_2\mathring\nabla_B\log\varphi_{-1}  - \mathring\nabla_A \tau_2  - \Phi^{-2}(\nu_0)_2(\overline W_A)_{-1}
 \nonumber
\\
&&- \Phi^{-2} D^{(\nu_0)}(\overline W_A)_0
 - \frac{1}{2}[(\tilde w_A{}^B)_1(\tilde w_B{}^C)_1 +(\tilde w_A{}^C)_2] (\overline W_C)_{-1}
 \nonumber
\\
 &&  - (\tilde w_A{}^B)_1\Big[(\overline W_B)_{0}  + \Phi^{-2} D^{(\nu_0)}(\overline W_B)_{-1}  \Big]
\label{nuA_cond}
 \;.
\end{eqnarray}
We express \eq{nuA_cond} in terms of $\ol W{}^A=\ol g^{AB} \ol W_B$.
It can be solved for $(\overline W{}^A)_1$, and thus provides a condition on $(\overline W{}^A)_1$ once $(\overline W{}^A)_{-1}$ and $(\overline W{}^A)_0$ have been chosen.

Assuming  that the gauge  functions $\ol W{}^A$ have an asymptotic behavior which fulfills (\ref{nuA_cond}), any solution of  (\ref{eqn_xiA2})  has the asymptotic  form $\nu_A \ourdoteq   \sum_{n=\hat\ell}^{\infty} (\nu_A)_n r^{-n}$. For suitable, globally defined  integration functions $D^{(\nu_A)}_A$ we find,
%
%
\begin{eqnarray}
 \nu_A &=& \frac{\Phi^{2} }{4}(\overline W_A)_{-1} r^2
+ \Big[ \frac{\Phi^2}{2}\big[(\overline W_A)_0    +(\sigma_A{}^B)_2(\overline W_B)_{-1}+\frac{\tau_2}{4}(\overline W_A)_{-1} \big]
\nonumber
\\
 &&
+\frac{1}{4}D^{(\nu_0)}(\overline W_A)_{-1}  \Big] r + D^{(\nu_A)}_A + \mathcal{O}(r^{-1})
 \;.
\end{eqnarray}

\subsubsection{Expansion of $\zeta$}
\label{sec_exp_zeta}

We consider the ODE \eq{zeta_constraint} which determines the auxiliary function $\zeta$,
\begin{eqnarray}
 (\partial_r + \tau + \kappa)\zeta + \coneR - \frac{1}{2}\xi_A\xi^A + \conenabla_A\xi^A =0
 \;.
\end{eqnarray}
%
It remains to compute the source terms.
From \eq{expansion_gAB} and \eq{solution_xiA} we find
\begin{eqnarray}
 \frac{1}{2}\overline g^{AB} \xi_A \xi_B &=& \mathcal{O}(r^{-4})
\\
 \overline g^{AB}\conenabla_A\xi_B
 &=& (\varphi_{-1})^{-2} \mathring\nabla^A(\xi_A)_1  r^{-3} + \mathcal{O}(r^{-4})
 \;,
\end{eqnarray}
with
\begin{equation}
 (\xi_A)_1 = 2 (\check\nabla_B \sigma_A{}^B)_2-\mathring\nabla_A\tau_2
\;.
\end{equation}
%
%
\begin{lemma} \label{lemma_ricciscalar}
The curvature scalar $\coneR$ satisfies
\begin{eqnarray*}
  \coneR &=& \coneR_2 r^{-2} + [\tau_2\coneR_2 -  (\varphi_{-1})^{-2} \mathring\nabla^A(\xi_A)_1  ] r^{-3} + \mathcal{O}(r^{-4})
 \;,
\end{eqnarray*}
with 
$
 \coneR_2 =  2 (\varphi_{-1})^{-2}(1  -\Delta_s\log\varphi_{-1}  )
$.
\end{lemma}
Since the computation of the curvature scalar is elementary though somewhat lengthy
 we leave the proof of Lemma~\ref{lemma_ricciscalar} to  the reader.

Putting everything together, we observe that the $\zeta$-constraint is of the form
%
\begin{eqnarray*}
 \Big[ \partial_r + \frac{2}{r} + \tau_2r^{-2} + \mathcal{O}(r^{-3})\Big] \zeta
 &=&  -\coneR_2r^{-2}- \tau_2\coneR_2r^{-3} + \mathcal{O}(r^{-4})
 \;.
\end{eqnarray*}
We express the ODE in terms of $x\equiv 1/r$,
\begin{eqnarray*}
 \left[x\partial_x- 2 - \tau_2x + \mathcal{O}(x^2)\right] \zeta  &=&  \coneR_2 x + \tau_2\coneR_2 x^2 + \mathcal{O}(x^3)
 \;.
\end{eqnarray*}
Comparison with Appendix~\ref{asymptotic_solutions} tells us (with $\lambda=2$ and $\hat\ell=1$) that, due to the specific form of the source,  condition \eq{cond_no_logs}, which
excludes the appearance of logarithmic terms, is
automatically satisfied, and the asymptotic expansion of $\zeta$ reads
\begin{equation}
 \zeta = -\coneR_2r^{-1}  +C^{(\zeta)} r^{-2} + \mathcal{O}(r^{-3})
 \;,
\label{final_expansion_zeta}
\end{equation}
where $C^{(\zeta)} =C^{(\zeta)} (x^A)$ can be identified with the  integration function. It thus \emph{follows} that $\zeta=\mathcal{O}(r^{-1})$,
supposing that $\kappa=\mathcal{O}(r^{-3})$, that $\gamma$ has the asymptotic form \eq{initial_data} with $\varphi_{-1}\ne 0$,
and  that the no-logs-condition holds.

\subsubsection{Expansion of $\overline g{}^{rr}$}

Let us compute $\overline g{}^{rr}$ which satisfies \eq{dfn_zeta},
\begin{equation}
  (\partial_r + \frac{1}{2}\tau + \kappa)\overline g^{rr} = \frac{1}{2}\zeta  - \overline W{}^r  -\ol{\hat\Gamma}^r
 \;.
\end{equation}
Recall that now where we have established \eq{final_expansion_zeta}, it follows from the a priori restrictions that  $\ol W{}^r=\mathcal{O}(r)$.
We further have,
\begin{eqnarray*}
 \ol{\hat\Gamma}^r &=& 2\nu^0\hat\Gamma^r_{0r} + \ol g^{rr}\hat\kappa - \ol g^{rA}\hat\xi_A + \ol g^{AB} \hat \Gamma^r_{AB}
\\
&=& \hat\kappa\ol g^{rr}  - 2(\varphi_{-1})^{-2}r^{-1} -2(\varphi_{-1})^{-2}\tau_2r^{-2}+ \mathcal{O}(r^{-3})
\;,
\end{eqnarray*}
 so the ODE for $\overline g{}^{rr}$ is of the form
\begin{equation*}
  (\partial_r + r^{-1} +\mathcal{O}(r^{-2}) )\overline g{}^{rr} =\mathcal{O}(r)
 \;,
\end{equation*}
or, expressed in terms of the  $x\equiv 1/r$-coordinate,
\begin{equation}
  (x\partial_x -1  + f^*)\overline g^{rr} = f^{**}
 \;,
\label{ODE_g11}
\end{equation}
where
\begin{eqnarray*}
 f^* &=& -\frac{1}{2}\tau_2 x -(\frac{1}{2}\tau_3 + \kappa_3+\hat\kappa_3)x^2-  (\frac{1}{2}\tau_4 + \kappa_4+\hat\kappa_4)x^3 + \mathcal{O}(x^4)
 \;,
\\
 f^{**} &=& (\overline W{}^r)_{-1}x^{-2} +  (\overline W{}^r)_{0} x^{-1} +  (\overline W{}^r)_{1} + \frac{1}{2}\coneR_2 -2(\varphi_{-1})^{-2}
\\
 &&  + \Big[ (\overline W{}^r)_{2} - \frac{1}{2}\zeta_2 - 2(\varphi_{-1})^{-2}\tau_2  \Big] x + \mathcal{O}(x^2)
 \;.
\end{eqnarray*}
We analyze the occurrence of logarithmic terms in the asymptotic solution of \eq{ODE_g11}.
In the notation of Appendix~\ref{asymptotic_solutions} we have $\lambda=1$ and $\hat\ell=-2$.
The considerations made there show that the asymptotic expansion of $\overline g^{rr}$  is $\mathcal{O}(r^2)$
if and only if the following condition is fulfilled (cf.\ \eq{cond_no_logs}),
\begin{eqnarray}
 f^*_1(\overline g^{rr})_0 + f^*_2(\overline g^{rr})_{-1} + f^*_3(\overline g^{rr})_{-2} = f^{**}_1
 \;.
 \label{no_logs_ChConeExistence}
\end{eqnarray}
The expansion coefficients $(\overline g^{11})_i$ can be derived from \eq{formalsolution2},
\begin{eqnarray*}
 (\overline g^{rr})_{-2} &=& -\frac{1}{3} f^{**}_{-2}
 \;,
\\
 (\overline g^{rr})_{-1} &=& -\frac{1}{2} f^{**}_{-1} - \frac{1}{6} f^*_1 f^{**}_{-2}
 \;,
\\
  (\overline g^{rr})_{0} &=& -f^{**}_0 - \frac{1}{3} f^*_2f^{**}_{-2} - \frac{1}{2} f^*_1f^{**}_{-1} - \frac{1}{6} (f^*_1)^2f^{**}_{-2}
 \;.
\end{eqnarray*}
A straightforward calculation  reveals that \eq{no_logs_ChConeExistence} is equivalent to
\begin{eqnarray}
 (\overline W{}^r)_2 &=& \frac{\zeta_2 }{2}+ (\varphi_{-1})^{-2}\tau_2 + \frac{\tau_2}{4}\coneR_2 + \frac{\tau_2}{2} (\overline W{}^r)_1
 +   \Big[ \frac{\tau_3 }{4}+ \frac{\kappa_3+ \hat\kappa_3}{2} - \frac{(\tau_2)^2}{8} \Big](\overline W{}^r)_0
 \nonumber
\\
 &&  \Big[ \frac{1}{48}(\tau_2)^3 - \frac{\tau_2}{4}(\frac{\tau_3 }{2}+ \kappa_3+\hat\kappa_3) + \frac{\tau_4 }{6}+ \frac{1}{3}(\kappa_4+\hat\kappa_4) \Big](\overline W{}^r)_{-1}
 \;.
 \label{g00_cond}
\end{eqnarray}
By an appropriate choice of the gauge source function $\ol W{}^r$, or  merely the expansion coefficient $(\overline W{}^r)_2$, one can always arrange that \eq{g00_cond}  holds and thereby get rid of all logarithmic terms in the expansion of $\ol g^{rr}$.
In that case
\begin{eqnarray*}
 \overline g^{rr} &=& -\frac{1}{3}(\overline W{}^r)_{-1}r^2 + \mathcal{O}(r)
\\
 \Longrightarrow \quad \overline g_{00}
 &= &\Phi^4\Big[ \frac{1}{16}(\varphi_{-1})^2 s_{AB}(\overline W{}^A)_1(\overline W{}^B)_1 + \frac{1}{3} (\overline W{}^r)_{-1} \Big] r^2 + \mathcal{O}(r)
 \;.
\end{eqnarray*}
The integration function is represented by
$(\ol g_{00})_1$, and its explicit form is not relevant here.

\subsection{The no-logs-condition}
\label{sec_no_log}

The no-logs-condition \eq{log-condition}
  \begin{eqnarray}
 2\tau_2[\conenabla_B\sigma_A{}^B]_2  &=& 2[\conenabla_B\sigma_A{}^B]_3 - \mathring\nabla_A\sigma_4
 \;,
 \label{no-log-conditions2}
 \label{first_no-log}
 \end{eqnarray}
needs to be imposed to exclude logarithmic terms in the asymptotic expansion of $\xi_A$.
Let us rewrite and simplify it.
In \eq{expansion_div_sigma2} and \eq{expansion_div_sigma3} we have computed $ [\tilde\nabla_B\sigma_A{}^B]_2$ and $ [\tilde\nabla_B\sigma_A{}^B]_3$.
%
%
Plugging this in, we observe that the $\mathring \nabla_A\sigma_4$-terms cancel out in (\ref{first_no-log}), which becomes
\begin{eqnarray}
 \mathring\nabla_B[ \tau_2 (\sigma_A{}^B)_2 -(\sigma_A{}^B)_3 ] + 2[\tau_2 (\sigma_A{}^B)_2 - (\sigma_A{}^B)_3\mathring\nabla_B  ]\mathring\nabla_B\log\varphi_{-1} =0
 \;.
 \label{nolog_rewritten}
\end{eqnarray}
This can be written as a divergence,
\begin{equation}
 \mathring\nabla_B \big[   (\varphi_{-1})^2(\tau_2 (\sigma_A{}^B)_2 - (\sigma_A{}^B)_3) \big] =0
 \;.
 \label{divergence_equation}
\end{equation}
One can view this equation as a linear first-order PDE-system on $S^2$.
Note that by definition of $\sigma_A^{\phantom{A}B}$ the expansion coefficients $(\sigma_A^{\phantom{A}B})_n$ are $s$-traceless tensors on $S^2$, whence the term in squared brackets in (\ref{divergence_equation})
is traceless.
It is known (cf.\ e.g. \cite{JJpeeling}) that for any smooth source $v_A$ a PDE-system of the form
$
 \mathring\nabla_B\zeta_A{}^B=v_A
$,
with $\zeta_A{}^B$ a traceless tensor on the unit sphere $S^2$, admits precisely one smooth solution.
In our case the source of the PDE vanishes, so we conclude that the no-logs-condition \eq{first_no-log} is equivalent to
(recall our assumption $\varphi_{-1}\ne 0$)
\begin{eqnarray}
 (\sigma_A{}^B)_3  = \tau_2 (\sigma_A{}^B)_2
 \label{necessary_condition}
\quad
 \Longleftrightarrow \quad   \breve h^{(2)}_{AB} = \frac{1}{2}(h^{(1)}+\tau_2)\breve h^{(1)}_{AB}
 \label{necessary_condition2}
 \;.
\end{eqnarray}

Since $\xi_A=-2\ol \Gamma^r_{rA}$, determined by \eq{eqn_nuA_general},
has a geometric meaning one should expect the no-logs-condition to be  gauge-invariant.
In \cite{ChPaetz2} it is shown that this is indeed the case. 
Although the ($\kappa=0$, $\ol W{}^0 =0$)-wave-map gauge  invariably produces logarithmic terms except for the flat case \cite{ChPaetz2},
one can decide whether they can be transformed away or not by checking  \eq{necessary_condition}.

In the current scheme, where conformal data $\gamma$ are prescribed on $\mcN$, together with the gauge functions $\kappa$ and $W^{\lambda}$,
this requires to determine $\tau_2$ by solving the  Raychaudhuri equation, which makes this scheme not practical for the purpose.
 In particular, it is not a priori clear within this scheme whether \emph{any} initial data satisfying this condition exist unless both $(\sigma^A{}_B)_2$ and $(\sigma^A{}_B)_3$ vanish ($\Longleftrightarrow \enspace \breve h^{(1)}_{AB} =0= \breve h^{(2)}_{AB}$).
On the other hand, in gauge schemes where $\check g$ is prescribed,   \eq{necessary_condition} is a straightforward condition on its  asymptotic
behavior (cf.\ Section~\ref{sec_mgauge} where a related scheme is used).

In \cite{ChPaetz2} we also provide a geometric interpretation of the no-logs-condition  \eq{necessary_condition} via the conformal Weyl tensor.
Moreover,  \eq{necessary_condition}  can be related to Bondi's ``outgoing wave condition'', cf.\ Section~\ref{overview_mgauge}, Remark~\ref{rem_wave_cond}.

\subsection{Summary and discussion}
\label{summary1}

The subsequent theorem summarizes our analysis of the asymptotic behavior of  solutions to Einstein's vacuum constraint equations
((i) and (ii) follow from~\cite{ChPaetz2}):
\begin{theorem}
\label{thm_asympt_exp}
Consider the characteristic initial value problem for Einstein's vacuum field equations with smooth conformal data $\gamma=\gamma_{AB}\mathrm{d}x^A\mathrm{d}x^B$ and gauge  functions $\kappa$ and $\ol W^{\lambda}$, all defined on  a smooth characteristic surface
$\mcN$ meeting $\scrip$ transversally in a smooth spherical cross-section
(and supplemented by certain data on the intersection manifold $S$ if $\mcN$ is one of two transversally intersecting null hypersurfaces).
The following conditions are necessary-and-sufficient for the trace of the metric $g=g_{\mu\nu}\mathrm{d}x^{\mu}\mathrm{d}x^{\nu}$ on $\mcN$,
obtained as  solution to the characteristic wave-map vacuum constraint equations \eq{constraint_phi}-\eq{dfn_zeta},
 to admit a smooth conformal completion at infinity, and for the connection coefficients $\ol \Gamma^r_{rA}$  to be smooth at $\scri^+$ (in the sense of  Definition~\ref{definition_smooth}), when imposing a generalized wave-map gauge condition $H^{\lambda}=0$:
\begin{enumerate}
\item[(i)] There exists a one-parameter family $\varkappa=\varkappa(r)$ of Riemannian metrics on $S^2$ such that $\varkappa$ satisfies \eq{initial_data} and is conformal to $\gamma$ (in particular we may assume $\gamma$ itself to be  of the form \eq{initial_data}).
\item[(ii)] The functions $\varphi$, $\nu^0$, $\varphi_{-1}$ and $(\nu_0)_0$ have no zeros on $\mcN$ and $S^2$, respectively
 (the non-vanishing of $(\nu^0)_0$ is equivalent to  \eq{nonvanishing_nu0}).
\item[(iii)] The gauge source functions are chosen in such a way that  $\kappa=\mathcal{O}(r^{-3})$,
  $\overline W{}^0=\mathcal{O}(r^{-1})$, $\overline W{}^A=\mathcal{O}(r^{-1})$, $\overline W{}^r=\mathcal{O}(r)$, and such that the conditions (\ref{0gaugecond}), (\ref{nuA_cond}) and (\ref{g00_cond}) hold.
\item[(iv)] The  no-logs-condition is satisfied, i.e.
  \begin{eqnarray}
   (\sigma_A{}^B)_3  = \tau_2 (\sigma_A{}^B)_2
\quad
 \Longleftrightarrow \quad   \breve h^{(2)}_{AB} = \frac{1}{2}(h^{(1)}+\tau_2)\breve h^{(1)}_{AB}
 \;.
 \label{no-log-conditions}
 \end{eqnarray}
\end{enumerate}
\end{theorem}

\begin{Remark}
\label{remark_special_case}
{\rm
For further reference let us explicitly list the conditions (\ref{0gaugecond}), (\ref{nuA_cond}) and (\ref{g00_cond}) in the case where the gauge source functions satisfy
$\overline W{}^0=\mathcal{O}(r^{-2})\;, \enspace \overline W{}^A=\mathcal{O}(r^{-3})\;, \enspace \overline W{}^r=\mathcal{O}(r^{-2})$:
%
%
\begin{eqnarray}
 (\overline W{}^0)_2 &=& \tau_2 (\varphi_{-1})^{-2}
\label{special_case_1}
 \;,
\\
 (\overline W{}^A)_3 &=& -(\varphi_{-1})^{-2}\mathring\nabla^A\tau_2 +4 (\varphi_{-1})^{-3}(\sigma_B{}^{A})_2\mathring \nabla^B\varphi_{-1}
 \;,
 \label{needed_for_nogo}
\\
 (\overline W{}^r)_2 &=& \frac{1}{2}\zeta_2 + \tau_2\Big[ (\varphi_{-1})^{-2} + \frac{1}{4}\coneR_2 + \frac{1}{2} (\overline W{}^r)_1  \Big]
 \;.
\end{eqnarray}
If, in addition, the no-logs-condition (\ref{no-log-conditions}) is fulfilled then the leading order terms of the non-vanishing metric components restricted to the cone read
%
\begin{eqnarray}
 \overline g_{00} &=& 
 -(1 + \Delta_s\varphi_{-1}) + D^{(\ol g_{00})} r^{-1} + \mathcal{O}(r^{-2})
 \;,
\\
 \nu_0  &=& (\varphi_{-1})^2 + D^{(\nu_0)}r^{-1} + \mathcal{O}(r^{-2})
 \;,
\\
 \nu_A &=& D^{(\nu_A)}_A + \mathcal{O}(r^{-1})
 \;,
\nonumber
\\
 \overline g_{AB} &=& (\varphi_{-1})^2 s_{AB}r^2 + (\varphi_{-1})^2[\breve h^{(1)}_{AB}  - \tau_2 s_{AB} ]r
\\
&& -\frac{1}{2}\tau_2 \breve h^{(1)}_{AB} + (\frac{1}{4}(\tau_2)^2 + \kappa_3 + \frac{1}{2}\sigma_4)s_{AB}
+ \mathcal{O}(r^{-1})
\label{special_case_f}
 \;.
\end{eqnarray}
The coefficients denoted by $D$  have a global character in that they are globally defined by the initial data and the gauge functions.
}
\end{Remark}

\begin{remark}
{\rm
\label{sec_other_settings}
 For  definiteness we have restricted attention  to the setting where the initial data are provided by the conformal data $\gamma=\gamma_{AB}\mathrm{d}x^A\mathrm{d}x^B$ together~with the gauge functions $\kappa$ and $W^{\lambda}$.
Theorem~\ref{thm_asympt_exp}, though,  is quite~independent of the particular setting that has been chosen.
This is discussed in  \cite{ChPaetz2}.
}
\end{remark}

\label{sec_discussion}

It is shown in Appendix~\ref{asymptotic_solutions} that when analyzing the asymptotic behavior of a linear first-order Fuchsian  ODE-system
whose indicial matrix
has only integer eigenvalues  one has to distinguish two different cases.
If the indicial matrix cannot be diagonalized, or if it can be but the condition \eq{app-no-logs} of   Appendix~\ref{asymptotic_solutions} is violated,
then  the appearance of logarithmic terms always depends on the boundary conditions and thereby on the globally defined integration functions.
If, however, \eq{app-no-logs} is fulfilled  their
appearance depends exclusively on the asymptotic behavior of the coefficients in the corresponding ODE.
In fact, all the Einstein wave-map gauge constraints are of the latter type.%
\footnote{Since the integration functions of some constraint equations appear as coefficients in other ones, the no-logs-condition \emph{does} depend on the   integration functions $\varphi_{-1}$ and $\varphi_0$.}



Due to this property and since many of the constraint equations have a source term which involves a gauge freedom,
many, though not all, logarithmic terms arising can be eliminated  for given data $\gamma$  by
a carefully adjusted leading-order-term-behavior
of the  gauge  functions.
These are precisely the conditions of Theorem~\ref{thm_asympt_exp} which involve a gauge source function  $\kappa$ or $\overline W{}^{\lambda}$.
Logarithmic terms which appear if these conditions are violated are pure gauge artifacts.

This does not apply to the $\xi_A$- and the $\zeta$-equation.
Recall that to solve the equation for $\xi_A$ both $\kappa$ and $\varphi$ need to be known. This requires a choice of the $\kappa$-gauge. Since the choice of $\overline W{}^0$ does not affect the $\xi_A$-equation, there is no gauge-freedom left. 
If the no-logs-condition \eq{no-log-conditions}, which is gauge-invariant~\cite{ChPaetz2}, does not hold, there is no possibility to get rid of the log terms that arise there,
whatever $\kappa$  has been chosen to be.
Similarly, there is no gauge-freedom left~when the equation for $\zeta$ is integrated but, due to the special asymptotic structure of its source term, no new log terms arise in the expansion of $\zeta$.

The $\xi_A$-equation is the reason why the conditions on $\overline W{}^{\lambda}$ have to be supplemented by
the no-logs-condition \eq{no-log-conditions} which involves two integration functions, $\varphi_{-1}$ and $\varphi_0$, globally determined by the initial data $\gamma$ and the gauge function $\kappa$.
A decisive grievance is that, at least in our current setting, 
their dependence  on  $\gamma$ and $\kappa$ is very intricate.
Thus the question arises for which class of initial data one finds a  function $\kappa=\mathcal{O}(r^{-3})$, such that the no-logs-condition
holds, and accordingly what the geometric restrictions are for this to be possible.
The only obvious fact  is that  \eq{no-log-conditions}
will be satisfied for sure whenever $\breve h^{(1)}_{AB} = \breve h^{(2)}_{AB}=0$.

For a ``generic''
choice of  $\gamma$ and $\kappa$ one should expect that the expansion coefficient $\tau_2=-2\varphi_0(\varphi_{-1})^{-1}$ will not vanish.
Equation~\eq{cond_Y0} then shows that a $\ol W{}^0=0$-gauge (in particular the harmonic gauge) is not adequate for our purposes, since logarithmic terms can only be removed by a non-vanishing gauge source function $\ol W{}^{0}\ne 0 $.
This is  illustrated best
by the no-go result \cite[Theorem~3.1]{ChPaetz}.
In order to fulfill (\ref{cond_Y0}) one needs a gauge choice for $\overline W{}^0$ which
depends on the globally defined integration functions $(\varphi_{-1},\varphi_0)$.
This indicates the need of an initial data-dependent gauge choice to get rid of the logarithmic terms.
(Note  that the higher-order terms in the expansion of $\overline W{}^0$ do not affect the appearance of log terms.)
We shall address this issue in the next sections.

\section{Metric gauge}
\label{sec_mgauge}

\subsection{Introduction}

Up to now we investigated the overall form of the asymptotic expansions of the trace of the metric on $\mcN$.
It turned out that it is not possible to manifestly eliminate all logarithmic terms just by imposing restrictions on the asymptotic behavior of the gauge functions.
Instead, an additional gauge-independent ``no-logs-condition'' \eq{no-log-conditions},  which depends on some of the globally defined integration functions, needs to be fulfilled to ensure expansions at conformal infinity in terms of power series.
Nonetheless, the current gauge scheme is unsatisfactory in that it seems hopeless to characterize those classes of initial data which  satisfy \eq{no-log-conditions}.

In the following  we will modify the  scheme to allow for a better treatment of these problems at hand.
We develop a gauge scheme which is adjusted to the initial data in such a way that we can solve at least some of the constraint equations
analytically, so that  the   values of the troublesome functions $\varphi_{-1}$ and $\varphi_0$, which appeared hitherto as ``integration functions''
in the asymptotic solutions and which are related to the appearance of log terms via \eq{no-log-conditions}, can be computed explicitly.
In view of the computation of all the fields appearing in the conformal field equations we will choose a gauge in which
the components of the metric tensor take preferably simple values
(at the expense of more complicated gauge  functions $\kappa$ and $W{}^{\lambda}$).
As before we shall adopt Rendall's point of view \cite{Rendall} and regard $[\gamma]$ (together with a choice of $\kappa$) as the free ``physical''   data.

\subsection{Gauge scheme}

The trace of certain components of the vacuum Einstein equations on $\mcN$
 together with the gauge condition $\ol H^{\lambda}=0$  can be used
to determine the metric  and some of its transverse derivatives on $\mcN$.
Similarly, the remaining components of the Einstein equations on $\mcN$ as well as their transverse derivatives,
$\ol{\partial_0^n R_{\mu\nu}} =0$,  $n\in\mathbb{N}$,
in combination with the gauge condition $H^{\lambda}=0$, provide a way to  determine higher-order transverse derivatives of the metric on $\mcN$ when necessary.
To compute all the fields which appear in the conformal field equations, such as e.g.\ the rescaled Weyl tensor,
this is what needs to be done.

However, it is very convenient to exploit the gauge freedom contained in the vector field  $W^{\lambda}$
to prescribe certain metric components and transverse derivatives thereof on  $\mcN$, and treat the corresponding equations
as equations for $W^{\lambda}$.
Indeed, proceeding this way the computations below can be significantly simplified.
It turns out that  the most convenient way here is a mixture of schemes described in  \cite{ChPaetz}:
We still regard the conformal class $[\gamma]$ of $\check g_{\Sigma_r}$ as the free ``non-gauge''-initial data.%
\footnote{
We do not regard e.g.\ $\ol g_{\mu\nu}$ as the free data as in \cite{ChPaetz} since we want to separate gauge degrees of freedom and physical initial data.
}
Recall that up to this stage we have regarded $\kappa$ and $W^{\lambda}$ as gauge degrees of freedom.
Now, instead of $\ol W^{\lambda}$ and $\ol{\partial_0 W^{\lambda}}$  we shall show that it is possible to prescribe the functions
$$
\ol g_{0\mu} \quad \text{and} \quad \ol{\partial_0 g_{0\mu }}
 \quad \text{with} \quad \nu_0=\ol g_{0r}\ne 0
\;.
$$
Of course there are some restrictions coming from regularity when $\mcN$ is a light-cone, and from the requirement of the metric to be continuous at the intersection manifold when $\mcN$ is one of two intersecting null surfaces.
Since we are mainly interested in prescribing these function  for large values of $r$, this issue can be ignored for our purposes.
Some comments are  given in the course of this~section.

Let us now explain how the above gauge scheme can be realized:
First of all we solve the Raychaudhuri equation to compute $\ol g_{AB}$, where, as before, we assume that a global solution $\tau$  exists.
Then we compute $\ol W^{\lambda}$ from the constraint equations \eq{constraint_nu0}, \eq{eqn_xiA} and \eq{dfn_zeta},
a procedure which was introduced in \cite{ChPaetz} as an alternative scheme for integrating the null constraint equations.
In order to make sure that we may prescribe $\ol{\partial_0 g_{0\mu }}$ rather than $\overline{\partial_0 W{}^{\lambda}}$
we need to analyze the remaining  Einstein equations  on $\mcN$ (those components which have not already been used to derive the
Einstein wave-map gauge constraints in \cite{CCM2}).

We impose a  wave-map gauge condition $H^{\lambda}=0$ with arbitrarily prescribed gauge functions $\kappa$ and $W^{\lambda}$.
From
$
\overline H{}^{\lambda} = 0
$
we obtain algebraic equations for  $\overline{\partial_0g_{rr}}$, $\overline{\partial_0g_{rA}}$ and $\overline g{}^{AB} \overline{\partial_0g_{AB}}$,
\begin{eqnarray}
 \overline {\partial_0 g_{rr}}
 &=&   \tau\nu_0 +(\nu_0)^2( \ol{\hat\Gamma}^0 + \ol W{}^0)
 \;,
\label{expression_H0}
\\
 \ol g^{AB}\ol{\partial_0 g_{rB} }
 &=& \check\nabla^A\nu_0   - \partial_r\nu^A    -\nu_0 \ol g^{CD}\check\Gamma^A_{CD}
-\ol g^{rA}(\ol{\partial_0 g_{rr}}-\tau\nu_0)
\nonumber
\\
&&
+\nu_0(\ol{\hat\Gamma}^A + \ol W{}^A)
\label{expression_HA}
\;,
\\
 \ol g^{AB}\ol{\partial_0 g_{AB}}
 &=&2\check\nabla_A\nu^A - (2\partial_r + 2\tau  - \nu^0\ol{\partial_0 g_{rr}})(\nu_0 \ol g^{rr})
-2\nu_0(\ol{\hat\Gamma}^r + \ol W{}^r)
\label{expression_Hr}
\;.
\end{eqnarray}

Recall that
\begin{equation}
 R_{\mu\nu} \,=\, \partial_{\alpha}\Gamma^{\alpha}_{\mu\nu} - \partial_{\mu}\Gamma^{\alpha}_{\nu\alpha}  + \Gamma^{\alpha}_{\mu\nu}\Gamma^{\beta}_{\alpha\beta} - \Gamma^{\alpha}_{\mu\beta}\Gamma^{\beta}_{\nu\alpha}
\;.
\label{Ricci_Christoffels}
\end{equation}

\paragraph*{Einstein equations $\breve{\overline R}_{AB}=0$:}

Using  the formulae in  \cite[Appendix~A]{CCM2} (which we shall make extensively use of for all  the subsequent  computations) we find
\begin{eqnarray*}
\breve{ \overline R}_{AB}
 &=& (\overline{\partial_0\Gamma^0_{AB}})\breve{}  - \frac{1}{2}\nu^0(\partial_r-\tau)(\overline{\partial_0 g_{AB}} )\breve{}
- \frac{1}{2}\nu^0\ol g_{AB} \sigma^{CD}(\overline{\partial_0 g_{CD}})\breve{}
\\
 && +2 \nu^0\sigma_{(B}{}^C(\overline{\partial_{|0|}g_{A)C}})\breve{} - (\nu^0)^2\overline{\partial_0g_{0r}}\sigma_{AB}  + \text{known quantities}
\\
&=&   -\nu^0\partial_r(\overline{\partial_0 g_{AB}})\breve{}
+ \frac{1}{2}( 2\tau\nu^0 +  \overline W{}^0 + \ol{\hat\Gamma}{}^0 )
 (\overline{\partial_0g_{AB}})\breve{}
+2 \nu^0\sigma_{(A}{}^C (\overline{\partial_{|0|}g_{B)C}})\breve{}
\\
 && + \text{known quantities}
 \;,
\end{eqnarray*}
where we used that
\begin{eqnarray}
\overline {\partial_0\Gamma^0_{AB}}
 &=& \frac{1}{2}(\nu^0)^2\overline {\partial_0g_{rr}}\,\overline{\partial_0 g_{AB}} +(\nu^0)^2\overline{\partial_0 g_{0r}} \chi_{AB}
 - \frac{1}{2}\nu^0 \partial_r\overline{\partial_0g_{AB}}
\nonumber
\\
 &&+ \text{known quantities}
\label{0Gamma0AB}
\;.
\end{eqnarray}
Note that $\overline{\partial_0g_{0r}}$, which appears in an intermediate step, cancels out at the end.
We observe that $\breve{\overline R}_{AB}=0$ provides a coupled linear ODE-system for $(\overline{\partial_0 g_{AB}})\breve{}$.

The relevant boundary condition if $\mcN$ is a light-cone is \cite[Section~4.5]{CCM2} $\lim_{r\rightarrow 0}(\overline {\partial_0 g_{AB}})\breve{}= 0$,
in the case of two transversally intersecting characteristic surfaces the boundary condition is determined by the shear of $\mcN_2$.

\paragraph*{Einstein equation $\overline R_{0r}=0$:}

We have
\begin{eqnarray*}
 \overline R_{0r}
 &=& - \overline{\partial_0\Gamma^r_{rr}} - \overline{\partial_0\Gamma^A_{rA}}  + \text{known quantities}
\\
 &=&  \frac{1}{2}\nu^0\overline{\partial^2_{00}g_{rr}} - \nu^0(\partial_r-\kappa)\overline{\partial_0 g_{0r}}   + \text{known quantities}
 \;,
\end{eqnarray*}
when taking into account that
\begin{eqnarray*}
 \overline{\partial_0\Gamma^r_{rr}}
 &=&  \nu^0(\partial_r\overline{\partial_0 g_{0r}} - \frac{1}{2}\overline{\partial^2_{00} g_{rr}})
- (\nu^0)^2 (\partial_r\nu_0 - \frac{1}{2}\overline{\partial_0 g_{rr}}) \overline {\partial_0g_{0r}}
 + \text{known quantities}
\;,
\\
  \overline{\partial_0\Gamma^A_{rB}}
 &=& \text{known quantities}
 \;.
\end{eqnarray*}
Moreover, employing
\begin{eqnarray*}
 \overline{\partial_0\Gamma^0_{0r}}
 &=& \frac{1}{2}\nu^0 \overline{\partial^2_{00} g_{rr}} - \frac{1}{2} (\nu^0)^2\overline{\partial_0 g_{rr}} \, \overline{\partial_0 g_{0r}} + \text{known quantities}
 \;,
\\
 \overline{\partial_0\Gamma^0_{ir}}
 &=&  \text{known quantities}
 \;.
\end{eqnarray*}
we find that
\begin{eqnarray*}
 \overline{\partial_0H^0} &=& \overline{\partial_0g^{\mu\nu}}( \overline \Gamma{}^0_{\mu\nu} - \ol{\hat\Gamma}{}^0_{\mu\nu}) + \overline g^{\mu\nu}(\overline{\partial_0\Gamma^0_{\mu\nu}}- \overline{\partial_0\hat\Gamma^0_{\mu\nu}})
  - \overline{\partial_0 W^0}
\\
 &=& 2\nu^0\overline{\partial_0\Gamma^0_{0r}} + \overline g^{rr}\overline{\partial_0\Gamma^0_{rr}} + 2\overline g^{rA}\overline{\partial_0\Gamma^0_{rA}} + \overline g^{AB}\overline{\partial_0\Gamma^0_{AB}}
 - 2(\nu^0)^3\overline{\partial_0 g_{rr}}\,\overline{\partial_0 g_{0r}}
\\
 && -2(\nu^0)^2\ol{\partial_0 g_{0r}}\hat\Gamma^0_{0r}-\overline{\partial_0 W^0} + \text{known quantities}
 \\
 &=& (\nu^0)^2\overline{\partial^2_{00}g_{rr}} - [2 (\nu^0)^2(\tau
+\hat\Gamma^0_{0r} )
+ 3\nu^0(\overline W{}^0+\ol{\hat\Gamma}{}^0)
  ] \overline{\partial_0 g_{0r}}  - \overline{\partial_0 W^0}
\\
&&
+ \text{known quantities}
 \;.
\end{eqnarray*}
Consequently, the gauge condition $\overline{\partial_0 H^0}=0$ can be used to rewrite
the Einstein equation $\overline R_{0r}=0$ as a linear ODE for $\overline{\partial_0g_{0r}}$,
or, depending on the setting, as an algebraic equation for the gauge source function   $ \overline{\partial_0 W^0}$,
\begin{equation}
 \Big[\partial_r - \tau - \kappa -\hat\Gamma^0_{0r} -\frac{3}{2}\nu_0 (\overline W{}^0 + \ol{\hat\Gamma}{}^0)   \Big] \overline{\partial_0g_{0r}} = \frac{1}{2}(\nu_0)^2\overline{\partial_0W^0} + \text{known quantities}
 \;.
 \label{trans_constraint1}
\end{equation}
The boundary condition if $\mcN$ is a light-cone is \cite[Section~4.5]{CCM2} $ \lim_{r\rightarrow 0}\overline {\partial_0 g_{0r}}= 0$;
in the case of two   characteristic surfaces intersecting transversally  it is determined by
$\lim_{u\rightarrow 0} \partial_u g_{ur}|_{\mcN_2}$.

Once $\overline{\partial_0g_{0r}}$ (or $ \overline{\partial_0 W^0}$) has been determined, we obtain $\overline{\partial^2_{00}g_{rr}}$ algebraically from the gauge condition $\overline{\partial_0 H^0}=0$.

\paragraph*{Einstein equations $\overline R_{0A}=0$:}

We find
\begin{eqnarray*}
 \overline R_{0A}
 &=&  - \overline{\partial_0\Gamma^r_{rA}} - \overline{\partial_0\Gamma^B_{AB}} + \nu^0\chi_A{}^B\overline{\partial_0g_{0B}} + \text{known quantities}
\\
 &=& \frac{1}{2}\nu^0\overline{\partial^2_{00}g_{rA}}-\frac{1}{2}\nu^0\partial_r\overline{\partial_0g_{0A}} +\nu^0\chi_A{}^B\overline{\partial_0g_{0B}} + \text{known quantities}
 \;,
\end{eqnarray*}
where we used that
\begin{eqnarray*}
 \overline{\partial_0\Gamma^r_{rA}}
 &=& -\nu^0 \chi_A{}^B \overline {\partial_0 g_{0B}} + \frac{1}{2}\nu^0 (\partial_r\overline{\partial_0 g_{0A}} - \overline{\partial^2_{00} g_{rA}})
 + \text{known quantities}
\;,
\\
 \overline{\partial_0\Gamma^C_{AB}}
 &=& \nu^0\chi_{AB} \overline g^{CD} \overline{\partial_0 g_{0D}} + \text{known quantities}
 \;.
\end{eqnarray*}
From
\begin{eqnarray*}
 \overline{\partial_0\Gamma^A_{0r}}
 &=& -\frac{1}{2} \nu^0 \overline g^{AB} \overline{\partial_0 g_{rr}}\,\overline{\partial_0 g_{0B}}  + \frac{1}{2}\overline g^{AB} (\overline{\partial^2_{00} g_{rB}} + \partial_r\overline{\partial_0 g_{0B}})
 + \text{known quantities}
\;,
\\
 \overline{\partial_0\Gamma^A_{rr}}
 &=& \text{known quantities}
 \;,
\end{eqnarray*}
we obtain for the angle-components of the $u$-differentiated wave-gauge vector,
\begin{eqnarray*}
  \overline{\partial_0H^A} &=& \overline{\partial_0g^{\mu\nu}}( \overline \Gamma{}^A_{\mu\nu} - \hat\Gamma{}^A_{\mu\nu}) + \overline g^{\mu\nu}(\overline{\partial_0\Gamma^A_{\mu\nu}}- \overline{\partial_0\hat \Gamma^A_{\mu\nu}}) - \overline{\partial_0 W^A}
\\
 &=& -(\nu^0)^2\overline{\partial_0g_{rr}}\overline g^{AB}\overline{\partial_0g_{0B}}
 -2\nu^0 \overline g^{BC}\overline{\partial_0g_{0C}}(\chi_B{}^A - \hat\chi_B{}^A)
 +2\nu^0\overline{\partial_0\Gamma^A_{0r}}
\\
 &&
 + \overline g^{ij}\overline{\partial_0\Gamma^A_{ij}}- \overline{\partial_0 W^A}
  + \text{known quantities}
\\
 &=& \nu^0\overline g^{AB}\overline{\partial^2_{00}g_{rB}}   + \nu^0\overline g^{AB}\partial_r\overline{\partial_0g_{0B}}  -(\tau\nu^0 +  2\overline W{}^0 + 2\ol{\hat\Gamma}{}^0 )\overline g^{AB}\overline{\partial_0g_{0B}}
\\
 && -2\nu^0 \overline g^{BC}\overline{\partial_0g_{0C}}(\chi_B{}^A - \hat\chi_B{}^A)   - \overline{\partial_0 W^A}
+ \text{known quantities}
 \;.
\end{eqnarray*}
The gauge condition $\overline{\partial_0 H^A} = 0$ can be used to rewrite $\overline R_{0A}=0$
as a coupled linear ODE-system for $\overline{\partial_0 g_{0A}}$ or,  depending on the setting,  as an algebraic equation  for  $\overline{\partial_0 W^A}$,
\begin{eqnarray}
  \Big[\partial_r - \frac{3}{2}\tau + \frac{1}{2}\hat \tau  -\nu_0(\overline W{}^0 + \ol{\hat\Gamma}{}^0) \Big] \overline{\partial_0g_{0A}} - (2\sigma_A{}^B- \ol g_{AD} \ol g^{BC} \hat\sigma_C{}^D) \overline{\partial_0g_{0B}}
\nonumber
\\
  = \frac{1}{2}\nu_0 g_{AB} \overline{\partial_0W^B} + \text{known quantities}
 \;.
 \label{trans_constraint2}
\end{eqnarray}
In the light-cone-case \cite[Section~4.5]{CCM2} as well as in the case of two transversally intersecting null hypersurfaces the boundary condition is
$ \lim_{r\rightarrow 0}\overline {\partial_0 g_{0A}} = 0$.

The gauge condition $\overline{\partial_0 H^A} =0$ then determines $\ol{\partial^2_{00}g_{rA}}$ algebraically.

\paragraph*{Einstein equation $\overline R_{00}=0$:}

Finally, we have
\begin{eqnarray*}
 \overline R_{00}
 &=&\nu^0 \Big( \frac{1}{2}\partial_r + \kappa + \frac{\tau}{2} - \nu^0\partial_r \nu_0 \Big) \overline{\partial_0g_{00}} - \overline{\partial_0\Gamma^r_{0r}} - \overline{\partial_0\Gamma^A_{0A}}
+ \text{known quantities}
\\
 &=& \frac{1}{2}\nu^0 \tau \overline{\partial_0g_{00}}  - \frac{1}{2}\overline g^{AB}\overline{\partial^2_{00}g_{AB}}  + \text{known quantities}
 \;.
\end{eqnarray*}
For that calculation we used that
\begin{eqnarray*}
 \overline{\partial_0\Gamma^r_{0r}}
 &=& -\frac{1}{2}(\nu^0)^2\overline{\partial_0 g_{rr}}\,\overline{\partial_0 g_{00}} + \frac{1}{2} \nu^0 \partial_r \overline{\partial_0 g_{00}} + \text{known quantities}
\;,
\\
 \overline{\partial_0\Gamma^A_{0B}}
 &=& \frac{1}{2} \overline g^{AC} \overline{\partial^2_{00}g_{BC}} + \text{known quantities}
 \;.
\end{eqnarray*}
Taking into account  that
\begin{eqnarray*}
 \overline {\partial_0 \Gamma^r_{AB}}
 &=& (\nu^0)^2 \chi_{AB}  \overline{\partial_0 g_{00}}- \frac{1}{2}\nu^0 \overline {\partial^2_{00} g_{AB}}  + \text{known quantities}
\;,
\end{eqnarray*}
we compute the $r$-component of the $u$-differentiated wave-gauge vector,
\begin{eqnarray*}
 \overline{\partial_0 H^r} &=&  \overline{\partial_0g^{\mu\nu}}( \overline \Gamma{}^r_{\mu\nu} - \hat\Gamma{}^r_{\mu\nu}) + \overline g^{\mu\nu}(\overline{\partial_0\Gamma^r_{\mu\nu}}-\overline{\partial_0\hat \Gamma^r_{\mu\nu}}) - \overline{\partial_0 W^r}
\\
 &=& -(\nu^0)^2\Big( \kappa -\hat\kappa + \frac{1}{2}\nu^0\overline{\partial_0 g_{rr}} \Big)\overline{\partial_0g_{00}}  +2\nu^0\overline{\partial_0\Gamma^r_{0r}}    + \overline g^{ij}\overline{\partial_0\Gamma^r_{ij}}
\\
 &&  - \overline{\partial_0 W^r}+ \text{known quantities}
\\
 &=&  (\nu^0)^2\Big[\partial_r - \frac{\tau}{2} - \kappa + \hat \kappa -  \frac{3}{2} \nu_0 (\overline W{}^0 + \ol{\hat\Gamma}{}^0)\Big]\overline{\partial_0g_{00}}
-\frac{1}{2}\nu^0 \overline g^{AB} \overline{\partial^2_{00}g_{AB}}
\\
 && - \overline{\partial_0 W^r} + \text{known quantities}
 \;.
\end{eqnarray*}
Once again we exploit the gauge condition,  here $\overline{\partial_0H^r}=0$, to rewrite $\overline R_{00} =0$ as a linear ODE for $\overline{\partial_0 g_{00}}$, or, alternatively, an algebraic equation for $\overline{\partial_0 W^r}$
\begin{equation}
  \Big[\partial_r -\tau - \kappa + \hat \kappa  - \frac{3}{2}\nu_0(\overline W{}^0 + \ol{\hat \Gamma}{}^0) \Big] \overline{\partial_0g_{00}} = (\nu_0)^2\overline{\partial_0W^r} + \text{known quantities}
 \;.
 \label{trans_constraint3}
\end{equation}
For both light-cones \cite[Section~4.5]{CCM2} and transversally intersecting null hypersurfaces, the boundary condition is
$  \lim_{r\rightarrow 0}\overline {\partial_0 g_{00}} =0$.
The gauge condition $\overline{\partial_0H^r}=0$ then determines $\overline g^{AB} \overline{\partial^2_{00} g_{AB}}$ algebraically.

The gauge scheme we want to use here now works as follows:
We prescribe $\gamma$, $\kappa$, $\ol g_{0\mu}$ and $\ol{\partial_0 g_{0\mu}}$. The Raychaudhuri equation and the vacuum Einstein
wave-map gauge constraints then determine $\ol g_{AB}$ and $\ol W^{\lambda}$.  We then solve the hierarchical system of equations derived  above to
compute $\ol {\partial_0 W^{\lambda}}$ from \eq{trans_constraint1}, \eq{trans_constraint2} and \eq{trans_constraint3}.
We choose a  smooth space-time vector field $W^{\lambda}$ which induces
the computed values for $\ol W^{\lambda}$ and $\ol {\partial_0 W^{\lambda}}$  on $\mcN$ (in the light-cone case
it is a non-trivial issue whether such an extension off the cone exists near the tip, we return to this issue below).
Then we solve the reduced Einstein equations (or the conformal field equations, cf.\ Section~\ref{sec_sufficient}). It is well-known that the solution
induces the prescribed data for the metric tensor on the initial surface and satisfies $H^{\lambda}=0$
with the prescribed  source vector field $W^{\lambda}$, and thus solves the full Einstein equations.
But then it follows from the equations \eq{trans_constraint1}-\eq{trans_constraint3},
together with the relevant boundary conditions at the vertex or the intersection manifold, respectively, that the desired values
for  $\ol{\partial_0 g_{0\mu}}$ are induced, as well.

Some comments in the light-cone-case concerning regularity at the vertex are in order:
What we have described above is exactly the strategy on two transverse characteristic surfaces
(some care is needed to obtain a $W^{\lambda}$ which is continuous at the intersection manifold, cf.\ \cite[Section~3.1]{ChPaetz}).
 On a null cone there is
a difficulty related to the behavior  near the tip:
In \cite{C1} it has been shown that if the metric  $\ol g$ is the restriction to the cone of a smooth space-time metric, then  $\ol W^{\lambda}$
does admit a smooth extension (and the known well-posedness result is applicable).
However,
we also want to prescribe $\ol{\partial_0 g_{0\mu}}$,  whence  $\ol{\partial_0 W^{\lambda}}$ needs to be of a specific form on the cone, as well.
It seems difficult to ensure that $\overline W{}^{\lambda}$ and $\ol{\partial_0 W^{\lambda}}$ arise from
a smooth vector field $W^{\lambda}$ in space-time if one proceeds this way.
To avoid dealing with the issue~of extendability near the tip of the cone, the gauge scheme will be somewhat modified: 

Since we are merely interested in the asymptotic behavior of the fields, it suffices for our purposes to prescribe $\ol g_{0\mu}$ and $\ol{\partial_0 g_{0\mu}}$
just for large $r$, say $r>r_2$. For small $r$, say $r<r_1<r_2$ we can  use a conventional scheme where for instance   $\gamma$, $\kappa$ and $W^{\lambda}$ are prescribed with e.g.\ $W^{\lambda}=0$. Then the regularity issue near the tip of the cone is well-understood  \cite{C1}.
A smooth transition in the regime $r_1<r<r_2$ is obtained via cut-off functions:

Let $\chi\in C^\infty(\R)$ be any non-negative non-increasing function satisfying $\chi(r)=0$ for $r\ge r_2$ and $\chi = 1$ for $r\le r_1$.
Let $( \ol g_{0\mu})\mathring{}$ and $(\ol{\partial_0 g_{0\mu}})\mathring{}$ denote the values of $\ol g_{0\mu}$ and $\ol{\partial_0 g_{0\mu}}$ computed from $(\gamma,\kappa)$ for say $W^{\lambda}=0$,
and denote by $ (\ol g_{0\mu})^\dagger$ and $(\ol{\partial_0 g_{0\mu}})^\dagger$ those values
we want  realize for $r>r_2$.
We then  set
\begin{eqnarray}
 \ol g_{0\mu} &=& (1-\chi)(\ol g_{0\mu})^\dagger+\chi (\ol{ g}_{0\mu})\mathring{}
 \;,
  \label{gauge_nu_0}
\\
 \ol{\partial_0 g_{0\mu}} &=& (1-\chi)(\ol{\partial_0 g_{0\mu}})^\dagger+ \chi (\ol{\partial_0 g_{0\mu}})\mathring{}
 \;.
  \label{gauge_nu_A}
\end{eqnarray}
For these new values for $\ol g_{0\mu}$ and $\ol{\partial_0 g_{0\mu}}$ we go through the above scheme again.
By construction we still obtain the desired values for $\ol g_{0\mu}$ and $\ol{\partial_0 g_{0\mu}}$ for $r>r_2$, but now we also have $\ol W^{\lambda}=0=\ol{\partial_0 W^{\lambda}}$ for $r<r_1$, and the extension problem becomes trivial:
Since there are no  difficulties in extending a vector field defined along the light-cone to a neighborhood of the cone away from the tip,
we conclude that $\overline W{}^{\lambda}$ and $\ol{\partial_0 W^{\lambda}}$  arise from the restriction to the cone of some smooth vector field $W^{\lambda}$.

The  coordinate $r$ which parameterizes the null rays generating the initial surface
is determined to a large extent by the choice of $\kappa$. We would like to choose an $r$ for which the expansion on $\mcN$ takes the form
$\tau=2/r$. On a light-cone this is straightforward, we simply  choose $\kappa=\frac{r}{2}|\sigma|^2$. It follows from \cite{C1} that this choice can be made up to the vertex, and it follows from the Raychaudhuri equation and regularity at the vertex that $\tau$ takes the desired form.

If $\mcN$ is one of two transversally intersecting characteristic surfaces we cannot achieve $\tau=2/r$ globally, for the expansion needs
to be regular on the intersection manifold.
We would like to construct a $\kappa$ such that $\tau=2/r$  for large $r$.
Let us therefore again modify our gauge scheme slightly to bypass this issue:

Instead of $\kappa$ (supplemented by the initial data for the $\varphi$-equation $\varphi|_S\ne 0$ and $\partial_r\varphi|_S$
in the case of two characteristic surfaces, cf.\ \cite{Rendall,ChPaetz}), we prescribe the function $\varphi$ on $\mcN$.
An analysis of the Raychaudhuri equation~\eq{constraint_phi}
 shows that, \emph{for given $\kappa$},  the existence of a nowhere vanishing $\varphi$  on $\mcN$ (which can then without loss of generality be taken to be positive), and which further satisfies  $\varphi_{-1}>0$  \emph{implies} $\partial_r\varphi>0$ on $\mcN$.%
\footnote{
Indeed, we observe that for  $\kappa=0$ any  globally positive $\varphi$ is concave in $r$.
Together with the required positivity of $\varphi_{-1}$ this implies  that
$\partial_r\varphi$ needs to be positive for all $r$. The case $\kappa\ne 0$ can be reduced to $\kappa=0$, cf.\ \cite[Section~2.4]{ChPaetz2}.
}
When prescribing $\varphi$ rather than $\kappa$ we therefore assume that $\varphi$ is a strictly increasing, positive function with   $\varphi_{-1}>0$.
Indeed, the function $\kappa$ is then determined algebraically via the $\varphi$-equation (cf.\  \cite{ChPaetz}).

In this section we have established the following setting:
On $\mcN$ we regard the conformal class
of the family $\gamma=\gamma_{AB}\mathrm{d}x^A\mathrm{d}x^B$ of Riemannian metrics as the ``physical'' initial data
(in the case of two transversally intersecting null hypersurfaces supplemented by data on $\mcN_2$ and $S$), while
 the functions
\begin{equation}
 \varphi\;, \quad \ol g_{0\mu}\;, \quad \ol{\partial_0 g_{0\mu}}
\;,
\label{gauge_functions}
\end{equation}
rather than $\kappa$, $\ol W^{\mu}$ and $\ol{\partial_0 W^{\mu}}$, are regarded
as gauge functions on $\mcN$, at least for $r>r_2$ (recall that $\varphi$ , $\partial_r\varphi$ and $\nu^0$ need to be  positive).


\subsection{Metric gauge and some conventions}

We now  impose specific values for the gauge functions \eq{gauge_functions}.
Guided by their Minkowskian values we  set, for $r>r_2$,
\bel{15VII11.2}
 \varphi =  r
 \;,
 \quad
 \nu_0 = 1
 \;,
 \quad
 \nu_A = 0
 \;,
 \quad
 \ol g_{00} = -1
\;,
\quad
 \ol{\partial_0 g_{0\mu}} =0
 \;.
\ee
Moreover, we assume a Minkowski target for $r>r_2$,
\begin{equation}
 \hat g  = \eta \equiv -\mathrm{d}u^2 + 2\mathrm{d}u\mathrm{d}r + r^2s_{AB}\mathrm{d}x^A\mathrm{d}x^B
 \;.
 \label{Minktarget}
\end{equation}

From now on everything will be expressed in this gauge, which we call \emph{metric gauge}.
Although it may not explicitly  be  mentioned each time, since \eq{15VII11.2} and \eq{Minktarget} are merely assumed to hold  for $r>r_2$ we will tacitly assume henceforth that all the equations are meant to hold in this regime.

The relation between  metric gauge and harmonic coordinates is discussed in Appendix~\ref{meaning_kappa},
where we address the  ``paradox'' that in the metric gauge we do not need to impose  conditions on $\gamma$ while in the harmonic gauge
we do need to do it to  make sure that the constraint equations admit a global solution.

We have not discussed here what in this gauge scheme (which may be used on $\mcN_2$ as well) the free data on the intersection manifold are, and we leave
it to the reader to work this out. We are interested in the asymptotic regime, 
and the main object of this section was to show
that the choice \eq{15VII11.2} can be made without any geometric restrictions  for large $r$, if one does not prescribe $\kappa$ and $W^{\lambda}$ anymore,
but treats them as unknowns determined by the constraints.



\subsection{Solution of the constraint equations in the metric gauge}

We solve the Einstein vacuum constraints in the metric gauge for  $\kappa$ and  $\overline W{}^{\lambda}$,
where we assume the initial data $\gamma$ to be of the form \eq{initial_data}.
Recall our convention that all equations are meant to hold for $r>r_2$.

Equation \eq{constraint_phi}
yields with $\varphi=r$
\begin{eqnarray}
  \kappa &=& (\partial_r\varphi)^{-1} \Big(\partial_{rr}^2\varphi + \frac{1}{2}|\sigma|^2\varphi \Big) \,=\, \frac{1}{2}r|\sigma|^2
\label{Bondi_kappa}
\\
 &=& \frac{1}{2}\sigma_4 r^{-3} + \mathcal{O}(r^{-4})
\,=\,  \frac{1}{8}|\breve h^{(1)}|^2 r^{-3} + \mathcal{O}(r^{-4})
 \;.
\label{expansion_kappa_mgauge}
\end{eqnarray}
We further note that $\varphi=r$  implies
\begin{equation}
 \tau =2/r
\;,
\end{equation}
as desired, in particular $\tau_2=0$.
We emphasize that the argument used for the No-Go Theorem in \cite{ChPaetz2}, that the coefficient $\tau_2$ 
vanishes only for Minkowski data,
does not apply when $\kappa\ne 0$. In our case $\kappa$ does not vanish unless $|\sigma|^2\equiv 0$.

It follows from \eq{expansion_gAB}, \eq{15VII11.2} and \eq{expansion_kappa_mgauge}
that
\begin{equation}
 \ol g_{AB}  \,=\, \varphi^2\sqrt{\frac{\det s}{\det \gamma}} \gamma_{AB}
\,=\, r^2s_{AB} + r\breve h^{(1)}_{AB} + \frac{1}{4}|\breve h^{(1)}|^2s_{AB} - (\sigma_A{}^C)_3 s_{BC} + \mathcal{O}(r^{-1})
 \;.
\end{equation}
Moreover,
\begin{eqnarray}
 \overline {W}{}^0 &=& -(2\partial_r  +\tau + 2\kappa)\nu^0 - \ol{\hat\Gamma}^0
\,=\,  -2r^{-1}  - r|\sigma|^2 +  r\overline g^{AB}s_{AB}
 \\
&=&\frac{1}{4}|\breve h^{(1)}|^2 r^{-3} + \mathcal{O}(r^{-4})
 \;,
\end{eqnarray}
where we used that
\begin{equation}
 \ol{\hat\Gamma}^0 \,= \, - r\overline g^{AB}s_{AB} \,=\, -2r^{-1} -\frac{1}{2}|\breve h^{(1)}|^2r^{-3} + \mathcal{O}(r^{-4})
 \;.
\end{equation}

The equation \eq{eqn_nuA_general}
which determines $\xi_A$ cannot be solved in explicit form. The asymptotic analysis of its solutions has already been done in Section~\ref{sec_exp_xiA}.
The asymptotic expansion of $\xi_A$ does not contain logarithmic terms if and only if the no-logs-condition is fulfilled, which in the metric gauge where $\tau_2=0$ reduces to
\begin{equation}
(\sigma_A{}^B)_3=0
 \;.
\label{no-logs_mgauge}
\end{equation}
%
%
Assuming \eq{no-logs_mgauge} one finds
\begin{eqnarray}
 \xi_A
 \,=\, -\mathring{\nabla}^B \breve h_{AB}^{(1)}r^{-1} + C_A^{(\xi_B)}+ \mathcal{O}(r^{-3})
 \;.
\label{mgauge_exp_xiA}
\end{eqnarray}
%
Now we can solve \eq{eqn_xiA}
for $\ol W^A$,
\begin{eqnarray}
 \overline W{}^A
\,=\,  \xi^A + \ol g^{CD}\check\Gamma^A_{CD} - \hat\Gamma^A
 \,=\, \mathcal{O}(r^{-4})
 \;,
\label{exp_WA}
\end{eqnarray}
where we used
\begin{equation}
  \hat\Gamma^A = \ol g^{CD}\mathring\Gamma^A_{CD} \,=\, s^{CD}\mathring\Gamma^A_{CD} r^{-2}
- \breve h^{(1)CD}\mathring\Gamma^A_{CD} r^{-3}
 + \mathcal{O}(r^{-4})
\;.
\end{equation}
Note that the two leading-order terms in \eq{exp_WA} cancel out.

The $\zeta$-equation \eq{zeta_constraint}
cannot be solved analytically, its asymptotic solution, which  has been determined in Section~\ref{sec_exp_zeta},
does not involve log terms,
%
\begin{eqnarray}
  \zeta
 &=& -2r^{-1} + C^{(\zeta)} r^{-2} + \mathcal{O}(r^{-3})
 \;.
\label{mgauge_exp_zeta}
\end{eqnarray}
Finally, we solve the constraint equation \eq{dfn_zeta}
for $\overline{W}{}^r$,
\begin{eqnarray}
 \overline W{}^r  &=&\frac{1}{2}\zeta - \big(\partial_1 + \frac{1}{2}\tau + \kappa\big)\overline g^{rr} -\ol{\hat\Gamma}^r
\nonumber
\\
&=& \frac{1}{2}\zeta - r^{-1} -\frac{r}{2}|\sigma|^2 +  r\overline g^{AB}s_{AB}
\,=\, \frac{1}{2}C^{(\zeta)} + \mathcal{O}(r^{-3})
\;,
\end{eqnarray}
as
\begin{eqnarray}
\ol{\hat\Gamma}^r \,=\, -r\ol g^{AB}s_{AB} \,=\,  -2r^{-1} -\frac{1}{2}|\breve h^{(1)}|^2r^{-3} + \mathcal{O}(r^{-4})
\;.
\end{eqnarray}

\subsection{Overview of the metric gauge I}
\label{overview_mgauge}

In  the metric gauge one treats, at least for large $r$, some of the metric components as gauge degrees of freedom rather than $\kappa$, $ \ol W{}^{\lambda}$ and $\ol{\partial_0 W^{\lambda}}$.
This provides  the decisive advantage that one obtains more explicit solutions of the constraint equations
since most of them are algebraic equations rather than ODEs in this setting,
so that the asymptotic expansions contain less integration functions,  whose values are not explicitly known.
Moreover, the  computations of Schouten tensor, Weyl tensor  etc.\ (as needed in Section~\ref{sec_sufficient} for Friedrich's equations) will be simplified significantly by the fact
 that several metric components  take their Minkowskian values.

We have seen in Section~\ref{solutions_constraints}
that many log terms are produced due to a bad choice of coordinates.
In the metric gauge  all the gauge-dependent conditions of Theorem~\ref{thm_asympt_exp} (cf.\ Remark~\ref{remark_special_case} and
compare \eq{metric_gauge_kappa}-\eq{com_metric_gaugeII} below with \eq{special_case_1}-\eq{special_case_f}),
are satisfied  and we are left with the gauge-invariant no-logs-condition, which
can be expressed as an explicit condition on $\gamma$ in this gauge,
\begin{equation}
0\,=\,  (\sigma_A^{\phantom{A}B})_3 \,=\,  \frac{1}{2}h^{(1)}\breve h_A^{(1)B} - \breve h^{(2)B}_{A}
 \;.
 \label{relation_h1_h2}
\end{equation}
Consequently we  can freely prescribe  all the $h^{(n)}_{AB}$'s except for $\breve h^{(2)}_{AB}$  which is determined
by \eq{relation_h1_h2}.

It is easy to find sufficient conditions on the initial data $\gamma$ such that the no-logs-condition~\eq{no-log-conditions}
in its general form is fulfilled (just take $\breve h^{(1)}_{AB}=\breve h^{(2)}_{AB} =0$), i.e.\ Theorem~\ref{thm_asympt_exp}
shows that there is a large class of initial data for which the metric admits a smooth conformal completion at infinity.
To find necessary conditions on $\gamma$ is more involved, since the expansion coefficient $\tau_2$ is in general not explicitly known.
In the metric gauge, though, $\tau_2$ \emph{is} explicitly known, whence it is easily possible to \emph{characterize} the initial data sets completely which lead to the restriction of a metric to $\mcN$
which admits a smooth conformal completion at infinity.
Moreover, it becomes obvious that for ``generic'' initial data the no-logs-condition will be violated.

Note that we have not studied yet the asymptotic  behavior of  the unknowns of the conformal field equations which also involve transverse derivatives of the metric.
The gauge functions are more complicated in the metric gauge and contain certain  integration functions, and one might wonder whether log-terms arise at some later stage.
Luckily, it turns out (cf.\ Section~\ref{sec_sufficient}) that this is not the case, and that the no-logs-condition is necessary-and-sufficient to guarantee smoothness of all the relevant fields
at conformal infinity .

 Of course, once we have determined the gauge functions it is possible to return to the original viewpoint and regard them as the relevant gauge degrees of freedom.
This leads to an \emph{initial data-dependent gauge}.
 Assuming  that the initial data $\gamma$ satisfy the no-logs-condition \eq{relation_h1_h2}
we give their values for   $r> r_2$ (the functions $\ol{\partial_0 W^{\lambda}}$ will be computed in Section~\ref{Computation of Transverse Derivatives}):
\begin{eqnarray}
 \kappa &=& \frac{1}{2} r |\sigma|^2 \, = \, \frac{1}{8}|\breve h^{(1)}|^2 r^{-3} + \mathcal{O}(r^{-4})
\label{metric_gauge_kappa}
\;,
\\
 \overline {W}{}^0 &=& - r|\sigma|^2 - \frac{2}{r} - \ol{\hat\Gamma}^0 \,=\,  \frac{1}{4}|\breve h^{(1)}|^2 r^{-3} + \mathcal{O}(r^{-4})
\;,
\label{metric_gauge_W0}
\\
 \overline W{}^A &=&  \xi^A + \ol g^{CD}\check\Gamma^A_{CD} - \hat\Gamma^A \,=\,  \mathcal{O}(r^{-4})
 \;,
\label{metric_gauge_WA}
\\
\overline W{}^r  &=& \frac{1}{2}\zeta - r^{-1} - \frac{1}{2}r|\sigma|^2- \ol{\hat\Gamma}^r
\,=\, \mathcal{O}(r^{-2})
\;,
\label{metric_gauge_W1}
\end{eqnarray}
with $\xi_A$ and $\zeta$ given by \eq{mgauge_exp_xiA} and \eq{mgauge_exp_zeta}, respectively.
The restrictions to $\mcN$ of the metric components then take the form (again for $r> r_2$):
\begin{eqnarray}
 &\overline g_{00} \,=\, -1\;, \quad \nu_0 \,=\, 1\;, \quad  \nu_A \,=\, 0\;, \quad \overline  g_{rr}\,=\, \overline  g_{rA} \,=\, 0 \;, &
 \label{com_metric_gaugeI}
\\
& \overline g_{AB}
 \,=\, r^2s_{AB}+r\breve h^{(1)}_{AB}+ \frac{1}{4}|\breve h^{(1)}|^2 s_{AB} + \mathcal{O}(r^{-1})
 \;.&
 \label{com_metric_gaugeII}
\end{eqnarray}

\begin{remark}
\label{rem_wave_cond}
\em{
The asymptotic expansion of $\ol g_{AB}$  depends only on the gauge choice for $\kappa$. Our $\kappa$, equation \eq{metric_gauge_kappa}, coincides with the $\kappa$
used to define Bondi coordinates (where one also requires $\tau=2/r$).
The  no-logs-condition which, in a $\kappa=\frac{r}{2}|\sigma|^2$-gauge is equivalent to the absence of $s$-trace-free terms in $(\ol g_{AB})_0$,
$$(\ol g_{AB})\breve{}_0=0\;, $$
recovers the ``outgoing wave condition'' imposed a priori by Bondi et al.\ and by Sachs to inhibit the appearance
of logarithmic terms (cf.\ e.g.\ \cite{bondi}).

For an arbitrary $\kappa=\mathcal{O}(r^{-3})$ the correct generalization of this condition, equivalent to the no-logs-condition \eq{no-log-conditions}, is
(cf. \eq{expansion_gAB}),
$$  (\ol g_{AB})\breve{}_0 \,= \,\tau_2 s_{AC}(\sigma_B{}^C)_2 \,=\, -\frac{1}{2}\tau_2\breve h^{(1)}_{AB} \;.$$
}
\end{remark}

\subsection{Transverse derivatives of the metric on $\mcN$}
\label{Computation of Transverse Derivatives}

In this section we  compute the $\ol{\partial_0 W^{\lambda}}$'s and  certain transverse derivatives of  the metric on  $\mcN$  (the asymptotic expansions thereof) in the metric gauge
using the vacuum Einstein equations $R_{\mu\nu}=0$ and assuming that the no-logs-condition is fulfilled, so that
 \eq{metric_gauge_kappa}-\eq{com_metric_gaugeII} and $\ol{\partial_0 g_{0\mu}}=0$ hold.
As before, all equalities are meant to be valid for $r> r_2$, even if this is not mentioned wherever relevant.

It follows from  \eq{expression_H0}-\eq{expression_Hr}  that
\begin{eqnarray}
 \overline{\partial_0g_{rr}} &=& -r|\sigma|^2 \,=\, -2\kappa
 \;,
\\
  \ol{\partial_0g_{rA}}   &=&    \xi_A
 \;,
\\
   \overline g^{AB} \overline{\partial_0g_{AB}}
 &=& -\zeta -\tau
 \;.
\end{eqnarray}
If we plug in the values for $\overline g_{\mu\nu}$ and $\ol{\partial_0 g_{\mu\nu}}$ we find
from \cite[Appendix~A]{CCM2}  the following expressions for the  Christoffel symbols restricted to $\mcN$, which we shall made frequently use of:
\label{christoffels_mgauge}
\begin{eqnarray}
 &\hspace{-3em}\ol\Gamma{}^{\mu}_{00}=\ol\Gamma{}^0_{rr}= \ol\Gamma{}^0_{rA}= \ol\Gamma{}^C_{rr}=0\;,\enspace \ol\Gamma{}^r_{rr}=- \ol\Gamma{}^0_{0r}= -\ol\Gamma{}^r_{0r}=\kappa\;, &
\label{christoffel_list1}
\\
&\hspace{-3em}\ol\Gamma{}^0_{0A}= \ol\Gamma{}^r_{0A}=-\ol\Gamma{}^r_{rA}=\frac{1}{2}\xi_A\;, \quad \ol\Gamma{}^C_{0r}=\frac{1}{2}\xi^C\;, \enspace \ol\Gamma^C_{rA}=\chi_A{}^C\;,\enspace \ol\Gamma{}^0_{AB}=-\chi_{AB}\;, &
\\
& \hspace{-3em} \ol\Gamma{}^r_{AB}=-\frac{1}{2}\ol{\partial_0 g_{AB}}-\chi_{AB}\;, \enspace \ol\Gamma{}^C_{0A}=\frac{1}{2}\ol g^{CD}\ol{\partial_0 g_{AD}}\;, \enspace \ol\Gamma{}^C_{AB}=\check\Gamma^C_{AB}\;.&
\label{christoffel_list3}
\end{eqnarray}

\paragraph*{Einstein equations $\breve{\ol R}_{AB}=0$:}

From \eq{Ricci_Christoffels} we obtain
\begin{eqnarray*}
\ol  R_{AB} &=&\check R_{AB} + \ol{\partial_0\Gamma^0_{AB}} + (\partial_r+ \ol  \Gamma{}^C_{rC})\ol \Gamma{}^r_{AB}
 + \ol \Gamma{}^0_{AB}(\ol \Gamma{}^0_{00}+ \ol \Gamma{}^r_{0r}+\ol  \Gamma{}^C_{0C})
\\
 &&  - 2\ol \Gamma{}^C_{0(A} \ol \Gamma{}^0_{B)C}
 -2 \ol \Gamma{}^r_{rA}\ol \Gamma{}^r_{rB} - 2\ol \Gamma{}^C_{r(A}\ol \Gamma{}^r_{B)C}
 \;.
\end{eqnarray*}
%
%
First of all we have to determine the $u$-differentiated Christoffel symbol,
\begin{eqnarray}
 \ol{\partial_0 \Gamma^0_{AB} }
 &=&  -\big(\frac{1}{2}\partial_r+\kappa\big)\ol{\partial_0g_{AB}} -2\kappa \chi_{AB}
 +\check\nabla_{(A}\xi_{B)}
 \;.
\end{eqnarray}
Employing \eq{christoffel_list1}-\eq{christoffel_list3} and the relation $\check R_{AB} = \frac{1}{2}\check R \ol g_{AB}$ we obtain after some simplifications that
\begin{eqnarray}
\breve{ \ol R}_{AB}
 &=& -\Big(\partial_r  -\frac{1}{r} + \frac{r}{2}|\sigma|^2 \Big) (\ol{\partial_0g_{AB} })\breve{}+ 2\sigma_{(A}{}^C(\ol{\partial_{|0|}g_{B)C}})\breve{}
 - (\partial_{r}\sigma_{AB})\breve{}  
 \nonumber
\\
 &&+\Big( \frac{\zeta}{2} + \frac{1}{r} - \frac{r}{2}|\sigma|^2\Big)\sigma_{AB}
 -\frac{1}{2}(\xi_A\xi_B)\breve{} +(\check\nabla_{(A} \xi_{B)})\breve{}
 \label{eqn_0RAB}
 \;,
\end{eqnarray}
which should vanish in vacuum.
The equations $\breve{\ol R}_{AB}=0$ form a closed linear ODE-system  for $(\ol{\partial_0 g_{AB}})\breve{}$.
Since it is generally hopeless to look for an analytic solution, we content ourselves with computing  the asymptotic solution.
For this it is convenient to rewrite the matrix equation \eq{eqn_0RAB}  as a vector equation,
\begin{eqnarray*}
\left[ \partial_r
 + \begin{pmatrix} \left(\frac{r}{2}|\sigma|^2 - \frac{1}{r}\right) - 2\sigma_2{}^2 & - 2\sigma_2{}^3 & 0 \\
 - \sigma_3{}^2 & \left(\frac{r}{2}|\sigma|^2 - \frac{1}{r}\right)  &  - \sigma_2{}^3 \\
 0 & -2\sigma_3{}^2 & \left(\frac{r}{2}|\sigma|^2-\frac{1}{r}\right) - 2\sigma_3{}^3
 \end{pmatrix}\right]
\begin{pmatrix} (\ol{\partial_0g_{22}})\breve{} \\ (\ol{\partial_0g_{23}})\breve{} \\ (\ol{\partial_0 g_{33}})\breve{} \end{pmatrix}
 = \begin{pmatrix}
    q_{22} \\ q_{23} \\ q_{33}
   \end{pmatrix}
 \;,
\end{eqnarray*}
where
\begin{eqnarray*}
 q_{AB} \,:= \, - (\partial_{r}\sigma_{AB})\breve{}  
 +\Big( \frac{\zeta}{2} + \frac{1}{r} - \frac{r}{2}|\sigma|^2\Big)\sigma_{AB}
-\frac{1}{2}(\xi_A\xi_B)\breve{} +(\check\nabla_{(A} \xi_{B)})\breve{} 
 =  \mathcal{O}(r^{-1})
 \;.
\end{eqnarray*}
In order to permit an easier comparison with Appendix~\ref{asymptotic_solutions}, we apply the transformation $r\mapsto x:=1/r$ (with all quantities treated as scalars).
Then the ODE adopts the following asymptotic form,
\begin{eqnarray*}
 \left[ x\partial_x
 + \begin{pmatrix}
    1 & 0 & 0 \\ 0 & 1 & 0 \\ 0 & 0 & 1
   \end{pmatrix}
 + M\right] \begin{pmatrix} (\ol{\partial_0g_{22}})\breve{} \\ (\ol{\partial_0g_{23}})\breve{} \\ (\ol{\partial_0 g_{33}})\breve{} \end{pmatrix}
 = \mathcal{O} (1)
 \;,
\end{eqnarray*}
where $ M = \mathrm{Mat}(3,3)=\mathcal{O}(x)$. In the notation of Appendix~\ref{asymptotic_solutions}
 we observe that $\lambda=-1$ and $\hat\ell=-1$,
and no logarithmic terms appear in the asymptotic solution.
Moreover, it is $\mathcal{O}(r)$ and the integration functions are represented by the leading order terms,
\begin{equation*}
( \ol{\partial_0g_{AB}})\breve{} = D_{AB} r + \mathcal{O}(1)
 \;.
\end{equation*}
Note that $D_{AB}$ is symmetric, $D_{AB}=D_{BA}$, and $s$-trace-free, $s^{AB}D_{AB}=0$.

\paragraph{Einstein equation $\ol R_{0r}=0$:}

\begin{eqnarray*}
 \ol R_{0r} \,=\, \partial_r\ol \Gamma^r_{0r} + \check\nabla_A\ol \Gamma^A_{0r} - \ol{\partial_0\Gamma^r_{rr}} -\ol{ \partial_0\Gamma^A_{rA}}
 + \ol \Gamma^0_{0r}\ol \Gamma^A_{0A} + \ol \Gamma^r_{0r}(\ol \Gamma^0_{0r} +\ol  \Gamma^A_{rA})
  - \ol \Gamma^A_{rB}\ol \Gamma^B_{0A}
\;.
\end{eqnarray*}
We determine the $u$-differentiated Christoffel symbols involved,
\begin{eqnarray}
 \ol{\partial_0\Gamma^r_{rr}}
 &=& - \frac{1}{2}\ol{\partial^2_{00}g_{rr}} + \frac{r^2}{2}|\sigma|^4  - \frac{1}{2}|\sigma|^2 - \frac{r}{2}\partial_r|\sigma|^2
\;,
\\
\ol{ \partial_0\Gamma^C_{rA}}
 &=& \frac{1}{2}\xi_A\xi^C + \frac{1}{2}\partial_r(\ol g^{BC}\ol{\partial_0g_{AB}} )+ \ol g^{BC} \check\nabla_{[A}\xi_{B]}
\\
 \Longrightarrow \quad
\ol{\partial_0\Gamma^A_{rA}}
&=&   \frac{1}{2}\xi_A\xi^A - \frac{1}{2}\partial_r\zeta  + \frac{1}{r^2}
 \;.
\nonumber
\end{eqnarray}
Note that $\ol{\partial^2_{00} g_{rr}}$ is the only unknown at this stage.
We obtain
\begin{eqnarray*}
 \ol R_{0r} &=&  \frac{1}{2}\ol{\partial^2_{00}g_{rr}}
  +\frac{1}{2}\Big(\partial_r+ r^{-1}+ \frac{r}{2}|\sigma|^2\Big)\zeta
 - \frac{r^2}{4}|\sigma|^4  -\frac{1}{2}|\sigma|^2
\\
 && + \frac{1}{2} (\check\nabla_A-\xi_A)\xi^A - \frac{1}{2}\sigma^{AB}\ol{\partial_0g_{AB}}
 \;,
\end{eqnarray*}
which, again, should vanish in vacuum. With \eq{zeta_constraint}
that yields
\begin{eqnarray}
\ol{ \partial^2_{00}g_{rr}}
 \,=\,  \frac{1}{r}\zeta + \check R
  + \frac{1}{2}\xi_A\xi^A
 + \frac{r^2}{2}|\sigma|^4  + |\sigma|^2     + \sigma^{AB}\ol{\partial_0g_{AB}}
\,=\, \mathcal{O}(r^{-3})
 \;.
\label{expansion_00grr}
\end{eqnarray}
The leading-order term
\begin{equation}
 \Xi := (\ol{\partial^2_{00}g_{rr}})_3 = \zeta_2 +[D_A{}^B-2\mathring\nabla_A\mathring\nabla^B](\sigma_B{}^A)_2
 \;.
\end{equation}
depends on certain integration functions and
is not explicitly known. We choose this special notation, though, since it will appear several times in the leading-order terms of other expansions.

The gauge condition then provides an  algebraic equation for $\ol{\partial_0 W^0}$,
\begin{eqnarray*}
0\,=\,  \ol{\partial_0 H^0}
 &=& -\ol{\partial_0 g_{rr}}\ol \Gamma{}^0_{00}  - 2\ol{\partial_0g_{rr}}\ol\Gamma{}^0_{0r} - 2\ol g^{AB}\ol{\partial_0g_{rB}}\ol \Gamma{}^0_{0A}
\\
 && + \ol{\partial_0g^{AB}}(\ol\Gamma{}^0_{AB} +r s_{AB}) +2\ol{\partial_0\Gamma^0_{0r} }+ \ol{\partial_0\Gamma^0_{rr} } + \ol g^{AB}\ol{\partial_0\Gamma^0_{AB}}
-\ol{\partial_0 W^0}
\\
 &=& \ol{\partial^2_{00}g_{rr}} +rs_{AB}\ol{\partial_0g^{AB}}
 +\check\nabla_A \xi^A - 2 \xi_A\xi^A
 -\frac{1}{r^2}  - \frac{3}{2}r^2|\sigma|^4
\\
 &&  - \frac{3}{2}|\sigma|^2 - \frac{r}{2}\partial_r|\sigma|^2  + \frac{1}{2}(\partial_r + r|\sigma|^2 )\zeta
-\ol{\partial_0 W^0}
\;,
\end{eqnarray*}
where we used that
\begin{eqnarray}
 \ol{ \partial_0\Gamma^0_{rr}}
 &=& \frac{r^2}{2}|\sigma|^4  - \frac{1}{2}|\sigma|^2 - \frac{r}{2}\partial_r|\sigma|^2
\;,
\\
\ol{ \partial_0\Gamma^0_{0r}}
 &=& \frac{1}{2}\ol{\partial^2_{00}g_{rr}} - \frac{r^2}{2}|\sigma|^4 - \frac{1}{2}\xi_A \xi^A
 \;.
\end{eqnarray}
Inserting \eq{expansion_00grr} and using again \eq{zeta_constraint}
we deduce that
\begin{eqnarray}
\ol{ \partial_0W^0} &=&-\Big(  \frac{1}{2}\partial_r +  \frac{1}{r}\Big)\zeta
 +(r s_{AB}-\sigma_{AB})\ol{\partial_0g^{AB}}  - \frac{1}{r^2} -\frac{r}{2}\partial_r|\sigma|^2
\nonumber
\\
 &&  - \frac{1}{2}|\sigma|^2 - r^2|\sigma|^4 - \xi_A\xi^A
\label{metric_gauge_transW0}
\\
 &=& 
\mathcal{O}(r^{-3})
 \;.
\end{eqnarray}

\paragraph*{Einstein equations $\ol R_{0A}=0$:}

%
\begin{eqnarray*}
 \ol R_{0A} &=& \partial_r\ol \Gamma{}^r_{0A} + \check\nabla_B\ol \Gamma{}^B_{0A} - \ol {\partial_0\Gamma^r_{rA}} -\ol { \partial_0\Gamma^B_{AB}}
   + \ol \Gamma{}^0_{0A}\ol \Gamma{}^B_{0B} +(\ol \Gamma{}^r_{rr} + \ol \Gamma{}^B_{rB} )\ol \Gamma{}^r_{0A}
\\
 && + \ol \Gamma{}^B_{0A}\ol \Gamma{}^r_{rB}
 - \ol \Gamma{}^0_{AB}\ol \Gamma{}^B_{00} - 2\ol \Gamma{}^r_{rA}\ol \Gamma{}^r_{0r} - \ol \Gamma{}^r_{AB}\ol \Gamma{}^B_{0r}
  -\ol  \Gamma{}^B_{rA}\ol \Gamma{}^r_{0B}
\;.
\end{eqnarray*}
For the $u$-differentiated Christoffel symbols we find,
\begin{eqnarray}
 \ol{ \partial_0\Gamma^r_{rA} }
 &=& - \frac{1}{2}\ol{\partial^2_{00}g_{rA}}
-  \chi_A{}^{B}\xi_B
   - \frac{r}{2}\partial_A|\sigma|^2  -\frac{1}{2}r |\sigma|^2\xi_A
 \;,
\\
\ol{\partial_0\Gamma^C_{AB} } &=& \ol g^{CD}\check\nabla_{(A}\ol{\partial_{|0|}g_{B)D}}
 -\frac{1}{2}\check\nabla^C\ol{\partial_0 g_{AB}}
+\frac{1}{2}\xi^C\ol{\partial_0g_{AB}} +\xi^C\chi_{AB}
\;,
\\
\hspace{-2em}\Longrightarrow \enspace  \ol{\partial_0 \Gamma^B_{AB}} &=&   \chi_A{}^{B}\xi_B
+ \frac{1}{2}\xi^B\ol{\partial_0g_{AB}}   -\frac{1}{2}\partial_A\zeta
 \;.
\nonumber
\end{eqnarray}
We insert these expressions into the formula for $\ol R_{0A}$ to obtain,
\begin{eqnarray*}
 \ol R_{0A} &=& \frac{1}{2}(\partial_r + r^{-1})\xi_A
+\frac{1}{2}(\check\nabla^B-\xi^B)\ol {\partial_0 g_{AB}} + \frac{r}{2}(\partial_A+\frac{1}{2}\xi_A)|\sigma|^2
+\frac{1}{2}\ol{\partial^2_{00}g_{rA}}
\\
 &&
 +\frac{1}{2}(\partial_A-\frac{1}{2}\xi_A)\zeta
 \;,
\end{eqnarray*}
which should vanish in vacuum, i.e.
\begin{eqnarray}
\hspace{-2.5em}\ol{\partial^2_{00}g_{rA}} &=&(\xi^B - \check\nabla^B)\ol {\partial_0 g_{AB}}
 -(\partial_r+ \frac{1}{r}-\frac{\zeta}{2})\xi_A
 - r(\partial_A+\frac{\xi_A}{2})|\sigma|^2
 -\partial_A\zeta
\\
&=&  - \mathring\nabla_B D_{A}{}^{B} r^{-1} + \mathcal{O}(r^{-2})
 \;.
\nonumber
\end{eqnarray}
We employ the gauge condition $\ol{\partial_0  H{}^A}=0$ to compute  $\ol{\partial_0W^A}$.
This  requires the knowledge of further $u$-differentiated Christoffel symbols,
\begin{eqnarray}
 \ol{\partial_0\Gamma^C_{0r} }
 &=& \frac{1}{2}\ol g^{CD}\ol{\partial^2_{00}g_{rD}} + \frac{r}{2}|\sigma|^2\xi^C + \frac{1}{2}\xi_D\ol{\partial_0g ^{CD}}
 \;,
\\
 \ol{\partial_0\Gamma^C_{rr}}
 &=& \ol g^{CD}\partial_r\xi_D  -\frac{1}{2}r|\sigma|^2 \xi^C + \frac{r}{2}\check\nabla^C|\sigma|^2
 \;.
\end{eqnarray}
We then obtain
\begin{eqnarray*}
0= \ol{\partial_0H^C}
 &=& -\ol{\partial_0g_{rr}}\ol\Gamma^C_{00} - 2\ol{\partial_0g_{rr}}\ol\Gamma^C_{0r}
- 2\ol g^{AB}\ol{\partial_0g_{rB}}\ol\Gamma^C_{0A}
- 2\ol g^{AB}\ol{\partial_0g_{rB}}(\ol\Gamma^C_{rA} - \hat\Gamma^C_{rA})
\\
 && + \ol{\partial_0g^{AB}}(\ol \Gamma^C_{AB}-\hat\Gamma^C_{AB}) + 2\ol{\partial_0\Gamma^C_{0r}} + \ol{\partial_0\Gamma^C_{rr}}
+ \ol g^{AB}\ol{\partial_0\Gamma^C_{AB}}
 - \ol{\partial_0 W^C}
\\
 &=&
\Big(\partial_r + \frac{3}{r} +  \frac{3}{2}r|\sigma|^2\Big)\xi^C
 + \ol g^{CD}\ol{\partial^2_{00}g_{rD}}
 +\frac{1}{2}(\check\nabla^C - \xi^C)\zeta
 + \frac{r}{2}\check\nabla^C|\sigma|^2
\label{transverse_WA_mgauge}
\\
 &&+\ol{\partial_0g^{AB}}(\check \Gamma^C_{AB}-\mathring\Gamma^C_{AB})
- (\check\nabla_{A}-2\xi_A)\ol{\partial_{0}g^{AC}}
 - \ol{\partial_0 W^C}
 \;.
\end{eqnarray*}
Combining this with the expression we found for $\ol{\partial^2_{00}g_{rA}}$ yields
\begin{eqnarray}
\ol{\partial_0 W^A} &=&
(\ol{\partial_0g^{AB}}-2\sigma^{AB})\xi_B
- \frac{r}{2}(\check\nabla^A-2\xi^A)|\sigma|^2
 -\frac{1}{2}\check\nabla^A\zeta
\nonumber
\\
 &&
+\ol{\partial_0g^{BC}}(\check \Gamma^A_{BC}-\mathring\Gamma^A_{BC})
\label{metric_gauge_transWA}
\\
  &=&  \mathcal{O}(r^{-4})
 \;.
\end{eqnarray}

\paragraph*{Einstein equation $\ol R_{00}=0$:}

The $(00)$-component of the Ricci tensor satisfies
\begin{eqnarray*}
 \ol R_{00}
 \,=\, (\partial_r+2\ol \Gamma^r_{rr} +  \ol \Gamma^A_{rA})\ol \Gamma^r_{00}  - \ol{\partial_0\Gamma^r_{0r}} - \ol{\partial_0\Gamma^A_{0A}}
   - \ol\Gamma^r_{0r}\ol\Gamma^r_{0r}
  - 2\ol \Gamma^r_{0A}\ol \Gamma^A_{0r} - \ol \Gamma^A_{0B}\ol \Gamma^B_{0A}
 \;.
\end{eqnarray*}
The $u$-differentiated Christoffel symbols appearing in this expression satisfy,
\begin{eqnarray}
 \ol{\partial_0\Gamma^r_{0r} }
&=&   \frac{1}{2r}\zeta +  \frac{1}{2}\check R
  - \frac{1}{4}\xi_A\xi^A
- \frac{r^2}{4}|\sigma|^4  +  \frac{1}{2} |\sigma|^2
   +  \frac{1}{2}\sigma^{AB}\ol{\partial_0g_{AB}}
\;,
\\
\ol{\partial_0\Gamma^A_{0A}} &=& \frac{1}{2}\ol g^{AB}\ol{\partial^2_{00}g_{AB}} + \frac{1}{2}\ol{\partial_0g^{AB}}\ol{\partial_0g_{AB}}  - \frac{1}{2}\xi_A\xi^A
 \;,
\end{eqnarray}
and we are led to
\begin{eqnarray*}
 \ol R_{00}
=
  -\frac{1}{2}\ol g^{AB}\ol{\partial^2_{00}g_{AB}}   -\Big(\frac{1}{2}\sigma^{AB} + \frac{1}{4}\ol{\partial_0g^{AB}}\Big)\ol{\partial_0g_{AB}}
    - \frac{1}{2r}\zeta -  \frac{1}{2}\check R
  + \frac{1}{4}\xi_A\xi^A  -  \frac{1}{2} |\sigma|^2
 \;,
\end{eqnarray*}
which should vanish in vacuum, solving for $  \ol g^{AB}\ol{\partial^2_{00}g_{AB}} $ we are led to
\begin{eqnarray}
\hspace{-2em} \ol g^{AB}\ol{\partial^2_{00}g_{AB}}
 &=&   -\big( \sigma^{AB} + \frac{1}{2}\ol{\partial_0g^{AB}}\big)\ol{\partial_0g_{AB}}
  - \frac{1}{r}\zeta - \check R
  + \frac{1}{2}\xi_A\xi^A  -  |\sigma|^2
\\
&=&  \frac{1}{2} |D|^2 r^{-2} + \mathcal{O}(r^{-3})
 \;.
\end{eqnarray}
To compute $ \ol{ \partial_0W^r} $  we employ the gauge condition
\begin{eqnarray}
\hspace{-1em} 0=  \ol{\partial_0 H^r}
 \hspace{-0.5em}&=& \hspace{-0.5em}
 -\ol{\partial_0g_{rr}}(\ol \Gamma^r_{00} - \ol \Gamma^r_{rr} ) - 2\ol g^{AB}\ol{\partial_0g_{rB}}\ol \Gamma^r_{0A}
 -2\ol g^{AB}\ol \partial_0g_{rB}\ol \Gamma^r_{rA}
  \nonumber
\\
 && + \ol{\partial_0g^{AB}}(\ol \Gamma^r_{AB} - \hat{\Gamma}^r_{AB}) + 2\ol{\partial_0\Gamma^r_{0r}} + \ol{\partial_0\Gamma^r_{rr}}
 + \ol g^{AB}\ol{\partial_0\Gamma^r_{AB}}
- \ol{ \partial_0W^r}
  \nonumber
\\
&=&  \frac{1}{2}( \check\nabla_A-\xi_A) \xi^A
 - \frac{1}{2}\ol g^{AB}\ol{\partial^2_{00}g_{AB} }
- \frac{1}{r^2}  - (1+ \frac{r}{2}\partial_r)|\sigma|^2     - \frac{3}{4}r^2|\sigma|^4
\nonumber
\\
&&
 -\frac{\zeta}{2}( \frac{1}{r}- \frac{r}{2}|\sigma|^2)
+(r s_{AB} - \frac{1}{2}\sigma_{AB} -\frac{1}{2}\overline{\partial_0g_{AB}})\ol{\partial_0g^{AB}} - \ol{ \partial_0W^r}.
 \label{eqn_d0H1}
\end{eqnarray}
For this calculation we used that
\begin{eqnarray*}
 \ol g^{AB}\ol{\partial_0\Gamma^r_{AB} }
= \check\nabla_A \xi^A
+\frac{1}{2}(\partial_r+\frac{2}{r}+r|\sigma|^2)\zeta
+ \frac{1}{r^2} - |\sigma|^2
 -\sigma^{AB}\ol{\partial_0g_{AB}}
 - \frac{1}{2}\ol g^{AB}\ol{\partial^2_{00}g_{AB} }
 \;.
\end{eqnarray*}
We solve (\ref{eqn_d0H1}) for $ \ol{ \partial_0W^r} $, and insert the expression we found for $\ol g^{AB}\ol{\partial^2_{00}g_{AB}} $,
\begin{eqnarray}
 \ol{ \partial_0W^r} &=&   \frac{1}{2}( \check\nabla_A- \frac{3}{2}\xi_A) \xi^A
 +\frac{1}{4}r |\sigma|^2\zeta +  \frac{1}{2}\check R  - \frac{1}{r^2}  -  \frac{1}{2} |\sigma|^2
  - \frac{r}{2}\partial_r|\sigma|^2
\nonumber
\\
&&      - \frac{3}{4}r^2|\sigma|^4
 -(\frac{1}{4}\overline{\partial_0g_{AB}} + \sigma_{AB} - r s_{AB})\ol{\partial_0g^{AB}}
\label{metric_gauge_transW1}
\\
&=& \frac{1}{4}|D|^2 r^{-2} + \mathcal{O}(r^{-3})
 \;.
\end{eqnarray}
%

\paragraph*{Einstein equations $(\ol{\partial_0R_{AB}})\breve{}=0$}

Assuming the gauge condition $H^{\lambda}=0$ the Einstein equations $(\overline{\partial_0R_{AB}})\breve{}=0$ provide  a linear ODE-system for $(\overline{\partial^2_{00}g_{AB}})\breve{}$. We have
\begin{eqnarray*}
 \ol{\partial_0R_{AB}} =\ol{ \partial_{\lambda}\partial_0\Gamma^{\lambda}_{AB}} - \partial_A\ol{\partial_0\Gamma^{\lambda}_{\lambda B}}
+ \ol \Gamma^{\lambda}_{\rho\lambda}\ol{\partial_0\Gamma^{\rho}_{AB}}
 +\ol \Gamma^{\rho}_{AB} \ol{\partial_0\Gamma^{\lambda}_{\rho\lambda}}
 - \ol \Gamma^{\lambda}_{\rho A} \ol{\partial_0\Gamma^{\rho}_{\lambda B}} - \ol \Gamma^{\rho}_{\lambda B}\ol{\partial_0\Gamma^{\lambda}_{\rho A}}
\;.
\end{eqnarray*}
The right-hand side contains several first-order $u$-differentiated Christoffel symbols and one of second-order, $\ol{\partial^2_{00}\Gamma^0_{AB}}$,
which we have not determined yet,
\begin{eqnarray}
\ol{ \partial_0\Gamma^0_{00}} &=& \ol{\partial_{00}^2g_{0r}}
\;,
\\
\ol{ \partial_0\Gamma^0_{0A} }
 &=& \frac{1}{2}\ol{\partial^2_{00}g_{rA}}  - \frac{1}{2}\xi^B\ol{\partial_0g_{AB}}  + \frac{r}{2}|\sigma|^2\xi_A
\\
 &=& -\frac{1}{2}\mathring\nabla_B D_A{}^B r^{-1} + \mathcal{O}(r^{-2})
 \;,
\\
 \ol{\partial_0\Gamma^0_{rA} }
 &=&  - \chi_{A}{}^B \xi_B   - \frac{r}{2}(\partial_A+\xi_A)|\sigma|^2
\\
 &=& -2\mathring\nabla_B(\sigma_A{}^B)_2 r^{-2} + \mathcal{O}(r^{-3})
 \;,
\\
 \ol{\partial_0\Gamma^r_{AB}}  &=&  - \frac{1}{2}\ol{\partial^2_{00}g_{AB}} -\frac{1}{2}(\partial_r + r|\sigma|^2)\ol{\partial_0g_{AB}} + \check\nabla_{(A}\xi_{B)}  - r|\sigma|^2\chi_{AB}
\\
 &=&  - \frac{1}{2}\ol{\partial^2_{00}g_{AB}} -\frac{1}{2}D_{AB}  + \mathcal{O}(r^{-1})
\;,
\\
 \ol{\partial_0\Gamma^C_{0A}} &=& -\frac{1}{2}\xi_A\xi^C + \frac{1}{2}\ol{\partial_0g^{CD}}\ol{\partial_0g_{AD}}
 + \frac{1}{2}\ol g^{CD}\ol{\partial^2_{00}g_{AD}}
\\
&=& \frac{1}{2}g^{CD}\partial^2_{00}g_{AD} -\frac{1}{2}D_{AD}D^{CD}r^{-2} + \mathcal{O}(r^{-3})
 \;.
\end{eqnarray}
To calculate the two-times $u$-differentiated Christoffel symbol we use that
\begin{eqnarray*}
 \ol{\partial^2_{00}g^{\mu\nu}} &=& -2\ol g^{\nu\sigma}\ol{\partial_0g^{\mu\rho}}\ol{\partial_0g_{\rho\sigma}}
 -\ol g^{\mu\rho}\ol g^{\nu\sigma}\ol{\partial^2_{00}g_{\rho\sigma}}
 \;,
\end{eqnarray*}
whence
\begin{eqnarray}
 \ol{\partial^2_{00}g^{00}}
 &=& - \ol{\partial^2_{00}g_{rr}} + 2\xi_A \xi^A +  2r^2|\sigma|^4
\\
 &=& -\Xi r^{-3} + \mathcal{O}(r^{-4})
 \;,
\\
 \ol{\partial^2_{00}g^{0r}}
&=&  - \ol{\partial^2_{00}g_{0r}} - \ol{\partial_{00}^2g_{rr} } +\underbrace{ 2 \xi_A \xi^A  + 2r^2|\sigma|^4}_{=\mathcal{O}(r^{-4})}
\;,
\\
\ol{\partial^2_{00}g^{0C}}
 &=&  - \ol g^{CD}\ol{\partial^2_{00}g_{rD}} -2r|\sigma|^2\xi^C
 -2 \xi_D\ol{\partial_0g^{CD}}
\\
 &=& \mathring\nabla_BD^{BC}r^{-3} + \mathcal{O}(r^{-4})
 \;.
\end{eqnarray}
That yields
\begin{eqnarray*}
 \ol{\partial^2_{00}\Gamma^0_{AB}}
 &=& \Big( \frac{1}{2}\ol{\partial^2_{00}g_{rr}}-r^2|\sigma|^4 - \xi_C \xi^C \Big)(\ol{\partial_0g_{AB}}+  2\chi_{AB} ) +\chi_{AB} \ol{\partial^2_{00}g_{0r}}
\\
 &&  -\Big(\frac{1}{2}\partial_r +r|\sigma|^2\Big) \ol{\partial^2_{00}g_{AB}}  -  r|\sigma|^2\partial_r\ol{\partial_0g_{AB}}
+ 2r |\sigma|^2\check\nabla_{(A}\xi_{B)}
\\
 && - \xi^C(2\check\nabla_{(A}\ol{\partial_{|0|}g_{B)C}}-\check\nabla_C\ol{\partial_0g_{AB}})
  + \check\nabla_{(A}\ol{\partial^2_{|00|}g_{B)r} }
\\
 &=&\chi_{AB}\ol{\partial_{00}^2g_{0r}} -\Big(\frac{1}{2}\partial_r + r|\sigma|^2\Big)\ol{\partial^2_{00}g_{AB}}
  -\mathring\nabla_{(A}\mathring\nabla_{|C|}D_{B)}{}^C r^{-1}  + \mathcal{O}(r^{-2})
 \;.
\end{eqnarray*}
%
%
Now  all quantities have been determined  which are needed to compute $(\ol{\partial_0R_{AB}})\breve{}$ (the $\ol{\partial^2_{00}g_{0r}}$'s cancel out);
we are just interested in the asymptotic behavior,
\begin{eqnarray*}
 (\ol{\partial_0R_{AB}})\breve{}
 &=& - \big(\partial_r - r^{-1}+ r|\sigma|^2\big)(\ol{\partial^2_{00}g_{AB}} )\breve{}
+2 \sigma_{(A}{}^C(\ol{\partial^2_{|00|}g_{B)C}})\breve{}
\\
 &&  -  \ \frac{1}{2}(\Delta_s-2) D_{AB} r^{-1}
 + \mathcal{O}(r^{-2})
 \;,
\end{eqnarray*}
and this expression should vanish in vacuum.

We need to check whether there are logarithmic terms in the asymptotic expansion of $(\ol{\partial^2_{00}g_{AB}})\breve{}$, solution
to $(\ol{\partial_0R_{AB}})\breve{}=0$.
However, this ODE is of exactly the same form as the ODE (\ref{eqn_0RAB}) for $\ol{\partial_0g_{AB}}$.
Therefore an identical argument applies and leads to the conclusion that the solution can be asymptotically expanded as a power series,
\begin{eqnarray}
 (\partial^2_{00}g_{AB})\breve{} = C_{AB}r + \mathcal{O}(1)
 \;,
\end{eqnarray}
where the symmetric and $s$-trace-free tensor $C_{AB}$ can be identified with  the integration functions.
%
%

For completeness let us also compute
\begin{eqnarray}
\ol{ \partial_0\Gamma^r_{0A}}
 &=& \frac{1}{2}\ol{\partial^2_{00}g_{rA}} - \frac{1}{2}\xi^B\ol{\partial_0g_{AB}} + \frac{r}{2}|\sigma|^2\xi_A
\\
 &=& -\frac{1}{2}\mathring\nabla_B D_A{}^B r^{-1} + \mathcal{O}(r^{-2})
 \;.
\end{eqnarray}

\subsection{Overview of the metric gauge II}
\label{overviewI}

We give an overview of the values of all those objects we have computed so far in the metric gauge.
Recall  the values for the gauge functions $\kappa$, $\ol W^{\lambda}$ and $\ol{\partial_0 W^{\lambda}}$ as given by \eq{metric_gauge_kappa}-\eq{metric_gauge_W1}, \eq{metric_gauge_transW0}, \eq{metric_gauge_transWA}
and \eq{metric_gauge_transW1}, needed to realize these values.  Recall further that equality is meant to hold  for $r> r_2$.

\vspace{0.5em}

\noindent
\textbf{Metric components}
%

\noindent
\begin{tabular}{ll}
  $\ol g_{00} = -1$, & $\ol g^{00} = 0$,
\\
 $\ol g_{0r} = 1 $, & $\ol g^{0r} =1$,
\\
 $\ol g_{0A} =  0 $, & $\ol g^{0A} = 0$,
\\
 $\ol g_{rr} = 0$, &  $\ol g^{rr} =1 $,

\\
$ \ol g_{rA} = 0$, & $\ol g^{rA} =0$,
\\
 $\ol g_{AB} = r^2\sqrt{\frac{\det s}{\det \gamma}} \gamma_{AB} = r^2s_{AB} + \mathcal{O}(r)$, & $ \ol g^{AB} = r^{-2}\sqrt{\frac{\det \gamma}{\det s}} \gamma^{AB} = r^{-2}s^{AB} + \mathcal{O}(r^{-3})$.
\end{tabular}

\vspace{0.5em}

\noindent
\textbf{First-order $u$-derivatives of the metric components}
\noindent

\begin{tabular}{ll}
  $\overline{\partial_0 g_{00}}   =0$, & $\overline{\partial_0 g^{00}} = r|\sigma|^2 = \mathcal{O}(r^{-3})$,
\\
 $\overline{\partial_0g_{0r}} =0 $, & $ \overline{\partial_0 g^{0r}} = r|\sigma|^2= \mathcal{O}(r^{-3})$,
\\
  $\overline{\partial_0g_{0A}} =0 $, & $ \overline{\partial_0 g^{0A}} =-\xi^A = \mathcal{O}(r^{-3})$,
\\
 $\overline{\partial_0g_{rr}} =-r|\sigma|^2 = \mathcal{O}(r^{-3})$, & $\overline{\partial_0 g^{rr}} = r|\sigma|^2= \mathcal{O}(r^{-3}) $,
\\
 $\overline{\partial_0g_{rA}}= \xi_A = \mathcal{O}(r^{-1})$, &  $\overline{\partial_0 g^{rA}} =-\xi^A = \mathcal{O}(r^{-3})$,
\\
 $\overline{\partial_0g_{AB}} = D_{AB} r + \mathcal{O}(1)$, &  $\overline{\partial_0 g^{AB}}=-D^{AB}r^{-3}+\mathcal{O}(r^{-4}) $.
\end{tabular}

\vspace{0.5em}

\noindent
\textbf{Second-order $u$-derivatives of the metric components}

\noindent
\begin{eqnarray*}
 \overline{\partial^2_{00} g_{rr}} &=& \Xi r^{-3} + \mathcal{O}(r^{-4})
 \;,
\\
 \overline{\partial^2_{00} g_{rA}} &=&  -\mathring\nabla_B D_A{}^B r^{-1} + \mathcal{O}(r^{-2})
 \;,
\\
 \overline g^{AB}\overline{\partial^2_{00} g_{AB}} &=&  \frac{1}{2}|D|^2r^{-2}  + \mathcal{O}(r^{-3})
 \;.
\end{eqnarray*}

\noindent
\textbf{Asymptotic behavior of the Christoffel symbols (cf.\ p.~\pageref{christoffels_mgauge})}

\noindent
\begin{tabular}{ll}
  $\ol \Gamma^0_{00} =  0$, &   $\ol \Gamma^r_{rr}  = \mathcal{O}(r^{-3})$,
\\
 $\ol \Gamma^0_{0r}=\mathcal{O}(r^{-3})$, & $\ol \Gamma^r_{rA}  = \frac{1}{2} \mathring\nabla_B h^{(1)B}_A  r^{-1}+  \mathcal{O}(r^{-2})$,
\\
 $\ol \Gamma^0_{0A}= -\frac{1}{2}\mathring\nabla_B\breve h^{(1)}_{A}{}^B r^{-1} + \mathcal{O}(r^{-2})$, &  $\ol \Gamma^r_{AB} =  -(s_{AB}+\frac{1}{2}D_{AB})r + \mathcal{O}(1)$,
\\
 $\ol \Gamma^0_{rr} = 0$, & $\ol \Gamma^C_{00} =0$,
\\
 $\ol \Gamma^0_{rA} =  0$, &  $\ol \Gamma^C_{0r} = -\frac{1}{2}\mathring\nabla^Bh^{(1)C}_B  r^{-3} + \mathcal{O}(r^{-4})$,
\\
 $\ol \Gamma^0_{AB} =-rs_{AB} + \mathcal{O}(1)$, & $\ol \Gamma^C_{0A} =\frac{1}{2}D_A{}^C r^{-1} + \mathcal{O}(r^{-2})$,
\\
 $\ol \Gamma^r_{00} =0 $, &  $\ol \Gamma^C_{rr} = 0$,
\\
 $\ol \Gamma^r_{0r} = \mathcal{O}(r^{-3})$, &  $\ol \Gamma^C_{rA} = \frac{1}{r}\delta_A{}^C +\mathcal{O}(r^{-2})$,
\\
 $\ol \Gamma^r_{0A} =  -\frac{1}{2}\mathring\nabla_B \breve h^{(1)B}_Ar^{-1} + \mathcal{O}(r^{-2})$, &  $\ol \Gamma^C_{AB}  = \mathring\Gamma^C_{AB} + \mathcal{O}(r^{-1})$.
\end{tabular}


\vspace{0.5em}

\noindent
\textbf{First-order $u$-derivatives of the Christoffel symbols}

\noindent
We give  a list of the asymptotic behavior of all the $u$-differentiated Christoffel symbols crucial e.g.\ for the computation of the Weyl tensor,
which can be  straightforwardly derived from our previous results,



\vspace{0.5em}
\noindent
\begin{tabular}{ll}
 $\overline{\partial_0\Gamma^0_{00}} =\text{not needed}$, & $\overline{\partial_0\Gamma^r_{rr}} = -\frac{1}{2}\Xi r^{-3} + \mathcal{O}(r^{-4})$,
\\
  $\overline{\partial_0\Gamma^0_{0r}} =\frac{1}{2} \Xi r^{-3} + \mathcal{O}(r^{-4})$, &  $ \overline{\partial_0\Gamma^r_{rA}} = \frac{1}{2}\mathring\nabla_B D_A{}^B r^{-1} + \mathcal{O}(r^{-2})$,
\\
 $\overline{\partial_0\Gamma^0_{0A}} = -\frac{1}{2}\mathring\nabla_B D_A{}^B r^{-1} + \mathcal{O}(r^{-2})$, & $ \overline{\partial_0\Gamma^r_{AB}} = -\frac{1}{2}C_{AB}r + \mathcal{O}(1)$,
\\
 $\overline{\partial_0\Gamma^0_{rr}} = \frac{3}{8}|\breve h^{(1)}|^2 r^{-4} + \mathcal{O}(r^{-5})$, &  $\overline{\partial_0\Gamma^C_{00}} = \text{not needed}$,
\\
 $\overline{\partial_0\Gamma^0_{rA}} = \mathring\nabla_Bh^{(1)B}_Ar^{-2} + \mathcal{O}(r^{-3}) $, & $ \overline{\partial_0\Gamma^C_{0r}} =-\frac{1}{2}\mathring\nabla_B D^{BC}r^{-3} + \mathcal{O}(r^{-4})$,
\\
 $\overline{\partial_0\Gamma^0_{AB}} =-\frac{1}{2}D_{AB} +\mathcal{O}(r^{-1})$, & $ \overline{\partial_0\Gamma^C_{0A}}= \frac{1}{2}C_A{}^C r^{-1} + \mathcal{O}(r^{-2})$,
\\
 $\overline{\partial_0\Gamma^r_{00}} = \text{not needed}$,  & $\overline{\partial_0\Gamma^C_{rr}} = \mathring\nabla^B h^{(1)C}_B r^{-4} + \mathcal{O}(r^{-5})$,
\\
 $\overline{\partial_0\Gamma^r_{0r}} =\frac{1}{2}\Xi r^{-3} + \mathcal{O}(r^{-4})$,  &  $\overline{\partial_0\Gamma^C_{rA}} =-\frac{1}{2}D_A{}^Cr^{-2} + \mathcal{O}(r^{-3})$,
\\
 $\overline{\partial_0\Gamma^r_{0A}} =-\frac{1}{2}\mathring\nabla_B D_A{}^B r^{-1} + \mathcal{O}(r^{-2})$, &  $\overline{\partial_0\Gamma^C_{AB}} =\big[\mathring\nabla_{(A}D_{B)}{}^C
 - \frac{1}{2}\mathring\nabla^C D_{AB} \big]r^{-1} + \mathcal{O}(r^{-2})$.
\end{tabular}

\section{Asymptotic Expansions of  the unknowns of the CFE on $\mcN$}
\label{sec_sufficient}

\subsection{Conformal field equations (CFE)}

As indicated in the introduction, one would like to establish an existence theorem for the characteristic initial value problem for the vacuum Einstein equations which guarantees  a ``piece of a smooth $\scrip$''.
This global existence problem, where one needs to control the limiting behavior near null infinity,
can be transformed into a local one via a conformal rescaling $g\mapsto \tilde g = \Theta^2 g$, where
 the conformal factor $\Theta$ has to vanish on the  hypersurface $\Scri^+$ which represents (future) null infinity  (with $\mathrm{d}\Theta\ne 0$ there).
Henceforth we use $\tilde{\enspace}$ to label objects related to the unphysical space-time metric $\tilde g$.

The  Einstein equations,  regarded as equations for $\tilde g$ and  $\Theta$, become singular wherever $\Theta$ vanishes, and are therefore not suitable to tackle the issue of proving  existence of a solution near $\Scri^+$.
However,  H.\ Friedrich was able to find a representation of the vacuum Einstein equations which
remains regular even where the conformal factor vanishes, cf.\ e.g.\ \cite{F3}.
These \textit{conformal field equations (CFE)} treat $\Theta$ as an unknown rather than a gauge function.
Its gauge freedom is reflected in the freedom to prescribe  the curvature scalar $\tilde R$. 
This still leaves the freedom to prescribe $\Theta$ on the initial surface. We shall take
\begin{equation}
\overline \Theta=1/r=x \quad \text{for $r>r_2$}
\label{gauge_choice_Theta}
\end{equation}
 as initial data on $\mcN$.
The CFE give rise to a complicated and highly overdetermined system of PDEs,
which, in 4~space-time dimensions,
can be represented as a symmetric hyperbolic system, supplemented by certain constraint equations.

\subsubsection{Metric conformal field equations (MCFE)}

There exist different versions of the CFE depending on which fields are regarded as unknowns.
Let us first introduce the \textit{metric conformal field equations (MCFE)} \cite{F3}.
Beside $\tilde g$ and $\Theta$ the unknowns are the Schouten tensor,
\begin{equation*}
 \tilde L_{\mu\nu} = \frac{1}{2}\tilde R_{\mu\nu} - \frac{1}{12} \tilde R\tilde g_{\mu\nu}
\;,
\end{equation*}
the rescaled Weyl tensor,
\begin{equation*}
 \tilde d_{\mu\nu\sigma}{}^{\rho} = \Theta^{-1} \tilde C_{\mu\nu\sigma}{}^{\rho}  = \Theta^{-1} C_{\mu\nu\sigma}{}^{\rho}
 \;,
\end{equation*}
and the scalar function (set $\Box_{\tilde g} = \tilde g^{\mu\nu} \tilde\nabla_{\mu}\tilde\nabla_{\nu}$)
\begin{equation*}
 \tilde s = \frac{1}{4} \Box_{\tilde g} \Theta + \frac{1}{24} \tilde R \Theta
 \;.
\end{equation*}
The MCFE  read
\begin{eqnarray}
 && \tilde R_{\mu\nu\sigma}{}^{\kappa}[\tilde g] = \Theta \tilde d_{\mu\nu\sigma}{}^{\kappa} + 2(\tilde g_{\sigma[\mu}\tilde L_{\nu]}{}^{\kappa}  - \delta_{[\mu}{}^{\kappa}\tilde L_{\nu]\sigma} )
 \label{conf1}
\\
 && \tilde \nabla_{\rho} \tilde d_{\mu\nu\sigma}{}^{\rho} =0\;,
 \label{conf2}
\\
 && \tilde\nabla_{\mu}\tilde L_{\nu\sigma} - \tilde\nabla_{\nu}\tilde L_{\mu\sigma} = \tilde \nabla_{\rho}\Theta \, \tilde d_{\nu\mu\sigma}{}^{\rho}\;,
 \label{conf3}
\\
 && \tilde\nabla_{\mu}\tilde\nabla_{\nu}\Theta = -\Theta\tilde L_{\mu\nu} + \tilde s \tilde g_{\mu\nu}\;,
 \label{conf4}
\\
 && \tilde\nabla_{\mu} \tilde s = -\tilde L_{\mu\nu}\tilde\nabla^{\nu}\Theta\;,
 \label{conf5}
\\
 && 2\Theta \tilde s - \tilde \nabla_{\mu}\Theta\tilde\nabla^{\mu}\Theta = 0\;,
 \label{conf6}
\;.
\end{eqnarray}
%
Equation (\ref{conf6}) is a consequence of \eq{conf4} and \eq{conf5} if it is known to hold  at one point (e.g.\ by an appropriate choice of the initial data).

\subsubsection{General conformal field equations (GCFE)}

Consider any frame field $e_k = e^{\mu}{}_k\partial_{\mu}$ such that the $\tilde g(e_i ,e_k)\equiv \tilde g_{ik}$'s are constants.
The \emph{general conformal field equations (GCFE)} \cite{F3} for the unknowns
\begin{eqnarray*}
 e^{\mu}{}_k\;, \quad \tilde\Gamma_i{}^k{}_j\;, \quad  \tilde d_{ijk}{}^l \;, \quad \tilde L_{ij}\;, \quad \Theta\;, \quad  \tilde s
\end{eqnarray*}
read (from now on Latin indices are used to denote frame-components)
\begin{eqnarray}
 && \hspace{-2em} [e_p,e_q] = (\tilde \Gamma_p{}^l{}_q-\tilde \Gamma_q{}^l{}_p)e_l \;,
 \label{oconfbegin}
\\
 &&  \hspace{-2em} e_{[p}(\tilde \Gamma_{q]}{}^i{}_j) - \tilde \Gamma_k{}^i{}_j\tilde \Gamma_{[p}{}^k{}_{q]} + \tilde \Gamma_{[p}{}^i{}_{|k|}\tilde \Gamma_{q]}{}^k{}_j
 = \delta _{[p}{}^i \tilde L_{q]j}  - \tilde g_{j[p} \tilde L_{q]}{}^i -\frac{1}{2} \Theta \tilde d_{pqj}{}^i \;,
\\
 && \hspace{-2em} \tilde\nabla_i \tilde d_{pqj}{}^i =0 \;,
\\
 && \hspace{-2em} \tilde\nabla_i\tilde L_{jk} - \tilde\nabla_j\tilde L_{ik} =\tilde\nabla_l\Theta \tilde d_{jik}{}^l \;,
\\
 && \hspace{-2em} \tilde\nabla_i\tilde\nabla_j\Theta = -\Theta \tilde L_{ij} + s\tilde g_{ij} \;,
\\
 && \hspace{-2em} \tilde\nabla_i \tilde s = -\tilde L_{ij} \tilde\nabla^j\Theta \;,
\\
 && \hspace{-2em}  2\Theta \tilde s - \tilde\nabla_j\Theta\tilde\nabla^j\Theta = 0
 \;,
 \label{oconfend}
\end{eqnarray}
where the  $\tilde\Gamma_i{}^j{}_k$'s denote the Levi-Civita connection coefficients in the frame $e_k$.
Again, \eq{oconfend} suffices to be satisfied at just one point.

\subsection{Asymptotic behavior of the fields appearing in the~CFE}

In this section we analyze the asymptotic behavior of the restriction to $\mcN$ of the fields
$\tilde g_{\mu\nu}$, $e^{\mu}{}_k$, $\tilde\Gamma^{\sigma}_{\mu\nu}$, $\tilde d_{\mu\nu\sigma}{}^\rho$, $\tilde L_{\mu\nu}$, $ \Theta$ and $ \tilde s$  and prove that they are  smooth at  infinity in the metric gauge and taking \eq{gauge_choice_Theta}
when constructed as solutions of the constraint equations induced by the wave-map gauge reduced vacuum Einstein equations,
supposing that the initial data $\gamma$ are of the form \eq{initial_data},
 and  the no-logs-condition \eq{no-log-conditions}
hold.
As we shall show that  the above fields are smooth at $\scrip$ w.r.t\ the $e_k$-frame
  if and only if they have this property in the coordinate frame defined by $\{u,x,x^A\}$, we shall end up with the result that \eq{initial_data} and  \eq{no-log-conditions}
lead to smooth initial data for both  MCFE and  GCFE.

Using our previous results summarized in Section~\ref{overviewI}
most of the computations will be straightforward.

\subsubsection{Asymptotic behavior of the metric tensor}

It follows from \eq{com_metric_gaugeI}-\eq{com_metric_gaugeII} that%
\footnote{
While up to now we used the variable $x=1/r$ mainly as an auxiliary quantity to facilitate the comparison of the constraint equations
with the formulae in Appendix~\ref{asymptotic_solutions}, we shall use it henceforth  as a coordinate.
}
\begin{eqnarray*}
\ol{\tilde g}_{00}= -x^2, \quad \ol{\tilde g}_{0x}=-1\;, \quad \ol{\tilde g}_{0A}=\ol{\tilde g}_{xx}=\ol{\tilde g}_{xA}=0\;,\quad \ol{\tilde g}_{AB}=s_{AB}  + \mathcal{O}(x)
\;,
\end{eqnarray*}
i.e.\ $\ol{\tilde g}$ has a smooth conformal completion at conformal infinity.

\subsubsection{Asymptotic behavior of the Weyl tensor}

Note that the rescaled Weyl tensor  $\ol{\tilde d}_{\mu\nu\sigma}{}^{\rho}=\mathcal{O}(1)$
will be smooth
 at $\scri^+$
if and only if the Weyl tensor is smooth at $\scrip$ and vanishes there, i.e.\ $\ol {\tilde C}_{\mu\nu\sigma}{}^{\rho} = \ol C_{\mu\nu\sigma}{}^{\rho}=\mathcal{O}(x)$.

In vacuum we have $R_{\mu\nu}=0$,  and  the Weyl tensor coincides
with the Riemann tensor,
\begin{equation*}
 \overline C_{\mu\nu\rho}{}^{\sigma} \,=\, \overline R_{\mu\nu\rho}{}^{\sigma} \,=\, \overline{\partial_{\nu}\Gamma^{\sigma}_{\mu\rho}} - \overline{\partial_{\mu}\Gamma^{\sigma}_{\nu\rho}}
 + \overline\Gamma{}^{\alpha}_{\mu\rho} \overline\Gamma{}^{\sigma}_{\alpha\nu} - \overline\Gamma{}^{\alpha}_{\nu\rho} \overline\Gamma{}^{\sigma}_{\alpha\mu}
 \;.
\end{equation*}
%
Due to its algebraic symmetries it suffices to consider the  components
\begin{eqnarray*}
&\ol  C_{0r0}{}^{0} = \mathcal{O}(r^{-3})\;,\quad\ol C_{0rA}{}^{0} = \mathcal{O}(r^{-3})\;,\quad
\ol C_{0A0}{}^{0} = \mathcal{O}(r^{-1})\;,&
\\
 &\ol  C_{0A0}{}^{B} = \mathcal{O}(r^{-1})\;, \quad \ol C_{AB0}{}^0 =  \mathcal{O}(r^{-1}) \;,\quad  \ol C_{rAB}{}^{0}
 =\mathcal{O}(r^{-3})\;, &
\end{eqnarray*}
as follows from the formulae in Section~\ref{overviewI}.
%
Remarkably all the leading order terms which would induce terms of constant order after carrying out  the coordinate transformation $r\mapsto x:=1/r$
cancel out, in particular those involving some of the integration constants whose explicit values are not known,
\begin{eqnarray*}
&\ol  C_{0x0}{}^{0} = \mathcal{O}(x)\;,\quad\ol C_{0xA}{}^{0} = \mathcal{O}(x)\;,\quad
\ol C_{0A0}{}^{0} = \mathcal{O}(x)\;,&
\\
 &\ol  C_{0A0}{}^{B} = \mathcal{O}(x)\;, \quad \ol C_{AB0}{}^0 =  \mathcal{O}(x) \;, \quad  \ol C_{xAB}{}^{0} =\mathcal{O}(x)\;.&
\end{eqnarray*}
(Recall that $\mathcal{O}(x)$ has been defined in Section~\ref{sec_notation}.)
To establish that  $\ol C_{rAB}{}^{0}  =\mathcal{O}(r^{-3})$ rather than $\ol C_{rAB}{}^{0}  =\mathcal{O}(r^{-2})$
one needs to employ the no-logs-condition and is led to the geometric interpretation described in \cite[Section~6.2]{ChPaetz}.
No further condition on the initial data needs to be imposed.

\subsubsection{Asymptotic behavior of $\ol{\partial_0 \Theta}$ and $\ol{\partial^2_{00}\Theta}$}

To compute the remaining fields on  $\mcN$ we first need to determine the trace of the first- and second-order $u$-derivative of the conformal factor $\Theta$ on $\mcN$.
However, the values of $\Theta$ away from $\mcN$ depend on the unphysical curvature scalar $\tilde R$, which is treated as a conformal gauge source function in the CFE \cite{F3}.
We  impose the gauge condition
\begin{eqnarray}
\ol{ \tilde R} \,=\, \mathcal{O}(1) \;,\quad \ol{\tilde\nabla_0\tilde R}=\mathcal{O}(1)
\label{tildeRgauge}
\end{eqnarray}
(which is no restriction since $\tilde R$ needs to be smooth at $\scrip$ anyway).
The Ricci scalars of $\tilde g =\Theta^2 g$ and $g$ are related via
\begin{eqnarray}
 \tilde R
 &=&  \Theta^{-2}(  R - 6  \Theta^{-1}g^{\rho\sigma}\partial_{\rho}\partial_{\sigma}\Theta
+ 6\Theta^{-1} g^{\rho\sigma}\Gamma^{\alpha}_{\rho\sigma}\partial_{\alpha}\Theta )
 \;.
 \label{tildeR}
\end{eqnarray}
With $R=0$ that yields an ODE for $\overline{\partial_0\Theta}$ on $\mcN$ where $\ol \Theta=1/r$,
\begin{eqnarray}
 \overline{\tilde R} 
 &=& 6r^2\Big[   -2r\Big( \partial_r+r^{-1}+\frac{r}{2}|\sigma|^2\Big)\overline{\partial_0\Theta}
+|\sigma|^2
+\frac{1}{2} r^{-1}\overline g^{AB}\ol{\partial_0 g_{AB}} \Big]
\nonumber
\\
 &=&  -12r^3\Big( \partial_r+r^{-1}+\mathcal{O}(r^{-3}) \Big)\overline{\partial_0\Theta}
+\mathcal{O}(r^{-1})
 \label{tildeRcone}
 \;.
\end{eqnarray}
Employing  \eq{tildeRgauge} it  takes the form
\begin{eqnarray}
 r \partial_r\overline{\partial_0\Theta}  + [1+\mathcal{O}(r^{-2})] \overline{\partial_0\Theta} &=& \mathcal{O}(r^{-2})
 \;.
\end{eqnarray}
Appendix \ref{asymptotic_solutions}
tells us (with $\lambda=\hat\ell=1$) that the asymptotic  expansion  does not contain logarithmic terms and  is of the form
\begin{equation}
 \overline{\partial_0\Theta}=  \mathcal{O}(r^{-1}) =  \mathcal{O}(x)
 \;.
 \label{partial0Theta}
\end{equation}
%

The second-order $u$-derivative of the conformal factor can be computed as follows:
From \eq{tildeR}
we deduce with $\ol{\partial_0 R}=0$ that
\begin{eqnarray*}
 \ol{\partial_0 \tilde R} &=& 6\overline{\partial_0(  \Theta^{-3} g^{\rho\sigma}\Gamma^{\alpha}_{\rho\sigma}\partial_{\alpha}\Theta
  -  \Theta^{-3}g^{\rho\sigma}\partial_{\rho}\partial_{\sigma}\Theta )}
\\
  &=& -3r\ol{\tilde R}\ol{\partial_0\Theta}  - 12r^3 \Big( \partial_{r}+r^{-1} + \mathcal{O}(r^{-3})\Big) \ol{\partial^2_{00}\Theta }
+\mathcal{O}(1)
 \;.
\end{eqnarray*}
Taking the gauge condition \eq{tildeRgauge} into account this ODE for $\overline{\partial^2_{00}\Theta}$  becomes
\begin{eqnarray}
 r\partial_r\overline{\partial^2_{00}\Theta}  + [ 1 + \mathcal{O}(r^{-2})]\overline{\partial^2_{00}\Theta}  &=&  \mathcal{O}(r^{-2})
 \;,
\end{eqnarray}
which is of the same form as \eq{partial0Theta}. Hence
\begin{equation}
 \overline{\partial^2_{00}\Theta}=  \mathcal{O}(r^{-1}) = \mathcal{O}(x)
\;.
\label{partial00Theta}
\end{equation}

\subsubsection{Asymptotic behavior of the Christoffel symbols}

We have computed the restriction to $\mcN$ of the Christoffel symbols in adapted null coordinates $(u,r,x^A)$ by
imposing the metric gauge condition, cf.~Section~\ref{overviewI}.
Using the well-known  behavior of Christoffel symbols under coordinate transformations we  determine their
asymptotic behavior in the $(u,x=1/r,x^A)$-coordinates,
\newline

\begin{tabular}{ll}
  $\overline \Gamma{}^0_{00} =  0$\;, &   $\overline \Gamma{}^x_{xx}  =  -2x^{-1} + \mathcal{O}(x)$\;,
\\
 $\overline \Gamma{}^0_{0x} =\mathcal{O}(x)$\;, & $\overline \Gamma{}^x_{xA} =  \mathcal{O}(x)$\;,
\\
 $\overline \Gamma{}^0_{0A}= \mathcal{O}(x)$\;, &  $\overline \Gamma{}^x_{AB} = (s_{AB}+\frac{1}{2}D_{AB})x + \mathcal{O}(x^2)$\;,
\\
 $\overline \Gamma{}^0_{xx} = 0$\;, & $\overline \Gamma{}^C_{00} =0$\;,
\\
 $\overline \Gamma{}^0_{xA} =  0$\;, &  $\overline \Gamma{}^C_{0x} =  \mathcal{O}(x)$\;,
\\
 $\overline \Gamma{}^0_{AB} =-x^{-1}s_{AB} + \mathcal{O}(1)$\;, &   $\overline \Gamma{}^C_{0A} =\frac{1}{2}D_A{}^{C}x + \mathcal{O}(x^2)$\;,
\\
 $\overline \Gamma{}^x_{00} =0 $\;,&  $\overline \Gamma{}^C_{xx} = 0$\;,
\\
 $\overline \Gamma{}^x_{0x} = \mathcal{O}(x^3)$\;, &  $\overline \Gamma{}^C_{xA} =-\delta_{A}{}^C x^{-1} +\mathcal{O}(1)$\;,
\\
 $\overline \Gamma{}^x_{0A} =\mathcal{O}(x^3)$\;, &  $\overline \Gamma{}^C_{AB} = \mathring \Gamma{}^C_{AB} + \mathcal{O}(x)$\;.
\end{tabular}
\newline

We compute the trace of the Christoffel symbols on $\mcN$ associated to the unphysical metric $\tilde g$.
The transformation law for Christoffel symbols under a conformal rescaling $g\mapsto \Theta^2 g$ of the metric reads,
\begin{eqnarray}
 \tilde\Gamma^{\rho}_{\mu\nu} &=& \Gamma^{\rho}_{\mu\nu} + \frac{1}{\Theta}\left(\delta_{\nu}^{\phantom{\nu}\rho}\partial_{\mu} \Theta + \delta_{\mu}^{\phantom{\mu}\rho}\partial_{\nu} \Theta -g_{\mu\nu}g^{\rho\sigma}\partial_{\sigma} \Theta\right)
 \;,
\label{christoffel_conformal_trafo}
\end{eqnarray}
which yields on $\mcN$, where   $\overline \Theta =x$ and $\ol {\partial_0\Theta} =\mathcal{O}(x)$,
\begin{eqnarray*}
&\overline {\tilde \Gamma}{}^0_{00} =\mathcal{O}(1)\;, \quad \overline {\tilde \Gamma}{}^0_{0x} =  \mathcal{O}(x)\;, \quad
\overline {\tilde \Gamma}{}^0_{0A}=  \mathcal{O}(x)\;, \quad  \overline {\tilde \Gamma}{}^0_{xx} = 0\;, \quad  \overline {\tilde \Gamma}{}^0_{xA} =  0\;,&
\\
&\overline {\tilde \Gamma}{}^0_{AB} = \mathcal{O}(1)\;, \quad \overline {\tilde \Gamma}{}^x_{00} = \mathcal{O}(x^2)\;, \quad
\overline {\tilde \Gamma}{}^x_{0x} =  \mathcal{O}(x)\;, \quad  \overline {\tilde \Gamma}{}^x_{0A} = \mathcal{O}(x^3)\;, &
\\
& \overline {\tilde \Gamma}{}^x_{xx}  =   \mathcal{O}(1)\;, \quad \overline {\tilde \Gamma}{}^x_{xA} = \mathcal{O}(x)\;, \quad
\overline {\tilde \Gamma}{}^x_{AB} =  \mathcal{O}(1)\;, \quad  \overline {\tilde \Gamma}{}^C_{00} =0\;, \quad
\overline {\tilde \Gamma}{}^C_{0x} =\mathcal{O}(x)\;,&
\\
&\overline {\tilde \Gamma}{}^C_{0A} = \mathcal{O}(1)\;, \quad \overline {\tilde \Gamma}{}^C_{xx} = 0\;, \quad
\overline {\tilde \Gamma}{}^C_{xA} = \mathcal{O}(1)\;, \quad  \overline {\tilde \Gamma}{}^C_{AB} =   \mathcal{O}(1)\;.&
\end{eqnarray*}
The Christoffel symbols are smooth without any further restrictions on $\gamma$.

\subsubsection{Asymptotic behavior of the Schouten tensor}

From now all tensors will be expressed in terms of the coordinates $(u,x,x^A)$.
We compute the Schouten tensor $\tilde L_{\mu\nu}= \frac{1}{2}\tilde R_{\mu\nu} - \frac{1}{12}\tilde R\tilde g_{\mu\nu}$, restricted to $\mcN$, for the conformally rescaled metric $ \tilde g= \Theta^2 g$.
The transformation law for the  Ricci tensor under conformal rescalings of the metric reads,
\begin{eqnarray}
 \tilde L_{\mu\nu}
 &=& L_{\mu\nu} - \Theta^{-1}(\partial_{\mu}\partial_{\nu}\Theta
 -\Gamma^{\alpha}_{\mu\nu}\partial_{\alpha}\Theta )
+2\Theta^{-2}(\partial_{\mu}\Theta\partial_{\nu}\Theta)\breve{}
 \;.
 \label{tildeRicci}
\end{eqnarray}
%
%
With $\overline L_{\mu\nu}=0$ we obtain on $\mcN$
\begin{eqnarray*}
\ol{ \tilde L}_{\mu\nu}
 &=& 2x^{-2}\ol{\partial_{\mu}\Theta}\,\ol{\partial_{\nu}\Theta}  -x^{-1}\ol{\partial_{\mu}\partial_{\nu}\Theta }
+ (x^{-1}\ol\Gamma{}^{0}_{\mu\nu}+\ol g_{\mu\nu})\ol{\partial_{0}\Theta} + x^{-1}\ol\Gamma{}^{x}_{\mu\nu}
 -   \frac{1}{2}x^{2}\ol g_{\mu\nu}
\;.
\end{eqnarray*}
Assuming  \eq{tildeRgauge}, so that \eq{partial0Theta} and \eq{partial00Theta} hold, we find
%
\begin{eqnarray*}
 \overline{\tilde L}_{00} &=&  2x^{-2}\overline{\partial_0\Theta}\,\overline{\partial_0\Theta}  - \overline{\partial_0\Theta} + \frac{1}{2}x^{2}  -x^{-1}\overline{\partial^2_{00}\Theta} \,=\, \mathcal{O}(1)
\;,
\\
  \overline{\tilde L}_{0x} &=& \frac{1}{2}
  -\partial_x(x^{-1}\overline{\partial_0\Theta}) + \mathcal{O}(1) \overline{\partial_0\Theta}+ \mathcal{O}(x^2) \,=\, \mathcal{O}(1)
\;,
\\
 \overline{\tilde L}_{0A}
 &=& -x^{-1}\partial_A\overline{\partial_0\Theta} + \mathcal{O}(1)\overline{\partial_0\Theta}   + \mathcal{O}(x^2) \,=\, \mathcal{O}(1)
\;,
\\
 \overline{\tilde L}_{xx} &=&   
  \mathcal{O}(1)
\;,
\\
 \overline{\tilde L}_{xA} &=& 
\mathcal{O}(1)\;,
\\
 \overline{\tilde L}_{AB}
 &=&  \mathcal{O}(x^{-1})\overline{\partial_{0}\Theta}  + \mathcal{O}(1)  \,=\, \mathcal{O}(1)
 \;,
\end{eqnarray*}
We conclude that the trace of the Schouten tensor on $\mcN$ is smooth at conformal infinity.

\subsubsection{Asymptotic behavior of the function $\tilde s$}

Let us  determine the asymptotic behavior of the function $\tilde s\equiv \frac{1}{4}\Box_{\tilde g}\Theta + \frac{1}{24}{\tilde R} \Theta$
on $\mcN$.
Using \eq{tildeR} with $\ol R=0$ and \eq{christoffel_conformal_trafo} we find that
\begin{eqnarray*}
 \overline{\tilde s}
 &=& \frac{1}{4} x^{-2}\overline g^{\mu\nu}\partial_{\mu}\partial_{\nu}\Theta
- \frac{1}{4}x^{-2}\overline g^{\mu\nu}\tilde\Gamma^{\kappa}_{\mu\nu}\overline{\partial_{\kappa}\Theta} + \frac{1}{24}\overline{\tilde R}x
\\
&=& \frac{1}{24} x^{-1} \ol R
+\frac{1}{4} x^{-2}  \ol g^{\mu\nu}(\ol \Gamma{}^{\alpha}_{\mu\nu} - \ol{\tilde  \Gamma}{}^{\alpha}_{\mu\nu})\partial_{\alpha}\Theta
\\
&=&\frac{1}{2} x^{-3}\ol g^{\alpha\beta}\ol{\partial_{\alpha}\Theta }\,\ol{\partial_{\beta} \Theta}
\;,
\end{eqnarray*}
which recovers \eq{conf5},
and which yields with \eq{partial0Theta} that
\begin{eqnarray}
  \overline{\tilde s}  \,=\, \frac{1}{2} x -x^{-1}\ol{\partial_{0}\Theta } \,=\, \mathcal{O}(1)
\;,
\end{eqnarray}
i.e.\ the function $\overline{\tilde s} $ is smooth at conformal  infinity.

\subsubsection*{Asymptotic behavior of the frame field}

The GCFE (\ref{oconfbegin})-(\ref{oconfend}) require a frame field $(e^{\mu}{}_k)$ w.r.t\ which the metric tensor is constant.
In our coordinates the  trace of $\tilde g$ on $\mcN$  looks  as follows (recall  that equality is meant to hold  for $r> r_2$):
\begin{equation*}
 \overline {\tilde g} = -x^2\mathrm{d}u^2 - 2\mathrm{d}u\mathrm{d}x + \tilde g_{AB}\mathrm{d}x^{A}\mathrm{d}x^B
 \quad \text{with} \quad  \tilde g_{AB} = x^2\overline g_{AB} = \mathcal{O}(1)
 \;.
\end{equation*}
We deduce that  we may take $(e^{\mu}{}_k)$ to be of the following form on~$\mcN$:
\begin{eqnarray*}
\ol  e_0 &=& \partial_u -\frac{x^2}{4}\partial_x\;,
\\
 \ol e_x &=&  \partial_x\;,
\\
\ol  e_{\tilde A} &=& e^{A}_{\phantom{A}\tilde A}\partial_{A} \quad \text{with} \quad  e^{A}_{\phantom{A}\tilde A} = \mathcal{O}(1) \quad \tilde A =2,3
\;,
\end{eqnarray*}
and its dual
\begin{eqnarray*}
 \ol\Theta^0 &=& \mathrm{d}u\;,
\\
 \ol\Theta^x &=& \mathrm{d}x + \frac{x^2}{4}\mathrm{d}u\;,
\\
 \ol\Theta^{\tilde A} &=& \hat e^{\tilde A}_{\phantom{A}A}\mathrm{d}x^A \quad \text{with} \quad  \hat e^{\tilde A}_{\phantom{A}A} = \mathcal{O}(1)\quad \tilde A =2,3
 \;.
\end{eqnarray*}
All the relevant fields are smooth at $\scrip$ w.r.t.\ this frame if and only if they have this property w.r.t.\ the coordinate frame defined by the adapted coordinates $\{u,x,x^A\}$, which we have shown to be the case.

\subsection{Main result}

%

Consider a space-time  $(\mcM,g)$
 which admits a smooth conformal completion at infinity \emph{\`a la Penrose}, and
 consider  a   null hypersurface $\mcN\subset\mcM$ whose closure in the conformally completed space-time $\mcM \cup \scrip$ is smooth and  meets $\scrip$ in a smooth spherical cross-section.
It  follows from the considerations in \cite{Tambourino66}, cf.\  \cite{ChPaetz2}, that then
one can introduce Bondi coordinates near  $\ol \mcN \cap \scrip$ w.r.t.\ which $\mcN$
 intersects $\scrip$ at the surface $\{u=0\}$, and  in which all the fields appearing in the CFE are smooth at $\scrip$
 in the sense of Definition~\ref{definition_smooth}.

The existence of adapted null coordinates in which the unknowns of the CFE  are smooth at $\scrip$
 is thus  a necessary condition for the existence  of a space-time which admits a smooth conformal completion \emph{\`a la Penrose}.
We are led to  the following result:
%
\begin{theorem}
 A necessary-and-sufficient condition
for the restrictions to $\mcN$ of all the fields appearing in the GCFE (\ref{oconfbegin})-(\ref{oconfend}), or the MCFE (\ref{conf1})-(\ref{conf6}),
constructed from initial data~$\gamma=\gamma_{AB}\mathrm{d}x^A\mathrm{d}x^B$ of the form \eq{initial_data}
to be smooth at $\scrip$,
 is that the initial data~$\gamma$  satisfy the no-logs-condition \eq{no-log-conditions} in some (and then all) adapted null coordinate systems.
In that case  the metric gauge provides a gauge choice where smoothness at $\scrip$  holds.
\end{theorem}

\vspace{1.2em}
\noindent {\textbf {Acknowledgments}}
I am grateful to my advisor Piotr T. Chru\'sciel for  many valuable comments and suggestions, and for reading a first draft of this article.
Supported in part by the  Austrian Science Fund (FWF): P 24170-N16.

\appendix

\section{Asymptotic solutions of Fuchsian ODEs}
\label{asymptotic_solutions}

The main object of  this appendix is to justify rigorously our use of  expansions as asymptotic  solutions to  Einstein's characteristic  constraint equations.
Smoothness of these solutions at infinity  is a crucial aspect of the analysis, which is why
we derive necessary-and-sufficient conditions for the asymptotic expansions to involve no logarithmic terms.
To do that we shall proceed as follows: Instead of using $r$ as the independent variable,  we introduce $x:=1/r$ as a new variable  and study the transformed ODE near $x=0$.
For this we make Taylor expansions of the coefficients appearing in the ODE at $x=0$, and write down the formal polyhomogeneous solutions.
The Borel summation lemma  guarantees that there exists a function whose polyhomogeneous expansion coincides with the formal series.
This function will be shown to approximate the exact solution
 around $x=0$,  from which we eventually conclude that the formal polyhomogeneous solution
is in fact an expansion of the exact solution at $x=0$.

To illustrate the procedure, we first show how it works for linear first-order scalar equations in full generality. We then show how this adapts to   linear first-order systems, under conditions corresponding to
those that arise in the main text, in order to avoid an uninteresting discussion of  several
special cases. Every dependence 
on further variables, which we assume to have compact support,  will be suppressed for convenience.

For the definition of polyhomogeneous functions we refer the reader to \cite[Appendix~A]{ChPaetz2}.

\subsection{Formal solutions}

\subsubsection{Scalar equation $x\partial_x f + h f = g$}

We consider the ODE
\begin{equation}
 x\partial_x f + h f = g
 \;,
\label{scalar_ODE}
\end{equation}
where $x^{-\ell}g=O(1)$, $\ell\in\mathbb{Z}$, and $h=O(1)$  (which clearly includes those cases where $h$ has a zero of any order at $x=0$) are assumed to be smooth functions on some interval $[0,x_0)$.

We represent  $x^{-\ell} g$   and $h$ via their Taylor expansions at $x=0$,
\begin{eqnarray*}
 g \ourdoteq  \sum_{n=\ell}^{\infty} g_n x^n
 \;, \quad h \ourdoteq  \sum_{n=0}^{\infty} h_n x^n
\end{eqnarray*}
(the symbol  $\sim$ has been defined in Section~\ref{sec_notation}).
We define the \emph{indicial exponent} to be
\vspace{-0.3em}
\begin{eqnarray}
 \lambda := -h(0) = -h_0
 \;.
\end{eqnarray}
When considering functions which further depend on angular variables, we will always make the supplementary hypothesis that
\bel{20VI11.1}
 \mbox{ $h_0$ is angle-independent.}
\ee

\paragraph*{1st case: $\lambda\not\in\mathbb{Z}\;.$}
We make the ansatz
\begin{eqnarray}
 f \ourdoteq  x^{\lambda} \sum_{n=\ell}^{\infty}f^{(0)}_{n+\lambda} x^n + \sum_{n=\ell}^{\infty}f^{(1)}_nx^n =: x^{\lambda}f^{(0)} + f^{(1)}
 \;.
\label{ansatz_1}
\end{eqnarray}
Here we use $f^{(a)}$ as short form for the corresponding \emph{formal} power series
(in Section~\ref{app_borel} we shall use Borel summation to obtain a proper function from these formal expansions).
In the course of this appendix it will become clear that any solution of \eq{scalar_ODE} admits an expansion of the form \eq{ansatz_1}, so this ansatz is not
restrictive.
It follows from \eq{scalar_ODE} that  $f^{(1)}$ needs to
satisfy
for any $N\in \N$
\begin{eqnarray}
 \sum_{n=\ell}^{N} nf^{(1)}_n x^n + \sum_{n=\ell}^{N} \sum_{k=0}^{n-\ell}h_k f^{(1)}_{n-k}x^n = \sum_{n=\ell}^{N}g_nx^n + o (x^N)
\nonumber
\\
 \Longleftrightarrow \quad (n-\lambda) f^{(1)}_n = g_n - \sum_{k=1}^{n-\ell}h_kf^{(1)}_{n-k} \quad \text{for $n=\ell,\ell+1,\dots$}
\end{eqnarray}
Since $n-\lambda\ne 0$ by assumption, this defines a unique formal solution  $f^{(1)}$ by solving the equations hierarchically. The formal series $f^{(0)}$ needs to satisfy
%
\begin{eqnarray}
 x^{\lambda}\sum_{n=\ell}^{N}(\lambda+n)f^{(0)}_{n+\lambda}x^n + x^{\lambda}\sum_{n=\ell}^{N} \sum_{k=0}^{n-\ell} h_kf^{(0)}_{n-k+\lambda}x^n =o(x^{\lambda+N})
 \nonumber
 \\
 \Longleftrightarrow \quad nf^{(0)}_{n+\lambda} = -\sum_{k=1}^{n-\ell} h_k f^{(0)}_{n-k+\lambda}  \quad \text{for $n=\ell,\ell+1,\dots$}
 \label{eqn_case_lambda}
\end{eqnarray}
We observe that $f^{(0)}_{n+\lambda}=0$ for $n<0$, while $f^{(0)}_{\lambda}$ can be chosen arbitrarily.
Once this has been done,  (\ref{eqn_case_lambda}) determines the higher-order coefficients.
Hence, our ansatz leads to a formal solution, where  $f^{(0)}_{\lambda}$ can be thought of as representing the integration constant, or function if angular variables are involved. For convenience we will just speak of an integration constant in what follows.

\paragraph*{2nd case: $\lambda\in\mathbb{Z}\;.$}
 
We start with the (again non-restrictive)  ansatz
\begin{eqnarray}
 f \ourdoteq \sum_{n=\hat\ell}^{\infty}f^{(0)}_nx^n + \log x \sum_{n=\hat\ell}^{\infty}f^{(1)}_nx^n =: f^{(0)} + f^{(1)} \log x
 \;,
\label{ansatz_2}
\end{eqnarray}
where we have set $\hat\ell:=\mathrm{min}(\lambda,\ell)$. 
Inserting \eq{ansatz_2} into \eq{scalar_ODE} yields
%
\begin{eqnarray}
 \log x &&\hspace{-1em}\sum_{n=\hat\ell}^{N}n f^{(1)}_nx^n +  \sum_{n=\hat\ell}^{N} f^{(1)}_nx^n + \log x \sum_{n=\hat\ell}^{N}\sum_{k=0}^{n-\hat\ell} h_k f^{(1)}_{n-k}x^n
 + \sum_{n=\hat\ell}^{N} nf^{(0)}_n x^n
 \nonumber
\\
  &&\qquad + \sum_{n=\hat\ell}^{N} \sum_{k=0}^{n-\hat\ell}h_k f^{(0)}_{n-k}x^n = \sum_{n=\hat\ell}^{N}g_nx^n + o(x^N \log x )
 \nonumber
\\
 \Longleftrightarrow &&\quad
  (n-\lambda) f^{(1)}_n + \sum_{k=1}^{n-\hat\ell} h_k f^{(1)}_{n-k} =0
 \label{formalsolution1}
\\
  \text{and} &&\quad f^{(1)}_n + (n-\lambda) f^{(0)}_n +  \sum_{k=1}^{n-\hat\ell}h_k f^{(0)}_{n-k} = g_n \quad \text{for any $n\geq\hat\ell$}
 \label{formalsolution2}
 \;.
\end{eqnarray}

If $n=\hat \ell $, \eq{formalsolution1} is understood as $( \hat \ell-\lambda) f^{(0)}_{\hat \ell}=0$,
 for  $\hat \ell\le  n<\lambda$ it leads to $f^{(1)}_n=0$.
Then \eq{formalsolution2} with   $\hat \ell\le  n<\lambda$  determines the coefficients $f^{(0)}_n$.
If $n=\lambda$ \eq{formalsolution1}, holds automatically, while \eq{formalsolution2} determines $f^{(1)}_{\lambda}$.
The coefficient
$f^{(0)}_{\lambda}$ can always be chosen arbitrarily. Once this has been done, all coefficients $f^{(1)}_n$ and $f^{(0)}_n$ with $\lambda>n$ are  determined by the \eq{formalsolution1} and \eq{formalsolution2},
respectively.
This way we obtain a formal solution with one free parameter, $f^{(0)}_{\lambda}$, which can be regarded as integration constant.

These considerations also reveal that the formal solution \eq{ansatz_2}
contains no logarithmic terms if and only if $f^{(1)}_{\lambda} = 0$ or, equivalently,
%
\begin{eqnarray}
 \sum_{k=1}^{\lambda-\hat\ell}h_k f^{(0)}_{\lambda-k} = g_{\lambda}
 \;.
 \label{cond_no_logs}
\end{eqnarray}
Indeed the vanishing of $f^{(1)}_{\lambda}$ enforces by (\ref{formalsolution1}) that all the $f^{(1)}_{n}$'s are zero for $n>\lambda$ as well, while those with $n<\lambda$ have to vanish anyway.
Note that (\ref{cond_no_logs}) is in fact a condition relating $h$ and $g$,
since the $f^{(0)}_{\lambda-k}$'s are determined hierarchically from (\ref{formalsolution2}) (with $f^{(1)}_n=0$), it does not depend
on the boundary conditions as captured by the integration constant. 


\subsubsection{ODE-system $x\partial_x f + h f = g$}

Let us now consider  a  first-order   linear ODE-system,
\begin{equation}
 x\partial_x f + h f = g
 \;,
\label{ODE_system}
\end{equation}
where $h=O(1)\in\mathrm{Mat}(n,n)$ and $g=O(x^{\ell})$, $\ell\in\mathbb{Z}$. The components of $h$ and $x^{-\ell}g$ are assumed to be smooth functions on the interval $[0,x_0)$. 
For convenience, and because it suffices for our purposes, we focus on the case $n=2$. Only at some points we add a comment how the general case looks like.


Again, we represent  $x^{-\ell}g$  and $h$ via their Taylor expansions at $x=0$,
\begin{eqnarray*}
 g \ourdoteq  \sum_{n=\ell}^{\infty} g_n x^n
 \;, \quad g_n\in\mathbb{R}^2
 \;, \quad h \ourdoteq  \sum_{n=0}^{\infty} h_n x^n
 \;, \quad h_n\in\mathrm{Mat}(2,2)
 \;.
\end{eqnarray*}
There exists a change of basis matrix $T\in\mathrm{GL}(2)$ such that $Th_0T^{-1}=:h_0^J$ adopts Jordan normal form.
Hence, it suffices to study the system
\begin{eqnarray*}
 x\partial_x Tf + \Big[ h_0^J + \sum_{n=1}^{\infty} Th_n T^{-1}x^n \Big] Tf = Tg
\;,
\end{eqnarray*}
or, by relabeling the symbols,
\begin{eqnarray}
 x\partial_x f + \Big(\sum_{n=0}^{\infty} h_nx^n\Big) f = g
\;,
\label{ODE_system2}
\end{eqnarray}
with either
\begin{eqnarray}
 h_0= \begin{pmatrix} -\lambda_1 & 0 \\ 0 & -\lambda_2 \end{pmatrix} \quad \text{or} \quad  h_0= \begin{pmatrix} -\lambda & 1 \\ 0 & -\lambda \end{pmatrix}
 \;.
\end{eqnarray}
We mentioned that a dependence on additional variables is permitted. However, as in the scalar case, we assume
\begin{equation}
 \mbox{ the \emph{indicial matrix} $h_0$ is angle-independent,}
\end{equation}
%
and thus is a truly constant matrix.
In addition, since this covers all cases we are interested in with regard to the main text, we assume that
\begin{equation}
 \lambda,\lambda_i \in \mathbb{Z}
\;.
\end{equation}
%

\paragraph*{1st case: $h_0= \begin{pmatrix} -\lambda_1 & 0 \\ 0 & -\lambda_2 \end{pmatrix}\;.$}
W.l.o.g.\ we assume $\lambda_1\leq\lambda_2$. Furthermore, we define
$$
 \hat\ell:=\mathrm{min}(\lambda_1,\ell)
  \;.
$$
We make the ansatz (again, later on it will be shown that any solution of \eq{ODE_system} admits an expansion of the form \eq{ansatz_system}) 
\begin{eqnarray}
 f_i \,\ourdoteq \, \sum_{k=0}^2 \log^k x \sum_{n=\hat\ell}^{\infty} (f^{(k)}_i)_n x^n
 \, =: \,
 f^{(0)}_i  + f^{(1)}_i\log x  + f^{(2)}_i\log^2 x
 \;,
 \label{ansatz_system}
\end{eqnarray}
where the symbols appearing in this definition are to be understood in the same way as above.
The upper index in brackets displays the order of the log term.
Whenever useful, the coefficients $f^{(a)}_k$ with $k<\hat \ell$ are defined as to be zero.

We insert \eq{ansatz_system} into \eq{ODE_system2}. The coefficients need to satisfy the following set of equations (for $n\geq \hat\ell$):
\begin{eqnarray*}
 &(1)&  (n-\lambda_i)(f^{(2)}_i)_n + F_i[(f^{(2)})_k,k<n] = 0
 \;, \quad i=1,2
 \;,
\\
 &(2)&  2(f^{(2)}_i)_n + (n-\lambda_i)(f^{(1)}_i)_n + G_i[(f^{(1)})_k,k<n]=0
 \;, \quad i=1,2
\;,
\\
 &(3)& (f^{(1)}_i)_n + (n-\lambda_i)(f^{(0)}_i)_n + H_i[(f^{(0)})_k,k<n]=(g_i)_n \;, \quad i=1,2
 \;,
\end{eqnarray*}
where $F_i$, $G_i$ and $H_i$ are multi-linear functions of the indicated quantities.
The explicit form of $H_i$, which will be needed later on, is
\begin{eqnarray}
 H_i= \sum_{k=1}^{n-\hat\ell} \Big[ h_k (f^{(0)})_{n-k} \Big]_i
 \;,
 \label{eqn_H_i}
\end{eqnarray}
analogously for the other functions.
Note that $H_i$ generally depends on both components of $f^{(0)}$ since there is no need for the $h_k$'s, $k\geq 1$, to be diagonal.

A solution to the equations (1)-(3) can be constructed as follows:
We describe  the case $\lambda_1<\lambda_2$, the case $\lambda_1=\lambda_2$ can be treated similarly.

$\mathbf{n<\lambda_1:}\enspace$
We have to choose $(f^{(2)}_i)_n=0$ and $(f^{(1)}_i)_n=0$ to fulfill (1) and (2). The coefficients $(f^{(0)}_i)_n$ will be generally non-zero and are determined
by~(3).

$\mathbf{n=\lambda_1:}\enspace$
Choose $(f^{(2)}_i)_{\lambda_1}=0$ and $(f^{(1)}_2)_{\lambda_1}=0$. The first component of (2) (i.e.\ the one with $i=1$) is automatically satisfied,
while the first component of (3) determines $(f^{(1)}_1)_{\lambda_1}$.
The coefficient $(f^{(0)}_1)_{\lambda_1}$ is free to choose, while
$(f^{(0)}_2)_{\lambda_1}$ follows   from the second component of (3) (the one with $i=2$).

$\mathbf{\lambda_1<n<\lambda_2:}\enspace$
(1) still requires $(f^{(2)}_i)_n=0$. The coefficients $(f^{(1)}_i)_n$ are determined by (2), while the $(f^{(0)}_i)_n$'s are determined by (3).

$\mathbf{n=\lambda_2:}\enspace$
Set $(f^{(2)}_1)_{\lambda_2}=0$. The second component of (1) holds automatically, no matter what the value of $(f^{(2)}_2)_{\lambda_2}$ is. The coefficient $(f^{(1)}_1)_{\lambda_2}$
is determined by the first component of (2). The coefficient $(f^{(2)}_2)_{\lambda_2}$ follows from (2).
The coefficient  $(f^{(1)}_2)_{\lambda_2}$ is  determined by   the second component of (3).
The coefficient $(f^{(0)}_1)_{\lambda_2}$ follows from (3), while $(f^{(0)}_2)_{\lambda_2}$ can be chosen arbitrarily.

$\mathbf{n>\lambda_2:}\enspace$
All coefficients $(f^{(j)}_i)_{n}=0$ are  determined by (1)-(3).

We remark that for $\lambda_1=\lambda_2$ the coefficients which can be viewed as integration constants are $(f^{(0)}_i)_{\lambda}$, $i=1,2$. Moreover, this case implies $f^{(2)}=0$.

Consequently, the ansatz (\ref{ansatz_system}) leads to a formal solution of 
\eq{ODE_system2} with two free parameters, which can be considered as integration constants $(f^{(0)}_i)_{\lambda_i}$.

In fact a similar ansatz, namely
\begin{eqnarray}
 f_i &\ourdoteq & \sum_{k=0}^N \sum_{n=\hat\ell}^{\infty} (f^{(k)}_i)_n x^n  \log^k x
 \;,
\end{eqnarray}
leads to a formal solution of the corresponding $N$-dimensional system with $h_0=\mathrm{diag}(-\lambda_1,\dots,-\lambda_N)$.
The integration constants can be identified with $(f^{(0)}_i)_{\lambda_i}$.

Logarithmic  terms do not appear in \eq{ansatz_system} if and only if 
\begin{eqnarray}
\sum_{k=1}^{\lambda_i-\hat\ell} [h_k(f^{(0)})_{\lambda_i-k}]_i = (g_i)_{\lambda_i}
 \;, \quad i=1,2
 \;.
 \label{no_log_terms_diagsystem}
\end{eqnarray}
Here one has to distinguish two cases: If
\begin{equation}
\text{\eq{no_log_terms_diagsystem} is independent of $(f^{(0)}_1)_{\lambda_1}$}  \;,
\label{app-no-logs}
\end{equation}
then
\eq{no_log_terms_diagsystem} is, as in the scalar case, a condition  involving exclusively $g$ and $h$, i.e.\ it just concerns the equations itself
and is independent of the boundary conditions.
In particular this case  occurs for $\lambda_1=\lambda_2$.
Otherwise
the appearance of log terms depends on the value of the integration constant
$(f^{(0)}_1)_{\lambda_1}$.


\paragraph*{2nd case: $h_0= \begin{pmatrix} -\lambda & 1 \\ 0 & -\lambda \end{pmatrix}\;.$}

We use the same ansatz \eq{ansatz_system} as  in the 1st case,
%
%
and with $\hat\ell:=\mathrm{min}(\lambda,\ell)$.
We insert this ansatz into \eq{ODE_system2} to end up with the following relations among the coefficients ($n\geq \hat\ell$):
\begin{eqnarray*}
 \hspace{-1em} &(1)&  (n-\lambda)(f^{(2)}_i)_n + \delta^1_{\phantom{1}i}(f^{(2)}_2)_n +  F_i[(f^{(2)})_k,k<n] = 0
 \;, \, i=1,2,
\\
 \hspace{-1em}  &(2)&  2(f^{(2)}_i)_n + (n-\lambda)(f^{(1)}_i)_n + \delta^1_{\phantom{1}i}(f^{(1)}_2)_n  + G_i[(f^{(1)})_k,k<n]=0
 \;, \, i=1,2,
\\
\hspace{-1em}  &(3)& (f^{(1)}_i)_n + (n-\lambda)(f^{(0)}_i)_n + \delta^1_{\phantom{1}i}(f^{(0)}_2)_n  + H_i[(f^{(0)})_k,k<n]=(g_i)_n
 \;, \, i=1,2,
\end{eqnarray*}
where $F_i$, $G_i$ and $H_i$ are again multi-linear functions of the indicated quantities whose explicit formulae look similar to (\ref{eqn_H_i}).
We describe how one obtains a solution of these equations:

$\mathbf{n<\lambda:}\enspace$ All coefficients are determined by (1)-(3), in particular $(f^{(1)}_i)_n=(f^{(2)}_i)_n=0$. 

$\mathbf{n=\lambda:}\enspace$
Equation (1) is fulfilled iff $(f^{(2)}_2)_{\lambda}=0$.
The second component of (2) holds since we have chosen $(f^{(2)}_2)_{\lambda}=0$. The first component of (2) enforces
$ 2(f^{(2)}_1)_{\lambda} + (f^{(1)}_2)_{\lambda}=0$.
In order to satisfy the second component of (3) the yet unspecified $(f^{(1)}_2)_{\lambda}$ (equivalently $(f^{(2)}_1)_{\lambda}$) has to be chosen such that
\begin{eqnarray}
 (f^{(1)}_2)_{\lambda} + H_2[(f^{(0)})_k,k<\lambda]=(g_2)_{\lambda}
 \;,
  \label{vanishinglog_system1}
\end{eqnarray}
whatever is taken for $(f^{(1)}_1)_{\lambda}$ and $(f^{(0)}_2)_{\lambda}$.
The first component of (3) can be fulfilled by an appropriate choice of $(f^{(1)}_1)_{\lambda}$, and is independent of $(f^{(0)}_1)_{\lambda}$,
\begin{eqnarray}
 (f^{(1)}_1)_{\lambda} + (f^{(0)}_2)_{\lambda}  + H_1[(f^{(0)})_k,k<{\lambda}]=(g_1)_{\lambda}
 \;.
 \label{vanishinglog_system2}
\end{eqnarray}

$\mathbf{n>\lambda:}\enspace$ All coefficients $(f^{(j)}_i)_n$ are uniquely determined by (1)-(3).

This way we get a formal solution for any choice of $(f^{(0)}_i)_{\lambda}$, $i=1,2$, which  may be regarded as representing the integration constants.
 
The above algorithm 
 shows that, in the current setting, logarithms are absent if and only~if
\begin{equation}
 (f^{(1)}_i)_{\lambda}=0 \;.
 \label{non-diag_no-log}
\end{equation}
According to (\ref{vanishinglog_system1}), $(f^{(1)}_2)_{\lambda}$  vanishes iff $H_2[(f^{(0)})_k,k<\lambda]=(g_2)_{\lambda}$, i.e.\ iff $h$ and $g$ satisfy appropriate
relations. However, (\ref{vanishinglog_system2}) shows that the vanishing of $(f^{(1)}_1)_{\lambda}$  depends on the integration constant $(f^{(0)}_2)_{\lambda}$,  i.e.\ on the boundary conditions,  and thus cannot be guaranteed to hold generally. 
Only specific  boundary data  lead to solutions without logarithmic  terms.

\subsection{Borel summation}
\label{app_borel}

We have seen that there exist formal solutions of the ODE $x\partial_x f + h f = g$ (in one or two dimensions) of the form
\begin{eqnarray}
 f_{\mbox{\scriptsize formal}} = \varkappa^{(0)} f^{(0)} + \varkappa^{(1)} f^{(1)} + \varkappa^{(2)} f^{(2)}
 \;, \quad \varkappa^{(i)}\in \{\log x,\log^2 x, x^{\lambda},0,1\}
 \;,
\end{eqnarray}
where the form of $\varkappa^{(i)}$ depends on the specific value of the indicial exponent $\lambda$. However, in any case all the $f^{(i)}$'s are formal power series.
We can therefore appeal to the \textit{Borel Summation Lemma} (see, e.g., \cite[Appendix~D]{ChJezierskiCIVP})
which states that for every formal power series $f$ one can find a smooth function $\hat f$ whose Taylor expansion, around $x=0$ say, coincides with $f$.
Applied to our case that means that there are smooth functions $\hat f^{(i)}$, $i=0,1,2$, such that
\begin{eqnarray}
 \hat f :=  \varkappa^{(0)} \hat f^{(0)} + \varkappa^{(1)} \hat f^{(1)} + \varkappa^{(2)} \hat f^{(2)} \ourdoteq  f_{\mbox{\scriptsize formal}}
 \;.
\end{eqnarray}

Finally we emphasize that the integration constants are $(f^{(0)})_{\lambda}$ in the scalar case, and $(f^{(0)}_i)_{\lambda_i}$ or $(f^{(0)}_i)_{\lambda}$, respectively, in the 2-dimensional
case, so that the corresponding expansion coefficients of  $\hat f$ can be specified arbitrarily.
This will be crucial for the subsequent argument.

\subsection{Approximation of the exact solution}
\label{approx_exact_sol}

In the final step we will show that $\hat f$ approximates the exact solution $f$ to arbitrary high order in $x$; equivalently, they have the same polyhomogeneous expansions. 
We  denote by $c$ (possibly supplemented by some index) a generic positive constant, while $C$ is supposed to be a constant with a specific value.
If angular variables are involved
$c$  and $C$ will still supposed to be constant.
We consider the ODE
\begin{eqnarray}
 x\partial_x f + hf = g
 \;, \quad h=O(1)\in\mathrm{Mat}(n,n)
 \;, \quad 0<x<x_0
 \label{originalODE}
 \;,
\end{eqnarray}
where, as before, the components of  $h$ and $x^{-\ell}g$ are smooth functions on $[0,x_0)$.
Set
\begin{eqnarray}
 \hat g := x\partial_x \hat f + h \hat f
 \;.
\end{eqnarray}
By construction of $\hat f$ the Taylor expansion of $g-\hat g$ at $x=0$ is zero,
\begin{eqnarray}
 &&\delta g := g - \hat g \ourdoteq  0
 \;, \quad \mathrm{i.e.}
\\
 &&\forall \enspace N\in\mathbb{N} \enspace \exists  \enspace c^{(N)}>0 : \|\delta g\| \leq c^{(N)} x^N \enspace \forall \enspace x<x_0
 \;,
\end{eqnarray}
where $\|\cdot \|:=\sqrt{\braket{\cdot,\cdot}}$.
Set $\delta f := f-\hat f$, then
\begin{eqnarray}
 x\partial_x\delta f + h \delta f = \delta g
 \;.
\end{eqnarray}
We want to show that for a given $f$ we can adjust the initial conditions of $\hat f$ (equivalently, of the formal solution), such that $\delta f$ is a solution of the ODE
and satisfies $\|\delta f \|\leq c^{(N)} x ^N$ for all $N$. We find
\begin{eqnarray*}
 |x\partial_x\|\delta f\|^2| &=& 2 |\braket{\delta f,x\partial_x\delta f}| = 2 |\braket{\delta f, \delta g - h\delta f}|
\\
 &\leq& 2 |\braket{\delta f, \delta g}| + 2|\braket{\delta f, h\delta f}| \leq (1+2 \|h\|)\|\delta f\|^2 + \|\delta g\|^2
 \;.
\end{eqnarray*}
Setting $\psi:=\|\delta f\|^2$, we thus have
\begin{eqnarray*}
  \pm x\partial_x \psi &\leq& (1+2\|h\|)\psi + \|\delta g\|^2
\\
 \Longrightarrow \quad \pm\partial_x(\chi^{\pm 1}\psi) &\leq& \frac{\|\delta g\|^2}{x}\chi^{\pm 1}
 \;,
\end{eqnarray*}
where
\begin{equation*}
\chi := \exp\left( \int_x^{x_0} \frac{1 + 2\|h\|}{y}\,\mathrm{d} y \right)
 \;.
\end{equation*}
The function $\chi$ satisfies the inequality (with $h_0\equiv h(0)$)
\begin{eqnarray}
 \chi \leq \exp\Big( \int_x^{x_0} \frac{1 + 2\|h_0\|}{y}\,\mathrm{d} y \Big) \cdot \underbrace{\exp\Big( 2\int_x^{x_0}\frac{\|h-h_0\|}{y}\,\mathrm{d} y}_{=:\hat\chi =\mathcal{O}(1)} \Big)
 = x^{-\mu}x_0^{\mu}\hat\chi
 \;,
 \label{eqn_chi}
\end{eqnarray}
where 
$$
  \mu:= 1+2\|h_0\| \geq 1 \;,
$$
and $\hat\chi>0$ is a smooth function bounded away from zero in $[0,x_0)$,
\begin{eqnarray}
 \hat\chi^{\pm 1} &\leq& c
 \label{eqn_chihat}
\\
 \Longrightarrow \quad \pm\partial_x(\chi^{\pm 1}\psi) &\leq& c \,\|\delta g\|^2 x_0^{-\mu} x^{\mu -1 } \leq c^{(N)}x^N
 \;.
\end{eqnarray}
%
The inequality with the minus sign yields
\begin{eqnarray}
&& \hspace{-8em} (\chi^{- 1}\psi)(x) \leq (\chi^{- 1}\psi)(x_0) + \int_x^{x_0} c^{(N)}y^N\,\mathrm{d}y \leq c
 \nonumber
\\
 &\overset{(\ref{eqn_chi}) \enspace\&\enspace (\ref{eqn_chihat}) \,\text{`+'} }{ \Longrightarrow}  &\psi \leq c \chi \leq c\, x ^{-\mu}
 \nonumber
\\
  &\Longrightarrow & \|\delta f\| \leq c\, x^{-\mu/2}
  \label{inequality_deltaf}
 \;.
\end{eqnarray}
We conclude that
\begin{eqnarray}
 \|x\partial_x \delta f + h_0 \delta f \| = \|\delta g - (h-h_0)\delta f \| \leq c \,x^{-\mu/2+1}
 \;.
\end{eqnarray}
This leads us to the study of the ODE
\begin{eqnarray}
 x\partial_x \tilde f + h_0 \tilde f = \tilde g
\end{eqnarray}
with source $\tilde g$ fulfilling $\|\tilde g\| \leq c\, x^{1-\mu/2 }$.
The general solution to this equation is
\begin{eqnarray}
\tilde f (x) = x^{-h_0}x_0^{h_0}\tilde f(x_0) - x^{-h_0} \int_x^{x_0} y^{h_0-\mathbf{1}}\tilde{g}(y)\,\mathrm{d}y
 \label{solution_ODE}
 \;.
\end{eqnarray}
Here $\mathbf{1}$ denotes the $n$-dimensional identity matrix.
When writing $x^{h_0}$ for a matrix $h_0$ we mean $\exp(h_0\log x)$.
In the following we will distinguish two cases, depending on whether $h_0$ can be diagonalized or not.


\paragraph*{1st case:}
We assume that $h_0$ can be diagonalized.
Clearly this case includes the 1-dim.\ case.
In fact, let us focus for the time being on that case and return to the general case later.
Equation (\ref{solution_ODE}) then implies
\begin{eqnarray*}
 |\delta f| &\leq& c^{(1)}x^{-h_0} + c^{(2)}x^{-h_0} \int_x^{x_0} y^{h_0-1} (|\delta g|+c^{(3)} y |\delta f|) \mathrm{d}y
\\
 &\overset{(\ref{inequality_deltaf})}{\leq}& c^{(1)}x^{-h_0} + c^{(2)}x^{-h_0} \int_x^{x_0} y^{h_0-\mu/2} \mathrm{d}y
 \;.
\end{eqnarray*}
Replacing $\mu$ by a slightly larger number if necessary, we may assume w.l.o.g.\ $h_0-\mu/2 \ne -1$. Then
\begin{eqnarray*}
 |\delta f| \leq c^{(1)}x^{-h_0} + c^{(2)}x^{-h_0} \left[ \frac{y^{h_0-\mu/2+1}}{h_0-\mu/2+1} \right]_x^{x_0}
 =  c^{(1)}x^{-h_0} + c^{(2)}x^{-\mu/2+1}
 \;.
\end{eqnarray*}
Suppose that $\mu/2-1<h_0$, then the inequality $|\delta f| \leq c\, x^{-h_0}$ follows.
If $\mu/2-1>h_0$, we can merely conclude $|\delta f| \leq c\, x^{-\mu/2 +1}$. However, this improves the estimate in (\ref{inequality_deltaf}) by a factor $x$.
Repeating the whole procedure $k$-times until $\mu/2-1-k<h_0$ we finally end up with the estimate
\begin{eqnarray*}
 |x^{h_0}\delta f| &\leq& c \quad \text{for all $x\in(0,x_0)$}
\;,
\end{eqnarray*}
which is 
independent of the  specific relation between $\mu$ and $h_0$.
Since
\begin{eqnarray*}
 |\partial_x(x^{h_0}\delta f)| = x^{h_0-1} |x\partial_x\delta f + h_0\delta f | = x^{h_0-1} |\delta g - (h-h_0)\delta f| \leq c
 \;,
\end{eqnarray*}
$x^{h_0} \delta f$ can be continued to a continuous function on $[0,x_0)$.
Hence, multiplying (\ref{solution_ODE}) (with $\tilde f=\delta f$) by $x^{h_0}$, we observe that $\delta F:= x^{h_0} \delta f$
is continuous even at $x=0$.
Performing the limit $x_0\rightarrow 0$, we find
\begin{eqnarray*}
 \delta F &=& C + \int_0^x y^{h_0-1}(\delta g -(h-h_0)\delta f)\,\mathrm{d}y
\\
 \Longrightarrow \quad \partial_x\delta F &=& x^{h_0-1}\delta g - \frac{h-h_0}{x} x^{h_0} \delta f
 \;,
\end{eqnarray*}
for a suitable constant $C$.
We read off that the function $\delta F$ is in fact  continuously differentiable at $x=0$.
Then, by Taylor's theorem,
\begin{eqnarray*}
 \delta F &=& (\partial_x\delta F)_0 + O(x)= C + O(x)
\\
\Longrightarrow \quad \delta f &=& x^{-h_0} C + O(x^{1-h_0})
 \;.
\end{eqnarray*}
Recall that in the polyhomogeneous expansion of  $\hat f$ the coefficient $\hat f^{(0)}_{\lambda}$, with $\lambda\equiv -h_0$,  can be chosen freely.
We choose it such that $\delta f_{\lambda} \equiv (x^{-\lambda} \delta f)_0$  vanishes, leading to
\begin{eqnarray}
 \delta f &=& x^{-h_0} \int_0^x y^{h_0-1}(\delta g- O(y)\delta f)\,\mathrm{d} y
 \label{deltaf_at_0}
\\
 \Longrightarrow \quad |\delta f| &\leq& c\,x^{-h_0+1}
 \;.
\label{important_estimate}
\end{eqnarray}
%
%
%
Inserting \eq{important_estimate} into \eq{deltaf_at_0}  improves the estimate in each step  by a factor of $x$. Repeating this as many times as necessary
(there is no disturbing term anymore which is proportional to $x^{-h_0}$ and
prohibits the improvement of the estimate), we eventually end up with the desired result,
\begin{eqnarray}
  |\delta f| \leq c^{(N)} x^N \quad \text{for all $N$, i.e.} \quad \delta f \ourdoteq  0
 \;.
\end{eqnarray}

When dealing with higher-dim.\ systems we can proceed in a similar manner (note that we have derived the formal solution only for $\lambda_i\in\mathbb{Z}$).
We give a sketch.
Denote by $-\lambda_1,\dots,-\lambda_n$
the eigenvalues of $h_0$ and assume w.l.o.g.\ $\lambda_1\leq \dots \leq \lambda_n$.
Equation (\ref{solution_ODE}) provides the estimate (for $\tilde g = \delta g-(h-h_0)\delta f$)
%
\begin{eqnarray}
 |\delta f_i| &\leq& c^{(1)} x^{\lambda_i} + c^{(2)} x^{\lambda_i} \int_x^{x_0} y^{-\lambda_i-1}(|\delta g_i| + c^{(3)}y\|\delta f\|)\mathrm{d}y
\;.
\label{estimate_deltaf_i}
\end{eqnarray}
%
%
Note that the integrand depends on $\|\delta f\|$ and not just on $|\delta f_i|$, because the higher-order terms in the expansion of $h$ do not need to be diagonal.

Again we assume without restriction $\mu/2 + \lambda_i\ne 1$.
We conclude from \eq{inequality_deltaf} and \eq{estimate_deltaf_i} that
if $\mu/2-1<-\lambda_1$ then $|\delta f_1| \leq c\, x^{\lambda_1}$. Otherwise $|\delta f_i|\leq c\, x^{-\mu/2+1}$ for all $i$, and thus
$\|\delta f \|\leq c\, x^{-\mu/2+1}$. This can be repeated until $\mu/2-1-k_1 < -\lambda_1$, which means $|\delta f_1| \leq c\, x^{\lambda_1}$.
By adjusting the initial conditions via the integration constant $(\hat f^{(0)}_1)_{\lambda_1}$ appearing in the expansion of $\hat f$ one achieves that the $\lambda_1$-th order term in the polyhomogeneous expansion of $\delta f_1$
vanishes. Then one proceeds in the same way until $\mu/2-1-k_{2}<-\lambda_{2}$. To continue this process the integration constant $(\hat f^{(0)}_{2})_{\lambda_{2}}$ has to be chosen suitably.
And so on. Eventually one obtains $\|\delta f\|\leq c\, x^{\lambda_n+1} $ and  a formula analog to (\ref{deltaf_at_0}) for higher dimensions. One then straightforwardly establishes the desired estimate,
\begin{eqnarray}
  \|\delta f\| \leq c^{(N)} x^N \quad \text{for all $N$, i.e.} \quad \delta f \ourdoteq  0
 \;.
\end{eqnarray}

\paragraph*{2nd case:}
It remains to deal with the case where $h_0$ cannot be diagonalized. For reasons of simplicity we restrict attention again to the two-dimensional case.
The matrix $h_0$ can be brought into Jordan normal form:
\begin{eqnarray}
 h_0 = \begin{pmatrix} -\lambda & 1 \\ 0 & -\lambda \end{pmatrix}
 \;.
\end{eqnarray}
Note that we have derived the formal solution only for $\lambda\in\mathbb{Z}$, which we assume here, as well.
First, we compute $x^{h_0}$,
\begin{eqnarray*}
 x^{h_0} &\equiv& e^{h_0 \log x} \equiv \sum_{m=0}^{\infty} \frac{h_0^m \log^m x}{m!} =
 x^{-\lambda} \begin{pmatrix}  1 & \log x \\ 0 & 1 \end{pmatrix}
 \;.
\end{eqnarray*}
From equation (\ref{solution_ODE}) we find
\begin{equation}
 \delta f (x) = x^{\lambda} x_0^{-\lambda}\begin{pmatrix} 1 & \log(x_0/x) \\ 0 & 1 \end{pmatrix} \delta f(x_0)
 - x^{\lambda} \int_x^{x_0} y^{-\lambda-1}\begin{pmatrix} 1 & \log (y/x) \\ 0 & 1 \end{pmatrix} \tilde{g}(y)\,\mathrm{d}y
 \;.
 \label{solution_ODE2}
\end{equation}
%
The second component of $\delta f$, $\delta f_2$, fulfills the integral equation
\begin{eqnarray}
 \delta f_2 (x) = x^{\lambda} x_0^{-\lambda} \delta f_2(x_0)
 - x^{\lambda} \int_x^{x_0} y^{-\lambda-1} \tilde{g}_2(y)\,\mathrm{d}y
 \;.
\label{expression_second_component}
\end{eqnarray}
Recall that $ \tilde{g}_i \equiv \delta g_i - [(h-h_0)\delta f]_i$.
As before, we may assume w.l.o.g.\ $\mu/2-1 \ne -\lambda$, and deduce
\begin{eqnarray}
 |\delta f_2| &\leq& c^{(1)} x^{\lambda} + c^{(2)} x^{\lambda} \int_x^{x_0} y^{-\lambda-1}(|\delta g_2| + c^{(3)}y\|\delta f\|)\mathrm{d}y
 \nonumber
\\
 &\leq& c^{(1)}x^{\lambda} + c^{(2)} x^{-\mu/2 +1}
 \label{estimate_f2}
\end{eqnarray}
by using (\ref{inequality_deltaf}).
Next, we consider the first component of (\ref{solution_ODE2}), which,
once we have an estimate for $\delta f_2$, supplies one for $\delta f_1$:
\begin{eqnarray}
 \delta f_1(x) &=&  x^{\lambda} x_0^{-\lambda}\left[ \delta f_1(x_0) + \log(x_0/x) \delta f_2(x_0) \right]
 \nonumber
\\
 &&- x^{\lambda} \int_x^{x_0} y^{-\lambda-1}\left[ \tilde{g}_1(y) + \tilde{g}_2(y)\log (y/x)\right] \mathrm{d}y
 \nonumber
\\
 &=& -\delta f_2(x)\log x + x^{\lambda} x_0^{-\lambda}\left[ \delta f_1(x_0) + \log x_0 \delta f_2(x_0) \right]
\nonumber
\\
 &&  - x^{\lambda} \int_x^{x_0} y^{-\lambda-1}\left[ \tilde{g}_1(y) + \tilde{g}_2(y)\log y\right] \mathrm{d}y
 \label{eqn_delta_f1}
 \;.
\end{eqnarray}
That yields (w.l.o.g.\ we assume $x_0<1$)
\begin{eqnarray}
 |\delta f_1(x)| 
 \leq |\delta f_2(x)\log x| + c^{(1)}x^{\lambda} -  c^{(2)} x^{\lambda} \int_x^{x_0}  \|\delta f\| \,y^{-\lambda}\log y\, \mathrm{d}y
 \;.
 \label{estimate_f1}
\end{eqnarray}
The  estimate \eq{inequality_deltaf}
supplemented by the one for $|\delta f_2|$,  (\ref{estimate_f2}), implies
\begin{eqnarray*}
 |\delta f_1(x)| &\leq& |\delta f_2\log x| + c^{(1)}x^{\lambda} +   \Big| x^{\lambda} \Big[ y^{-\lambda-\mu/2 +1} (c^{(2)} + c^{(3)}\log y ) \Big]_x^{x_0} \Big|
\\
 &\leq&  c^{(1)}x^{\lambda}|\log x| +   c^{(2)}  x^{-\mu/2 +1}|\log x|
 \;.
\end{eqnarray*}
If $\mu/2-1 < -\lambda$ we are immediately led to $|\delta f_2(x)| \leq c\, x^{\lambda}$ and $|\delta f_1(x)| \leq c\, x^{\lambda}|\log x|$, whence
$\|\delta f\| \leq cx^{\lambda}|\log x|$.

If the reverse inequality holds, we find
$|\delta f_2(x)| \leq  c\, x^{-\mu/2 +1}$ and
$|\delta f_1(x)| \leq  c\, x^{-\mu/2 +1}|\log x|$.
Combined, that gives $\|\delta f(x)\| \leq  c\, x^{-\mu/2 +1}|\log x|$.
Repeating this procedure $k$-times as long as $\mu/2-k > 1-\lambda$
the  estimates for $\delta f_i$  improve in each round by a factor of $x$, accompanied possibly by the appearance
of higher order powers of $\log x$,
\begin{eqnarray*}
 |\delta f_2(x)| \leq  c\, x^{-\mu/2 +k}|\log x|^{k-1}
 \;, \quad
 |\delta f_1(x)| \leq  c\, x^{-\mu/2 +k}|\log x|^k
 \;,
\end{eqnarray*}
with $k$ a positive integer. Anyway, the log terms do not cause any troubles here, because as soon as $k>\mu/2 + \lambda -1$ becomes true,
the inequalities
\begin{eqnarray}
|\delta f_2|\leq cx^{\lambda} \;, \quad |\delta f_1|\leq cx^{\lambda} |\log x|
\quad  \Longrightarrow \quad \|\delta f(x)\| \leq  c\, x^{\lambda}|\log x|
\label{intermediate_estimates}
\end{eqnarray}
replace the estimates with $\mu$. This is owing to the fact that a term of the form $x^q$, $q>0$, kills any log term, $x^q \log^m x \rightarrow 0$ if $x\rightarrow 0$.

We  proceed in a similar way as above to show that one can improve the estimates by adjusting the integration constants contained in $\hat f$.
Recall that
\begin{equation}
  \delta F_2(x) :=x^{-\lambda}\delta f_2 (x) =  x_0^{-\lambda} \delta f_2(x_0)
 -  \int_x^{x_0} y^{-\lambda-1} \tilde{g}_2(y)\,\mathrm{d}y
 \;.
\end{equation}
From the preceding considerations  we conclude that the integrand is $O(\log y)$.
 Thus we can perform the limit $x\rightarrow 0$. The function $\delta F_2$ is continuous on
$[0,x_0)$ and we rewrite it as
\begin{equation}
 \delta F_2(x) \,=\,  C +  \int^x_0 y^{-\lambda-1} \tilde{g}_2(y)\,\mathrm{d}y
\,=\, C + O(x|\log x|)
 \;.
\label{expression_deltaF2}
\end{equation}
%
By choosing the value of $(\hat f^{(0)}_2)_{\lambda}$ appropriately (this coefficient was free to choose in our analysis above) one achieves that
$C$ vanishes,
\begin{eqnarray}
 |\delta f_2| \leq c\, x^{\lambda+1}|\log x|
\;.
\label{improved_est_deltaf2}
\end{eqnarray}
%
Combined with $|\delta f_1|\leq c x^{\lambda}|\log x|$, this inequality can be used to improve the estimate for $|\delta f_1|$: Starting from (\ref{estimate_f1})
one establishes
\begin{equation}
 |\delta f_1| \leq c x^{\lambda} \quad \overset{\eq{improved_est_deltaf2}}{\Longrightarrow} \quad \|\delta f\| \leq  c x^{\lambda}
 \;.
\end{equation}
%

It follows  from \eq{eqn_delta_f1} that there exists a constant $C$ with
\begin{equation}
 \lim_{x\rightarrow 0} x^{-\lambda}\delta f_1(x) = C
 \;,
\end{equation}
so one can perform the limit $x_0\rightarrow 0$,
\begin{eqnarray}
\hspace{-3em}
 x^{-\lambda}\delta f_1(x) &=& C -x^{-\lambda}\delta f_2(x)\log x
  +  \int_0^{x} y^{-\lambda-1}\left[ \tilde{g}_1(y) + \tilde{g}_2(y)\log y\right] \mathrm{d}y
\\
&=& C + O(x|\log x|)
\nonumber
 \;.
\end{eqnarray}
We choose the (yet unspecified) value  $(\hat f^{(0)}_1)_{\lambda}$ in the expansion of $\hat f_1$ such that $C$ vanishes. Then
\begin{equation}
 | \delta f_1| \,\leq\, c x^{\lambda+1}|\log x|
\;,
\end{equation}
and \eq{expression_deltaF2} yields
\begin{equation}
   | \delta f_2| \,\leq\, c x^{\lambda+1}
\;,
\end{equation}
i.e.\ we have improved \eq{intermediate_estimates} by a factor of $x$.



If we continue this process, an analysis of (cf.\ \eq{expression_second_component} and \eq{eqn_delta_f1}),
\begin{eqnarray*}
 \delta f_2 (x) &=&x^{\lambda} \int_0^x y^{-\lambda-1} \tilde{g}_2(y)\,\mathrm{d}y
\;,
\\
 \delta f_1(x) &=&  -\delta f_2(x)\log x + x^{\lambda} \int_0^x y^{-\lambda-1}\left[ \tilde{g}_1(y) + \tilde{g}_2(y)\log y\right] \mathrm{d}y
\end{eqnarray*}
reveals that the estimates  improve in each step by a factor of $x$, 
and, since  the log term in $\delta f_1$ does not matter in the end, we arrive at
\begin{eqnarray}
  |\delta f_i| \leq c^{(N)} x^N \quad \text{for all $N$} \quad \Longleftrightarrow \quad \|\delta f\| \leq c^{(N)} x^N \quad \text{for all $N$}
 \;.
\end{eqnarray}

Summarizing, we have proved:

\begin{theorem}
 \label{t21VI11.1}
 Consider  the linear  ODE
\begin{equation*}
 x\partial_x f + hf = g \quad \text{on} \quad (0,x_0) 
 \;,
\end{equation*}
where $x^{-\ell} g=\Oinfty (1)$, $\ell \in\mathbb{Z}$, and $h=\Oinfty (1)$ are assumed to be smooth maps on $[0,x_0)$, and where $f$and $g$ have values in $\R$ or $\R^2$,
with $h$ having values in the space of corresponding linear maps.
Let $\hat f$ be a solution of the ODE specified  by  boundary conditions, that is by a choice
of the expansion coefficients  $\hat f_{\lambda}$ in the scalar case, and $(\hat f_i)_{\lambda_i}$ or $(\hat f_i)_{\lambda}$, respectively, in the 2-dimensional
case.
Denote by $g_{\text{formal}}$ and $h_{\text{formal}}$ the Taylor expansions of $g$ and $h$, respectively, at $x=0$.
Then there exists a formal solution $f_{\text{formal}}$ of
\begin{equation*}
 x\partial_x f + h_{\text{formal}} f = g_{\text{formal}}
 \;,
\end{equation*}
such that $f_{\text{formal}}$ is the polyhomogeneous expansion of $\hat f$ at $x=0$,
\[
 \hat f \ourdoteq  f_{\text{formal}}
 \;.
\]
\end{theorem}
We have further indicated that the theorem remains true in arbitrary dimensions if $h_0$ is a diagonal matrix with integer entries.

\section{Relation between $\kappa=0$- and $\kappa=\frac{r}{2}|\sigma|^2$-gauge}

\label{meaning_kappa}
\label{Coordinates on the Cone}

The metric gauge turned out to be very convenient to construct  data for the~CFE which are smooth at $\scrip$:
Global existence and positivity of $\varphi$ and $\nu^0$ are 
implicitly  contained in the gauge condition, while the no-logs-condition
reduces to a simple algebraic condition on the expansion coefficients of $\gamma$.

Using an affine parameterization, i.e.\ a $\kappa=0$-gauge, we have to \textit{assume} that the initial data $[\gamma]$  are chosen in such a way that the function $\varphi$ is positive  on $\mcN$ and that $\varphi_{-1}$ is positive on $S^2$.
These assumptions \emph{do} impose geometric restrictions on  $\gamma$ in that they exclude data producing conjugate
points or even space-time singularities on the initial surface.
On the other hand, the gauge choice $\kappa=\frac{r}{2}|\sigma|^2$, on a light-cone say,  \textit{implies} that the Raychaudhuri equation admits a global
solution with all the required properties without any additional assumptions.
To resolve this  ``paradox''
it is  useful to understand in which way a $\kappa=\frac{r}{2}|\sigma|^2$-gauge is related to
other choices of the $r$-coordinate parameterizing the null rays generating the cone,
such as affine parameterizations.
A similar problem arises for $\nu^0$, and can be resolved in the same way.

If the gauge function $\kappa$ depends  on the initial data, the physical/geometrical interpretation of the parameter $r$ (its deviation from an affine one) w.r.t.\ which the $\gamma$ is given, depends on  $\gamma$ itself.
Due to this ``implicit definition'' of $r$, the choice  $\kappa=\frac{r}{2}|\sigma|^2$
conceals
geometric restrictions, which we want to discuss~now.

We  consider initial data $\mathring\gamma$ given in a $\mathring\kappa$-gauge on an initial surface $\mcN$, which we assume for definiteness to be a light-cone,
 and analyze under which conditions
a transformation to a $\kappa=\frac{r}{2}|\sigma|^2$-gauge is possible, by which we mean
that both solutions  differ by a coordinate change only. 
The relevant  coordinate transformations $(\mathring r,\mathring x^A)\mapsto ( r, x^A)$ are
angle-dependent transformations of the $r$-coordinate
\begin{equation*}
  r =  r(\mathring r,\mathring x^A)\;, \quad  x^A = \mathring x^A\;.
\end{equation*}
It follows from the transformation behavior of connection coefficients that
\begin{eqnarray}
 \kappa  =\ol \Gamma^r_{rr}  =  \underbrace{\ol \Gamma^{\mathring r}_{\mathring r\mathring r}}_{=\mathring\kappa}\frac{\partial \mathring r}{\partial r} - \left(\frac{\partial \mathring r}{\partial r}\right)^2 \frac{\partial^2 r}{\partial \mathring r^2}
\;.
 \label{tildekappa}
\end{eqnarray}
The function $|\sigma|^2$ which contains partial derivatives of $r$ (cf.\  \eq{formula_sigma})
transforms~as
\begin{equation}
 |\sigma|^2(r) = \left(\frac{\partial \mathring r}{\partial   r}\right)^2 |\mathring  \sigma|^2(\mathring r( r))
 \;.
\end{equation}
With  $\kappa=\frac{r}{2}|\sigma|^2$ \eq{tildekappa} becomes
\begin{eqnarray*}
 \frac{\partial^2 r}{\partial \mathring r^2}  - \mathring \kappa\frac{\partial  r}{\partial \mathring r} + \frac{1}{2} r |\mathring\sigma|^2
 \,=\, 0
 \;.
\end{eqnarray*}
We observe that $r(\mathring r,\mathring x^A)$ and $\mathring\varphi(\mathring r,\mathring x^A)$ satisfy the same ODE.
Imposing the boundary conditions
$r|_{\mathring r=0}=0$ and $\partial_{\mathring r} r|_{\mathring r=0}=1$,
we conclude that
\begin{equation}
  r(\mathring r,\mathring x^A)=\mathring\varphi(\mathring r,\mathring x^A)
\;.
\label{relation_phi_r}
\end{equation}
Since
$ g_{AB}(r) = \mathring  g_{AB}( r(\mathring r))$  the function $\varphi\equiv(\frac{\det \check g_{\Sigma_r}}{\det s})^{1/4}$ transforms as a scalar,~and
%
\begin{equation}
  r=\mathring\varphi(\mathring r(r,x^A), x^A) = \varphi(r,x^A)
\;,
\end{equation}
as expected and required by the metric gauge.

To transform from a $\mathring\kappa$-gauge to a $\kappa=\frac{r}{2}|\sigma|^2$-gauge one simply identifies
$\mathring\varphi$ as the new $r$-coordinate.
However, this is only possible when  $(\mathring r,\mathring x^A)\mapsto ( r=\mathring\varphi, x^A=\mathring x^A)$  defines a diffeomorphism.
Globally this happens if and only if  $\mathring\varphi$ is a strictly increasing function, which is equivalent to
the existence of a global solution to the Raychaudhuri equation.
Another requirement on $r$ should be that
\begin{equation*}
 \lim_{\mathring r \rightarrow \infty}  r = \infty \quad \Longleftrightarrow \quad \lim_{\mathring r \rightarrow \infty}\mathring  \varphi = \infty \quad \Longleftrightarrow \quad \mathring \varphi_{-1}>0
\;.
\end{equation*}

This derivation clarifies in which way  the assumptions on $\mathring\varphi$ in say a $\mathring\kappa=0$-gauge enter:
Prescribing smooth data 
in a $\kappa=\frac{r}{2}|\sigma|^2$-gauge one implicitly excludes data violating these assumptions and thereby the existence of conjugate points up-to-and-including conformal infinity.
In this work, though, we are only interested in those cases where $\mathring \varphi$ \emph{is} strictly increasing with  $\mathring\varphi_{-1}>0$, in which case  a transition from a $\mathring\kappa=0$- to a $\kappa=\frac{r}{2}|\sigma|^2$-gauge  is  always possible.


Let us   take a look at the reversed direction.
It follows from \eq{tildekappa} and the requirement $\partial_r\mathring r|_{r=0}=1$ that
\begin{eqnarray*}
  \frac{r}{2}|\sigma|^2=  - \left(\frac{\partial \mathring r}{\partial r}\right)^2 \frac{\partial^2 r}{\partial \mathring r^2}
 =    \left(\frac{\partial \mathring r}{\partial r}\right)^{-1} \frac{\partial^2 \mathring r}{\partial  r^2}
\quad
\Longleftrightarrow \quad
    \frac{\partial \mathring r}{\partial r} = e^{\int_0^r \frac{\hat r}{2}|\sigma|^2 \mathrm{d}\hat r}>0
\;,
\end{eqnarray*}
which defines a  diffeomorphism.
Consequently,  a transformation from  a $\kappa=\frac{r}{2}|\sigma|^2$-gauge  to  a $\mathring\kappa=0$-gauge  is  possible
without any restrictions.

We conclude that the metric gauge is a reasonable and convenient gauge condition whenever the light-cone is supposed to be globally smooth.


\begin{thebibliography}{[10]}


\bibitem{ChoquetBruhat73}
Y.~Choquet-Bruhat, \emph{Un th\'eor\`eme d'instabilit\'e pour certaines
  \'equations hyperboliques non lin\'eares}, C.R. Acad. Sci. Paris \textbf{276}
  (1973), 281--284.


 \bibitem{CCM2} Y. Choquet-Bruhat, P.T. Chru\'sciel, and J.M. Mart\'in-Garc\'ia, \textit{The Cauchy problem on a characteristic cone for the Einstein equations in arbitrary dimensions},  Ann. Henri Poincar\'e \textbf{12} (2011), 419--482.
\bibitem{C1}  P.T. Chru\'sciel, \textit{The existence theorem for the general relativistic Cauchy problem on the light-cone}, (2012), arXiv:1209.1971 [gr-qc].
 \bibitem{ChJezierskiCIVP} P.T. Chru\'sciel, and J. Jezierski, \textit{On free general relativistic initial data on the light cone},
J. Geom. Phys. \textbf{62} (2012) 578-–593.
  \bibitem{bondi} P.T. Chru\'sciel, and M.A.H. MacCallum, D. Singleton: \textit{Gravitational Waves in General Relativity. XIV: Bondi Expansions and the ``Polyhomogeneity'' of Scri},  Phil. Trans. Royal Soc. of London A \textbf{350}, 113--141 (1995).
\bibitem{ChPaetz} P.T. Chru\'{s}ciel and T.-T. Paetz, \emph{{The many ways of the characteristic Cauchy problem}}, Class.\ Quantum Grav. \textbf{29} (2012), 145006.
\bibitem{ChPaetz2} P.T. Chru\'{s}ciel, and T.-T. Paetz, \emph{Characteristic initial data and smoothness of Scri. I. Framework and results}, preprint (2014).
 \bibitem{F3} H.\ Friedrich, \textit{Conformal Einstein Evolution}, in: \textit{The Conformal Structure of Space-Time -- Geometry, Analysis, Numerics}, J. Frauendiener, H. Friedrich (eds.), Berlin, Heidelberg, Springer, 2002, pp.\ 1--50.
 \bibitem{JJpeeling}  J. Jezierski, \textit{`Peeling property' for linearized gravity in null coordinates}, Class. Quantum Grav. \textbf{19} (2002), 2463–-2490.
 \bibitem{Rendall}  A.D. Rendall, \textit{Reduction of the characteristic initial value problem to the Cauchy problem and its applications to the Einstein equations},
                 Proc. Roy. Soc. London A \textbf{427} (1990), 221–-239.
 \bibitem{Tambourino66} L.A. Tambourino, and J.H. Winicour, \textit{Gravitational fields in finite and conformal Bondi frames}, Phys. Rev. \textbf{150}, 1039--1053 (1966).
\bibitem{TorrenceCouch}
R.J. Torrence, and W.E. Couch, \emph{Generating fields from data on
  ${\mcH}^-\cup{\mcH}^+$ and ${\mcH}^-\cup{\scri}^-$}, Gen.\ Rel.\ Grav.
  \textbf{16} (1984), 847--866.
\end{thebibliography}
\end{document}